\RequirePackage{fix-cm}

\documentclass[twocolumn,final,natbib]{svjour3}          

\smartqed  

\usepackage{geometry}
\usepackage{graphicx}
\usepackage{stfloats}

\graphicspath{{./Figures_papers/}}
\usepackage[font=normalsize]{caption}
\usepackage[font=normalsize]{subcaption}

\usepackage{enumitem}

\usepackage{color}
\usepackage{hyphenat}

\usepackage[utf8x]{inputenc} 

\usepackage[fleqn]{mathtools}
\usepackage{amsmath}
\usepackage{amssymb}
\usepackage{amsbsy}
\usepackage{mathrsfs}
\usepackage{bm}
\usepackage{relsize}
\usepackage{array}

\usepackage[flushmargin]{footmisc} 

\usepackage[misc]{ifsym} 

\usepackage[title]{appendix}

\usepackage{listings}

\usepackage[table]{xcolor}
\usepackage{booktabs}
\newcolumntype{C}[1]{>{\centering\arraybackslash}m{#1}<{}} 

\usepackage{empheq}

\usepackage{tikz}
\usetikzlibrary{shapes,arrows,chains}
\usetikzlibrary{patterns}
\usepackage{pgfplots}
\pgfplotsset{compat=1.17}

\usepackage{floatrow} 
\usepackage{multirow}
\usepackage{cuted}

\newcommand{\bx}{{\bf x}}
\newcommand{\bbx}{{(\bf x)}}
\newcommand{\bxhat}{{\hat{\bf x}}}
\newcommand{\bbxhat}{{(\hat{\bf x})}}

\newcommand{\bchi}{{(\chi)}}

\newcommand{\desvar}{{{\zeta}}}

\newcommand{\domega}{{ \, d\Omega }}
\newcommand{\dgamma}{{ \, d\Gamma }}
\newcommand{\mcolon}{{ \, , }}
\newcommand{\mdot}{{ \, . }}

\newcommand{\methodabb}{VARTOP}
\newcommand{\methodname}{Variational Topology Optimization}

\usepackage{widetext}

\setcitestyle{numbers,square,comma,sort&compress}
\usepackage[colorlinks,pdfpagelabels,bookmarksopen,bookmarksnumbered,citecolor=blue,urlcolor=blue,linkcolor=blue]{hyperref}

\journalname{Archives of Computational Methods in Engineering}

\usepackage[scale=5,opacity=0.75,color=gray]{background}
\backgroundsetup{contents={\sffamily Accepted manuscript}}

\begin{document}

\title{Topology optimization methods for 3D structural problems: a comparative study}

\titlerunning{Topology optimization methods for 3D structural problems: a comparative study}        

\author{Daniel$\ $Yago$^{1,2}$         \and
       Juan$\ $Cante$^{1,2}$           \and
       Oriol$\ $Lloberas-Valls$^{2,3}$ \and 
       Javier$\ $Oliver$^{2,3}$
       }

\authorrunning{Daniel Yago         \and
       		   Juan Cante           \and
       		   Oriol Lloberas-Valls  \and
       		   Javier Oliver
       		   } 

\institute{\begin{itemize}
           \item[\Letter] J. Oliver \at
                      	  \email{oliver@cimne.upc.edu} \\
           \item[1] Escola Superior d'Enginyeries Industrial, Aeroespacial i Audiovisual de Terrassa (ESEIAAT)\\
                         Technical University of Catalonia (UPC/Barcelona Tech), Campus Terrassa UPC, c/ Colom 11, 08222 Terrassa, Spain
           \item[2]      Centre Internacional de M\`{e}todes Num\`{e}rics en Enginyeria (CIMNE)\\ 
                         Campus Nord UPC, M\`{o}dul C-1 101, c/ Jordi Girona 1-3, 08034 Barcelona, Spain
           \item[3]     E.T.S d'Enginyers de Camins, Canals i Ports de Barcelona (ETSECCPB)\\
                         Technical University of Catalonia (UPC/Barcelona Tech), Campus Nord UPC, M\`{o}dul C-1, c/ Jordi Girona 1-3, 08034 Barcelona, Spain                   
           \end{itemize}	  	             	                  
}

\date{Received: 22nd April 2021 / Accepted: 25th June 2021 }

\date{}
\def\makeheadbox{\relax}
\journalname{}

\maketitle


\begin{abstract}
	
	The work provides an exhaustive comparison of some representative families of topology optimization methods for 3D structural optimization, such as the \emph{Solid Isotropic Material with Penalization} (SIMP), the \emph{Level-set}, the \emph{Bidirectional Evolutionary Structural Optimization} (BESO), and the \emph{\methodname{}} (\methodabb{}) methods. The main differences and similarities of these approaches are then highlighted from an algorithmic standpoint. 
	
	The comparison is carried out via the study of a set of numerical benchmark cases using industrial-like fine-discretization meshes (around 1 million finite elements), and Matlab as the common computational platform, to ensure fair comparisons. Then, the results obtained for every benchmark case with the different methods are compared in terms of  computational cost, topology quality, achieved minimum value of the objective function, and robustness of the computations (convergence in objective function and topology). Finally, some quantitative and qualitative results are presented, from which, an attempt of qualification of the methods, in terms of their relative performance, is done. 
	
	
\end{abstract}

\section{Introduction} \label{sec_Introduction}

In the past three decades, topology optimization has become an active research field to seek new optimal counterintuitive designs in a wide range of problems governed by different physics, i.e., solid mechanics \cite{Yulin2004,Wang2003,Allaire2004,Bruns2001,Wang2004b,Coelho2007}, fluid dynamics \cite{Borrvall2002,GersborgHansen2005,Guest2006}, thermal dynamics \cite{Li1999,Gao2008,Yamada2011}, acoustics \cite{Sigmund2003,Du2007,Dong2017,Li2016,Lu2013,Roca2019} and electromagnetism \cite{Huang2012,Zhou2010,Zhou2011}, among others. Furthermore, topology optimization of coupled multiphysics problems has been addressed in recent works, combining structural-thermal interaction \cite{Sigmund1997a,Sigmund2001a,Deng2012}, structural-fluid interaction \cite{Yoon2007,Maute2004,Andreasen2013,Jenkins2015} or even thermal-fluid interaction \cite{Alexandersen2016,Bruns2007,Yaji2015,Marck2013}. As a result of this substantial effort, the optimal design obtained from the minimization of a given topology optimization problem can be used by engineers as a first approximation in the development of new products in a wide range of applications. 

All up-to-date approaches exhibit certain strengths and weaknesses. As a first approximation\footnote{For further discussion about the classification, the reader is referred to \cite{Rozvany2008,Eschenauer2001,Deaton2013,Sigmund2013,Dijk2013,Munk2015}.}, optimization techniques can be grouped into two main blocks: (I) \emph{methods based on trial-and-error schemes}, e.g.,~\emph{Genetic Algorithms} or \emph{Ant Colony Algorithms} \cite{Hajela1993,Adeli1994,Chapman1994,Hare2013,Wang2005,Luh2009}, and (II) \emph{methods relying on the gradient computation} \cite{Bendsoe1989,Xie1993,Osher2001,Allaire2004,Allaire2005,Bourdin2003,Wang2004a}. The main disadvantage of the former group is their extremely high computational cost as the number of unknowns increases. This computational cost may become prohibitive for current computational systems since thousands of different layouts must be tested to find the optimal configuration. Consequently, the algorithms included in the second set are the most widespread algorithms, e.g.,~(a) \emph{topology optimization within homogenization theory} \cite{Bendsoe1988}, (b) \emph{density-based optimization} (SIMP) techniques \cite{Bendsoe1989,Mlejnek1992,Bendsoe2004}, (c) \emph{evolutionary methodologies} (ESO)\footnote{The \emph{evolutionary approaches} appear also in the literature as \emph{Sequential Element Rejection and Admission} (SERA), and henceforth they should be considered as synonyms.} \cite{Xie1997,Yang1999}, (d) \emph{Level-set approaches} \cite{Allaire2002,Allaire2004,Wang2003}, (e) \emph{Topological Derivative method} \cite{Sokolowski1999}, (f) \emph{Phase field approach} \cite{Bourdin2003,Wang2004,Takezawa2010}, and (g) \emph{\methodname{} (\methodabb{}) method} \cite{Oliver2019,Yago2020}, among others. 

Starting with the seminal paper of \citet{Bendsoe1988}, numerical methods for topology optimization have been extensively developed. In particular, this article was the basis for a stream of \emph{density-based approaches}, such as the \emph{Solid Isotropic Material with Penalization} (\emph{SIMP}) method \cite{Bendsoe1989,Mlejnek1992,Bendsoe2004}, being nowadays one of the most widely used topology optimization methods. Simple continuous \emph{element-wise design variables} are used in the formulation and resolution of the \emph{topology optimization problem} in a fixed design domain, for which Young's modulus is defined as a polynomial function of the \emph{element-wise density}, $\rho_e\in\left(0,1\right)$. The \emph{design variable} must be first penalized\footnote{Although existing other interpolation schemes, such as the \emph{Rational Approximation of Material Properties} (RAMP) material interpolation \cite{Stolpe2001} or the SINH method \cite{Bruns2005}, the SIMP scheme \cite{Bendsoe1989,Zhou1991} is the most popular one in structural optimization due to its simplicity. However, these alternative schemes may be of interest for topology optimization problems involving alternative physics, e.g.,~in dynamic problems.} (normally a penalty exponent $p\geq3$ is used) and later regularized, thus providing almost \emph{black-and-white solutions} (with semi-dense elements) not ensuring manufacturability. On the other hand, a large number of regularization schemes have been suggested to be used regarding topology optimization, including: (1) \emph{filtering}, via the classical sensitivity \cite{Sigmund1994,Sigmund1997,Sigmund1998} or density \cite{Bourdin2001,Bruns2001} filters, projection techniques \cite{Guest2004,Guest2011}, morphology-based filters \cite{Sigmund2007,Wang2010} or Helmholtz-type filters \cite{Lazarov2010,Kawamoto2010}, among others, and (2) \emph{geometric constraint techniques}, e.g.,~perimeter constraint \cite{Haber1996,Fernandes1999} or gradient constraints \cite{Petersson1998}. Through these mathematical techniques, significant numerical instabilities \cite{Sigmund1998,Jog1996} resulting from the ill-posedness of unconstrained topology optimization problems for continuous structures, including gray areas (semi-dense intermediate elements), checkerboard patterns, and mesh-dependency issues are alleviated. Finally, the solution to the topology optimization problem is obtained via the update of the \emph{design variable}, through the \emph{classical optimality criteria (OC) method} \cite{Bendsoe1988,Suzuki1991}\footnote{Other heuristic techniques have been proposed to tackle non-convex topology optimization problems, such as the \emph{modified optimality criteria technique} (MOC) \cite{Ma1993}.}, the \emph{moving asymptotes (MMA) algorithm} \cite{Svanberg1987} or other\emph{ mathematical programming-based optimization algorithms}, among others. 

Besides the \emph{SIMP method}, the \emph{Evolutionary Structural Optimization} (ESO) method, firstly introduced by \citet{Xie1997,Xie1993}  although similar ideas were presented earlier \cite{Schnack1988,MATTHECK1990}, is also one of the most used for industrial applications. In the recent years, the ESO approach has gained widespread popularity due its simplicity and ease of implementation in commercial FE codes. \emph{ESO}, considered as a \emph{hard-kill method}\footnote{The removed elements are not included in the subsequent finite element analysis. Consequently, no criterion function is computed for the void elements.}, relies on a simple \emph{heuristic criterion} to gradually remove inefficient material. The elements with low \emph{rejection criterion} are gradually removed starting from a full stiff design domain, thus evolving towards an optimum. Contrary to \emph{SIMP}, a \emph{discrete element design variable}, $\chi\in\{0,1\}$, is used to define the topology layouts, which are free of gray elements. This change in design variable results in convergence issues and a high dependency on the initial configuration, thus leading to local optimal solutions \cite{Zhou2001}. Despite these numerical issues, \emph{ESO} has been applied to a large range of problems, from the well-known structural problems \cite{Xie1994,Xie1994a,Zhao1997,Chu1996}, including non-linear problems \cite{Querin1996,Manickarajah1998}, to thermal problems \cite{Li2000,Li1999,Li2001a}, and contact problems \cite{Li1998,Li2005}. In addition, the above mentioned issues are mitigated in its later version, the \emph{Bi-directional Evolutionary Structural Optimization} (BESO) method \cite{Querin1998,Yang1999,Querin2000}, which extends the approach to allow for new elements to be added, while inefficient elements are removed at the same time. The new material is added either in the locations near to those elements with a \emph{high criterion function value} \cite{Querin1998,Querin2000} or in those void areas with higher criterion function values, computed via a linear interpolation of the displacement field \cite{Yang1999}. However, these extrapolation techniques are not consistent with those used for the solid elements. As any other \emph{density-based method}, the sensitivity is commonly regularized via a \emph{checkerboard suppression filter} \cite{Li2001}, a \emph{mesh-independent filter} \cite{Huang2007} or a \emph{perimeter control} \cite{Yang2002}, similar to the ones used in \emph{SIMP}, in order to reduce mesh-dependencies. Nevertheless, such \emph{hard-kill BESO methods} fail to obtain convergent solutions. For that reason, in later revisions, \citet{Huang2007} proposed a \emph{modified hard-kill BESO} to enhance the time-convergence via a \emph{stabilization algorithm}, where the historical information is used, in addition to a new mesh-independent filter and nodal sensitivities. As an alternative to \emph{hard-kill ESO/BESO methods}, \emph{soft-kill approaches} retain the void elements as very soft elements, thus allowing the computation of the \emph{criterion function} within the entire domain. \citet{Zhu2006} and \citet{Huang2008} proposed independent approaches that used a \emph{penalized density variable}. In particular, \citet{Huang2008} combined the \emph{modified BESO method with SIMP material interpolation scheme}, improving the numerical stability and the potential of the methodology. An excellent overview of some recent developments in topology optimization using \emph{ESO} is provided in \cite{Huang2010}.

The last major stream is constituted by \emph{Level-set-based methods}. In contrast to the previous topology optimization approaches, the optimal layout is implicitly defined by a scalar function $\phi$ (with the sign). Additionally, the structural boundary of the design $\Gamma$ is represented by the \emph{zero-level iso-contour} (or iso-surface) of the \emph{level-set function} (LSF) \cite{Osher1988,Sethian1999,Osher2003}. As a result, optimal designs with sharp and smooth edges are obtained, thus avoiding \emph{semi-dense (gray)} elements, like those observed in \emph{density-based methods}. Many formulations of \emph{Level-set-based approaches} have been proposed over the years since \citet{Haber1998} suggested its applications with topology optimization techniques, e.g.,~\cite{Sethian2000,Allaire2002,Wang2003,Allaire2005}. The most important ones relying on a \emph{level-set function} are the \emph{Level-set} (based on shape derivative), the \emph{Topological Derivative}, and the \emph{Phase-field methods}.

Regarding the original \emph{Level-set}, \citet{Osher2001}, \citet{Allaire2002,Allaire2004} and \citet{Wang2003} took up the idea of using the \emph{level-set function} for topology optimization and combined it with a \emph{shape-derivative-based} topology optimization framework. Therefore, only material boundaries are altered via shape-sensitivity analysis\footnote{The basis of this approach lies in computing the sensitivity of the functional when a normal infinitesimal deformation is applied on the boundaries of the domain \cite{Sokolowski1992}.} to seek the optimal design. As a consequence, this set of techniques can not nucleate new holes in the interior of the domain. Therefore, the resulting optimal solutions are heavily dependent on the initial layouts, which must be made up of many small holes evenly distributed throughout the domain \cite{Allaire2004}. This initial configuration ensures the merging and cancellation of holes via the propagation of the boundaries, while avoiding local optimal solutions. To overcome this limitation, a hole nucleation algorithm \cite{Eschenauer1994,Schumacher1996}, referenced as the \emph{bubble technique}, was introduced in the \emph{topology optimization approach}. This technique allows the creation of new holes in the material domain. Instead of the density variable used in \emph{density-based methods} to define the topology design, the \emph{level-set function} is used here, with the material domain being those points where the LSF is positive. The geometry in a fixed mesh is typically mapped to a mechanical model using either \emph{immersed boundary techniques} (e.g.,~via X-FEM \cite{Duysinx2006}) or \emph{density-based mapping} (e.g.,~via relaxed Heaviside functions \cite{Wang2003}) \cite{Wein2020}. When using the second mapping, the stiffness coefficients are expressed in terms of the \emph{level-set function} via approximated Heaviside functions, i.e., the \emph{ersatz material approximation} is used \cite{Allaire1997,Dambrine2009}, where the void material is replaced with a soft material. In addition, the LSF is commonly updated using a \emph{Hamilton-Jacobi (HJ) equation} \cite{Osher2001,Wang2003,Allaire2004}\footnote{However, other updating schemes have been proposed over the years, such as the resolution of a set of ordinary differential equations in the approaches using \emph{radial basis functions} (RBFs) \cite{Wang2006,Wang2007} or a \emph{system of algebraic equations using compactly supported RBFs} (CSRBFs) \cite{Luo2007,Luo2008}, among others.}, thus requiring the solution of a \emph{pseudo-time PDE equation}. Although this updating scheme tends to converge to smooth topologies, it may require a huge number of iterations\footnote{When an explicit scheme is performed, the time-step is limited by the CFL condition.}, as well as the regular application of reinitialization algorithms to a signed-distance function. These reinitializations must be applied each time there are significant shape-changes or after a hole nucleation process \cite{Allaire2004,Wang2003}, thus reducing the efficiency of the approach. Nevertheless, it has been extensively used for a broad range of design problems, including structural problems \cite{Allaire2005a}, vibration problems \cite{Osher2001,Allaire2005a}, thermal problems \cite{Ha2005}, among others. As an alternative to \emph{HJ equations}, updating procedures based on mathematical programming are used, e.g.,~using the parameters of the discretized \emph{level-set function} as optimization variables (for instance \emph{radial basis functions} (RBF) and \emph{spectral methods}) \cite{Wang2006,Luo2007}. Finally, as described for \emph{density-based approaches}, the topology optimization problem must be regularized to ensure mesh-independence and improve convergence. It can be achieved either by a filtering procedure or a constraint equation, e.g.,~perimeter constraint \cite{Allaire2004,MichaelYuWang2004}.

Alternatively to \emph{Level-set} using the shape-derivative framework, and after the mathematical development of the \emph{topological derivatives} \cite{Sokolowski1999,Cea2000,Garreau2001}, some researchers incorporated the concept of the \emph{topological derivative} into a \emph{shape-sensitivity-based Level-set method}, thus leading to the \emph{Topological Derivative approach} \cite{Burger2004,Allaire2005}. A similar algorithm as in classical \emph{Level-set} is performed. However, in contrast to those prior methods, this topology optimization technique can nucleate holes in the interior of the material domain by using the \emph{topological gradient} or \emph{topological derivative}. The sensitivity is defined as the variation of the objective function due to the insertion of an infinitesimal spherical void at any point $\bx$ in the design domain $\Omega$, thus avoiding the stagnation in local optimal solutions. Nevertheless, the \emph{topological gradient} must be analytically derived through rather complex mathematics for each of the topology optimization problems, e.g.,~structural linear optimization \cite{Garreau2001,Giusti2016,Novotny2007}, thermal orthotropic optimization \cite{Giusti2009,Marczak2007}, microstructure topology optimization \cite{Barbarosie2009,Amstutz2010}, among others. This mathematical operator represents an extra step required to proceed with the optimization, burdening its potential against other techniques with sensitivities easier to compute. The \emph{topological derivative} was first incorporated in conjunction with shape-derivative as a way to systematically nucleate holes \cite{Burger2004,Allaire2005,Challis2009,Fulmanski2007}, similar as in \emph{Level-set} with the \emph{bubble technique}. In later revisions, the \emph{topological derivative} was used exclusively to update the \emph{level-set function} \cite{Cea2000,Norato2007}. Until then, only stiff material could be removed from the material domain, making it impossible to add new material. It was not until \citet{Amstutz2006} and \citet{He2007} that fully bi-directional \emph{Topological Derivative approaches} were introduced. In these techniques, the optimal layout, expressed in terms of the LSF, is defined as a function of the \emph{topological gradient}. To improve stability, reaction and diffusive terms can be added to the classical \emph{HJ-equation}, leading to the so-called \emph{Generalized HJ-equation}. The diffusive term smooths out the design and suppresses sharp corners, avoiding the ill-posedness of the topology optimization problem.

The last group of interest is the \emph{Phase field topology optimization approach}, where the theory of phase transitions is adapted to the resolution of topology optimization problems \cite{Allen1979,Cahn1958,Eyre1993}. The design variable corresponds to the \emph{density}, as other \emph{density-based approaches}, but, in this case, a linear material interpolation is considered, without any exponent factor. In addition, an extra term is added to the objective function that controls the interface thickness while penalizing intermediate values, thus solving one of the main disadvantages of \emph{SIMP-like approaches}. The optimal solutions present smooth material domains, almost \emph{black-and-white designs} separated by sharp thin finite thickness interfaces. The modified functional is minimized based on the \emph{Cahn-Hilliard equation}, leading to the resolution of two coupled second-order equations without requiring a volume constraint, i.e., the volume stays constant through the optimization procedure.  However, some researchers have solved directly the modified topology optimization problem with the inclusion of a volume constraint \cite{Bourdin2003,Wang2004a,Burger2006,Wang2004}, resembling \emph{SIMP} with an explicit penalization in the density and a gradient regularization. The gradient regularization results in a smoothing similar to that obtained by the inclusion of diffusive terms in the \emph{level-set-based methods}\footnote{Alternatively, \citet{Takezawa2010} used a \emph{time-dependent reaction-diffusion equation} instead, called the \emph{Allen-Cahn equation}, to evolve the phase function.}. Connected with this concept, \citet{Yamada2010} suggested a \emph{Phase field approach} based on a \emph{level-set function}, used as the design variable, and a \emph{topological derivative} incorporating a \textit{fictitious interface energy}. This last mathematical technique allows to control the complexity of the optimal layout. Although being applied to other problems \cite{Yamada2011,Lim2011}, it still resorts to a \emph{Hamilton-Jacobi equation} to update the topology design, which may entail high computation resources to achieve convergence.

As an alternative to all these well-established techniques, the \emph{\methodname{}} (\methodabb{})\footnote{The method abbreviation \emph{UNVARTOP} used in previous papers has been here rebranded as \emph{\methodabb{}}, which should be taken as equals.} approach \cite{Oliver2019} combines the mathematical simplicity of \emph{SIMP-based methods}, while considering the \emph{characteristic function} $\chi$ as the design variable. Thus a binary configuration (\emph{black-and-white} design) is obtained. The domain, and so the \emph{characteristic function}, is implicitly represented through a \emph{0-level-set function}, termed as \emph{discrimination function}, as in \emph{level-set-based methods}. Nevertheless, the topology design is not updated neither via a \emph{Hamilton-Jacobi} equation nor a \emph{Reaction-Diffusion} equation, but via a \emph{fixed-point, non-linear, closed-form algebraic system} resulting from the derivation of the topology optimization problem. In addition, an \emph{approximated topological derivative}, in contrast to the \emph{exact Topological Derivative methods}, is used in the formulation within an \emph{ersatz material approach}, highly reducing the mathematical complexity independent of the tackled problem. The topology optimization problem is subjected to a volume constraint expressed in terms of a \emph{pseudo-time variable}. This constraint equation is iteratively increased until the desired volume is achieved, thus obtaining converged topologies for intermediate volumes.
By means of this procedure, referred to as \emph{time-advancing scheme}, the corresponding \emph{Pareto Frontier} is obtained. For each time-step, the \emph{closed-form optimality criteria} has to be solved to compute both the \emph{Lagrange multiplier} that fulfills the volume constraint and the optimal \emph{characteristic function}. As for the regularization, a \emph{Laplacian regularization}, similar to those used in SIMP and Phase-field approaches \cite{Bourdin2003,Lazarov2010,Kawamoto2010,Yamada2010}, is applied to the \emph{discrimination function}, providing not only smoothness in the optimal design but also mesh-size control. The technique has been already applied to linear static structural \cite{Oliver2019} and steady-state thermal \cite{Yago2020a} applications, considering the volume constraint as a single constraint equation, with promising results. 

In the literature, there are plenty of articles that either theoretically compare several topology optimization methods as the ones presented above \cite{Deaton2013,Eschenauer2001,Munk2015,Rozvany2008,Rozvany2001,Sigmund2013,Dijk2013}, or compare the results obtained with the topology optimization approach proposed by the corresponding authors with those computed using a recognized topology optimization approach. However, not many articles compare in a practical way a set of cases with a wide range of techniques under the same convergence criteria. This is one of the most relevant aspects of this work since the results of a set of widely used techniques are compared with each other: (1) \emph{SIMP} (briefly described in Section \ref{sec_theroy_SIMP}), (2) \emph{BESO using a soft-kill criterion} (detailed in Section \ref{sec_theroy_SOFTBESO}), (3) \emph{\methodabb{}} using a pure variational topological approach (presented in Section \ref{sec_theroy_UNVARTOP}), and (4) \emph{Level-set} (detailed in section \ref{sec_theroy_LevelSet}). These three well-known methods have been selected among all existing ones due to their wide use both at the professional and research level, as well as for the convenience of implementation and their documentation, thus facilitating their verification and assuring a fair comparison. Following the implementations proposed by the different methods' authors, the studied topology optimization techniques have been implemented including as few modifications as possible in order to match the original approaches. The modifications are detailed throughout the document for each of the addressed methods. Although the chosen topology optimization techniques have been applied to a wide spectrum of different applications, the comparison in this article focuses at \emph{minimizing the static structural problem}. The comparison of the results is addressed through a set of well-known benchmark cases, whose optimal layouts are easily recognized. Specifically, \emph{minimum mean compliance}, \emph{multi-load mean compliance}, and \emph{compliant mechanism topology optimization problems} are carried out.

The remainder of this paper is organized as follows. The structural static problem as well as the three addressed topology optimization problems are defined in Section \ref{sec_theroy}, while in Section \ref{sec_theroy_methods}, the considered approaches are reviewed in terms of their formulation and algorithms. In addition, specific comments are provided to address the studied topology optimization problems with each of the techniques. These techniques are compared with each other in terms of the objective function, the quality of topology design, and the computational cost via a set of benchmark cases detailed in Section \ref{sec_benchmark} and analyzed in Section \ref{sec_comparison_results}.

\section{Theoretical aspects} \label{sec_theroy}

	\subsection{Domain definition} \label{sec_domain_def}
	
\begin{figure*}[!h]
		\centering
		\includegraphics[width=14cm]{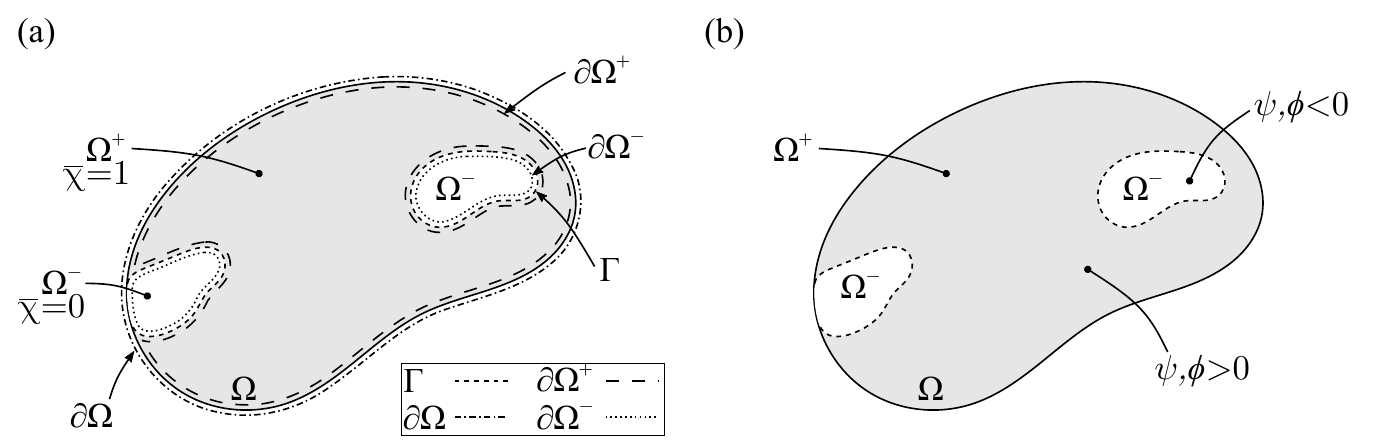}
		\caption{Domain representation: (a) Representation of the design domain, $\Omega$, comprising two disjoint sub-domains $\Omega^+$ and $\Omega^-$ and (b) Implicit representation via the \emph{level-set function} $\phi$ or the \emph{discrimination function} $\psi$. }
		\label{fig_design_domain}
	\end{figure*}

	Let the design domain, $\Omega$, denote a fixed smooth open domain of $\mathbb{R}^n$ for $n=\{2,3\}$, composed by two smooth subdomains $\Omega^+,\Omega ^-\subset\Omega$, with $\overline{\Omega}^+\cup\overline{\Omega}^-=\overline{\Omega}$ and $\Omega^+\cap\Omega^-=\emptyset$, as displayed in Figure \ref{fig_design_domain}-(a) \footnote{$\overline{(\cdot)}$ corresponds to the closure of the open domain $(\cdot)$.}. The boundary of the design domain, termed as $\partial\Omega$, is also composed of the boundaries corresponding to the two subdomains $\partial\Omega^+$ and $\partial\Omega^-$, satisfying $\partial\Omega^+\cap\partial\Omega^-=\Gamma$. The material domain, $\Omega^+$, consists of a stiff material with a high Young's modulus, while the second subdomain, $\Omega^-$, is formed by a soft material with a low Young's modulus. The stiffness ratio between both materials is given by the \emph{contrast factor}, $\alpha\ll1$.
	
	The topology layout of the design domain can be defined via a \emph{characteristic function} $\overline{\chi}\bbx:\Omega\rightarrow\{0,1\}$ as
	\begin{equation} \label{eq_characteristic_function_general}
		\left\{
		\begin{split}
			&\Omega^{+}\coloneqq\{\mathbf{x}\in \Omega \;/\; \overline{\chi}\bbx=1\} \\
			&\Omega^{-}\coloneqq\{\mathbf{x}\in \Omega \;/\; \overline{\chi}\bbx=0\} 
		\end{split}
		\right. \mcolon
	\end{equation}
	where $\overline{\chi}$ corresponds to the \emph{Heaviside function} of $(\rho-\overline{\rho})$ in \emph{density-based approaches} (\emph{SIMP}), the \emph{Heaviside function} of the \emph{level-set function} $\phi$ in \emph{Level-set-based methods}, and the \emph{characteristic function} $\chi$ itself in \emph{VARTOP}. Notice that the term $\overline{\rho}$ must be computed in \emph{density-based methods} so that the constraint equation is satisfied (normally the volume), thus obtaining a \emph{white-and-black design}.

	In particular, for \emph{Level-set-based} and \emph{VARTOP approaches}, the subdomains can be defined through a continuous function, $\varphi\bbx : \Omega \rightarrow \mathbb{R}, \varphi \in H^1(\Omega)$ (a \emph{level-set function} $\phi$ or a \emph{discrimination function} $\psi$, respectively) such that
	\begin{equation} \label{eq_domain_splitting}
		\left\{
		\begin{split}
			&\Omega^{+}\coloneqq\{\mathbf{x}\in \Omega \;/\; \varphi\bbx>0\} \\
			&\Omega^{-}\coloneqq\{\mathbf{x}\in \Omega \;/\; \varphi\bbx<0\} 
		\end{split}
		\right. \mcolon
	\end{equation}
	as illustrated in Figure {\ref{fig_design_domain}}-(b). The \emph{characteristic function} $\overline{\chi}$ can be then obtained as $\overline{\chi}\bbx={\cal H}(\varphi\bbx)$, where ${\cal H}(\cdot)$ stands for the \emph{Heaviside function}.

	\subsection{The Topology Optimization problem. Contextual introduction} \label{sec_theroy_problems}
	
		\begin{figure*}[pb]
			\centering
			\includegraphics[width=14cm, height=4.5cm]{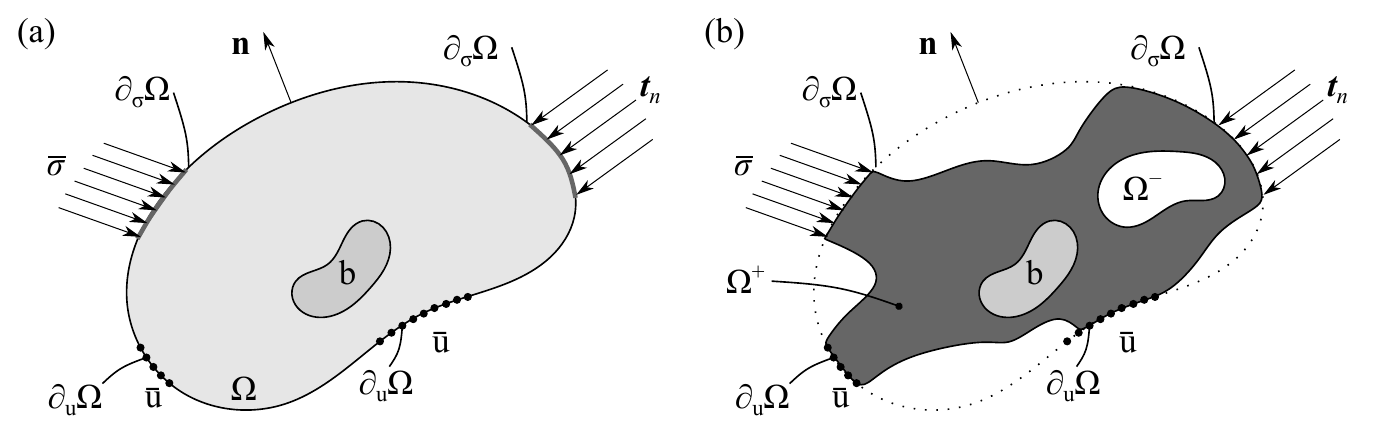}
			\caption{Elastic problem sketch: (a) fixed analysis domain $\Omega$ with boundary conditions (in which the displacement $\overline{\mathbf{u}}\bbx$ or the normal traction $\bm{t}_n\bbx$ can be prescribed at $\partial_{u}\Omega$ and $\partial_\sigma\Omega$, respectively) and (b) Stiff and soft material domains, $\Omega^+$ and $\Omega^-$, respectively, with the same boundary conditions.}
			\label{fig_elastic_problem}
		\end{figure*}
	
		\emph{Topology optimization methods} look for the optimal material distribution that minimizes a given \emph{objective function}, $\cal{J}$, subjected to one or more \emph{constraints} ${\cal{C}}_k$ (e.g.,~a volume constraint, ${\cal{C}}_0\le 0$, and possibly other $N$ design variable constraints, ${\cal{C}}_k\le 0, \; k:1\dots N$) and governed by a linear or non-linear state equation. The material distribution is described by the \emph{density variable} $\rho\bbx$, the \emph{characteristic function} $\chi\bbx$ or the \emph{level-set function} $\psi\bbx$, depending on the topology approach used to carry out the optimization. To keep the definition of the topology optimization problem as general as possible, let us define $\desvar\bbx$ as the \emph{design variable} at point $\bx$, which will be considered as $\rho\bbx$, $\chi\bbx$ or $\phi\bbx$ for the \emph{density-based}, \emph{\methodabb{}}, and \emph{level-set-based methods}, respectively. Based on this concept, the classical mathematical formulation of the corresponding \emph{topology optimization problem} is given by
		\begin{align} \label{eq_minimization_restricted}
			\left[\begin{aligned}
				&\underset{\desvar\in{\mathscr{U}}_{ad}} {\operatorname{min}}\;{\cal J}\left({\mathbf{u}}(\desvar),\desvar\right)\equiv\int_{\Omega}{j({\mathbf{u}}(\desvar),\desvar,\bx)}\domega &&(a) \\
				&\text{subject to:} &\\
				&\hspace{0.5cm}{\cal C}_0(\desvar)\equiv\int_{\Omega}{c_0(\desvar,\bx)\domega}\le0 &&(b-1) \\
				&\hspace{0.5cm}{\cal C}_k(\desvar)\le0, \;\;  k:1\dots N &&(b-2) \\
				&\text{governed by:} &\\
				&\hspace{0.5cm}{\text{State equation} } &&(c) 
			\end{aligned}\right. \mcolon
		\end{align}
		where the \emph{objective function} $\cal{J}$ can be expressed as a volume integral of a local function $j({\mathbf{u}}(\desvar),\desvar,\bx)$ over the entire domain, and the constraint functional ${\cal C}_0$ represents the \emph{volume constraint} in terms of the design variable $\desvar$. Additional \emph{constraint equations} ${\cal C}_k(\desvar)$ can be incorporated into the topology optimization problem to explicitly include constraints to the design variables, particular to each approach.

		The \emph{state equation} gives as a solution the unknown field $\mathbf{u}(\desvar)$ for a specific optimal design ${\desvar}$ included in the admissible set of solutions, ${\mathscr{U}}_{ad}$. This unknown field must satisfy the boundary conditions applied to the design domain. In particular, the linear elasticity equilibrium problem, formulated as 
		\begin{empheq}[left=\empheqlbrack,right=\hspace{-0.2cm}]{align}
			&\text{Find ${\mathbf{u}}(\desvar,\bx)$ such that} \notag \\
			&\hspace{0.25cm}\left\{\begin{aligned}
				&\bm{\nabla}\cdot\bm{\sigma}(\desvar,\bx)+\mathbf{b}(\desvar,\bx)= \mathbf{0} \ &&in \; \Omega\\
				&\bm{\sigma}(\desvar,\bx)\cdot\mathbf{n}={{\bm{t}}_n}\bbx \ &&on \; \partial_{\sigma}\Omega\\
				&\mathbf{u}(\desvar,\bx)=\overline{\mathbf{u}}\bbx \ &&on \; \partial_{u}\Omega\\
			 \end{aligned} \right. \mcolon \label{eq_strong_elastic_problem}
		\end{empheq}		
		is considered as the state equation for all the topology optimization problems addressed in this paper. In the preceding equation, $\bm{\sigma}(\desvar,\bx)$ and $\mathbf{b}(\desvar,\bx)$ stand for the \emph{second-order stress tensor field} and the \emph{volumetric force}, respectively, which both depend on the topology layouts. Additionally, ${{\bm{t}}_n}\bbx$ and $\overline{\mathbf{u}}\bbx$ are respectively the boundary tractions applied on $\partial_{\sigma}\Omega\subset\partial\Omega$ and the displacements prescribed on $\partial_{u}\Omega\subset\partial\Omega$, and $\mathbf{n}$ corresponds to the unit outward normal. As for the material behavior, the elastic material is governed by the Hooke's law, i.e., $\pmb{\sigma}=\mathbb{C}_\desvar:\pmb{\varepsilon}$, with $\pmb{\varepsilon}$ being the \emph{strain tensor} ($\pmb{\varepsilon} = {\bm{\nabla}^S \mathbf{u}}_{\desvar}\bbx$) and $\mathbb{C}_\desvar$ being the \emph{fourth-order, elastic constitutive tensor}. The \emph{constitutive tensor} depends on the \emph{design variable} $\desvar$ via the corresponding material interpolation of each topology optimization approach. 
		
		As depicted in Figure \ref{fig_elastic_problem}, the boundary $\partial\Omega$ of the analysis domain $\Omega$ is made of two mutually disjoint subsets, $\partial_u\Omega$ and $\partial_\sigma \Omega$, where the displacements and tractions are prescribed, respectively, as detailed in the previous equation.
		
		Alternatively, the variational form of the \emph{linear elasticity problem} (\ref{eq_strong_elastic_problem}) becomes
		\begin{empheq}[left=\empheqlbrack,right=\hspace{-0.2cm}]{align}
			&\text{Find the displacement field  ${\mathbf{u}}_\desvar\in{\cal{U}}(\Omega)$ such that} \notag \\
			&\hspace{0.25cm} a(\mathbf{w},\mathbf{u}_\desvar) = l(\mathbf{w}) \quad \forall \mathbf{w}\in {\cal V}(\Omega) \label{eq_weak_problem2}\\
   			&\text{where} \notag \\
			&\hspace{0.25cm}a(\mathbf{w},\mathbf{u}_{\desvar}) = \int_{\Omega}{\bm{\nabla}^S \mathbf{w}\bbx:\mathbb{C}_{\desvar}\bbx:\bm{\nabla}^S\mathbf{u}_\desvar\bbx}\domega\mcolon \label{eq_lhs_structural_problem} \\
			&\hspace{0.25cm}\begin{aligned}
				l(\mathbf{w}) = &\int_{\partial_{\sigma}\Omega}{\mathbf{w}\bbx\cdot{\bm{t}_n}\bbx}\dgamma \\ 
								&+ \int_{\Omega}{\mathbf{w}\bbx \cdot \mathbf{b}_\desvar\bbx}\domega \mcolon
			\end{aligned} \label{eq_rhs_structural_problem}
		\end{empheq}
		with $\mathbf{u}_\desvar$ and $\mathbf{w}$ being the \emph{displacement field} and the \emph{virtual displacement field}, respectively.

		The \emph{linear elasticity problem} (equations (\ref{eq_weak_problem2}) to (\ref{eq_rhs_structural_problem})), discretized using the standard \emph{finite element method}, reads
		\vspace{-\baselineskip}
		\begin{empheq}[left=\empheqlbrack,right=\hspace{-0.2cm}]{align}
			&{\mathbb K}_{\desvar}\hat{\mathbf{u}}_{\desvar}=\mathbf{f}_\desvar \label{eq_equilibrium} \\
			&\text{with} \notag \\
			&\hspace{0.25cm}{\mathbb K}_{\desvar}=\int_{\Omega}\mathbf{B}^{\text{T}}\bbx\ {\mathbb{C}}_{\desvar}\bbx\ \mathbf{B}\bbx \domega\mcolon \label{eq_equilibrium_stiff_force_K} \\
			&\hspace{0.25cm}\begin{aligned}
				\mathbf{f}_\desvar= &\int_{\partial_{\sigma}\Omega}\mathbf{N_{u}}^{\text{T}}\bbx{\bm{t}_n}\bbx\dgamma \\
									&+ \int_{\Omega} \mathbf{N_{u}}^{\text{T}}\bbx \mathbf{b}_{\desvar}\bbx\domega \mcolon \end{aligned} \label{eq_equilibrium_stiff_force_f}
		\end{empheq}
		where the stiffness matrix and the external force vector are denoted by ${\mathbb K}_{\desvar}$ and $\mathbf{f}$, respectively. For the sake of clarity, the dependence of the force vector on the design variable will be neglected, considering a constant force scenario $\mathbf{f}$, independent of the topology layout, with null \emph{volumetric forces}. The displacement field, $\mathbf{u}_{\desvar}\bbx$, and its gradients, ${\bm{\nabla}^S \mathbf{u}}_{\desvar}$, are approximated as follows
		\begin{align}
			&\mathbf{u}_{\desvar}\bbx\equiv\mathbf{N}_{u}\bbx\hat{\mathbf{u}}_\desvar \label{eq_shape_disp}\\
			&{\bm{\nabla}^S \mathbf{u}}_{\desvar}\bbx\equiv\mathbf{B}\bbx\hat{\mathbf{u}}_\desvar \label{eq_grad_disp}
		\end{align}
		where $\mathbf{N}_{u}\bbx$ and $\mathbf{B}\bbx$ stand for the \emph{displacement, shape function matrix} and the \emph{strain-displacement matrix}, respectively, and $\hat{\mathbf{u}}_\desvar$ corresponds to the \emph{nodal displacement vector}. Notice that the dependence on the design variable $\desvar$ is highlighted by the subscript $(\cdot)_\desvar$.
		
		\subsubsection{Minimum mean compliance} \label{sec_theroy_prob_compliance}
		
			The \emph{minimum mean compliance topology optimization problem} seeks the optimal topology layout that minimizes the global stiffness of the structure, or equivalently, maximizes the external work on the structure. The objective function ${\cal J}^{(I)}(\mathbf{u}_\desvar,\desvar)$, in variational form, is given by
			\begin{equation} \label{eq_compliance_cost_function}
				{\cal J}^{(I)}\left({\mathbf{u}}(\desvar),\desvar\right)\equiv l(\mathbf{u}_{\desvar}) \mcolon
			\end{equation}
			and the corresponding discretized \emph{topology optimization problem} can be written as
			
			\par \vspace{3\baselineskip} %
			\begin{widetext}
			\begin{equation} \label{eq_minimization_restricted_compliance}
				\left[\begin{split}
					&\underset{\desvar\in{\mathscr{U}}_{ad}} {\operatorname{min}}\;{\cal J}^{(I)}\left({\mathbf{u}}(\desvar),\desvar\right)\equiv \int_{\Omega} \hat{\mathbf{u}}_{\desvar}^{\text{T}} \mathbf{B}^{\text{T}}\bbx\ {\mathbb{C}}_{\desvar}\bbx\ \mathbf{B}\bbx \hat{\mathbf{u}}_{\desvar} \domega \hspace{5.3cm} &&(a) \\
					&\text{subject to:} \\
					&\hspace{0.95cm}{\cal C}_0(\desvar)\equiv\int_{\Omega}{c_0(\desvar,\bx)\domega}\le0 &&(b-1) \\
					&\hspace{0.95cm}{\cal C}_k(\desvar)\le0, \quad k:1\dots N  &&(b-2) \\
					&\text{governed by:} \\
					&\hspace{0.95cm}{\mathbb K}_{\desvar}\hat{\mathbf{u}}_{\desvar}=\mathbf{f} \text{, with } \ {\mathbb K}_{\desvar}=\int_{\Omega}\mathbf{B}^{\text{T}}\bbx\ {\mathbb{C}}_{\desvar}\bbx\  \mathbf{B}\bbx \domega \text{ and } \mathbf{f}=\int_{\partial_{\sigma}\Omega}\mathbf{N_{u}}^{\text{T}}\bbx{\bm{t}_n}\bbx\dgamma &&(c) 
				\end{split}\right. \mcolon
			\end{equation}
			\end{widetext}
			with $j({\mathbf{u}}(\desvar),\desvar,\bx)$ being ${\bm{\nabla}^S\mathbf{u}_{\desvar}\bbx :{\mathbb{C}}_{\desvar}\bbx:\bm{\nabla}^S\mathbf{u}_{\desvar}}\bbx$. The \emph{mean compliance} can be also defined as $\mathbf{f}^{\text{T}}\hat{\mathbf{u}}_{\desvar}$, when nodal variables are used.
			
		\subsubsection{Multi-load mean compliance} \label{sec_theroy_prob_multicompliance}
		
			\emph{Multi-load compliance topology optimization problems} are a subfamily of \emph{minimum mean compliance problems} (Section \ref{sec_theroy_prob_compliance}), in which a set of elastic problems with different loading conditions are solved independently, instead of a single one with all the external loads applied at the same time. As a result, the topology optimization procedure aims at a trade-off between the optimal topology layouts for each specific loading state. Hence, the objective function ({\ref{eq_minimization_restricted}}-a) is computed as the weighted average sum of all individual compliances, i.e.,
			\begin{align}
				\begin{aligned}
					&{\cal J}^{(II)}\left({\mathbf{u}}(\desvar),\desvar\right) \equiv \sum_{i=1}^{n_l}  l\left(\mathbf{u}_{\desvar}^{(i)}\right) \\
					&\equiv \sum_{i=1}^{n_l}  \int_{\Omega} \bm{\nabla}^S\mathbf{u}_{\desvar}^{(i)} \bbx :{\mathbb{C}}_{\desvar}\bbx:\bm{\nabla}^S\mathbf{u}_{\desvar}^{(i)}\bbx\domega \mcolon
				\end{aligned}
			\end{align} 
			where $i$ and $n_l$ corresponds to the index of the \emph{i-th loading state}\footnote{The displacements $\mathbf{u}_{\desvar}$ and the actual energy density ${{\cal U}_\desvar }$ for the i-th loading state are designated with the superscript $(\cdot)^{(i)}$.} and the \emph{number of loading states}, respectively. Consequently, the \emph{topology optimization problem} (\ref{eq_minimization_restricted_compliance}) becomes
			\begin{widetext}
			\begin{equation} \label{eq_minimization_restricted_multi}
				\left[\begin{split}
					&\underset{\desvar\in{\mathscr{U}}_{ad}} {\operatorname{min}}\;{\cal J}^{(II)}\left({\mathbf{u}}(\desvar),\desvar\right)\equiv \sum_{i=1}^{n_l} \int_{\Omega}{
					\hat{\mathbf{u}}_{\desvar}^{{(i)}\text{T}} \mathbf{B}^{\text{T}}\bbx\ {\mathbb{C}}_{\desvar}\bbx\ \mathbf{B}\bbx \hat{\mathbf{u}}_{\desvar}^{(i)}
					}\domega = \sum_{i=1}^{n_l} \mathbf{f}^{{(i)}\text{T}}\hat{\mathbf{u}}_{\desvar}^{(i)} \hspace{2cm} &&(a) \\
					&\text{subject to:} \\
					&\hspace{0.95cm}{\cal C}_0(\desvar)\equiv\int_{\Omega}{c_0(\desvar,\bx)\domega}\le0 &&(b-1) \\
					&\hspace{0.95cm}{\cal C}_k(\desvar)\le0, \quad k:1\dots N   &&(b-2) \\
					&\text{governed by:} \\
					&\hspace{0.95cm}{\mathbb K}_{\desvar}\hat{\mathbf{u}}_{\desvar}^{(i)}=\mathbf{f}^{(i)} \text{, with } 
					\begin{aligned}
						{\mathbb K}_{\desvar}&=\int_{\Omega}\mathbf{B}^{\text{T}}\bbx\ {\mathbb{C}}_{\desvar}\bbx\ \mathbf{B}\bbx \domega \\ \mathbf{f}^{(i)}&=\int_{\partial_{\sigma}\Omega^{(i)}}{\hspace{-10pt}\mathbf{N_{u}}^{\text{T}}\bbx{\bm{t}_n}^{(i)}\bbx\dgamma}, \quad i:1\dots n_l
					\end{aligned}  &&(c) 
				\end{split}\right. \mdot
			\end{equation}
			\end{widetext}
			Notice that $n_l$ independent \emph{linear elastic problems} must be solved in order to obtain the displacement field $\hat{\mathbf{u}}_{\desvar}^{(i)}$ for each of the loading states.

		\subsubsection{Compliant mechanism synthesis} \label{sec_theroy_prob_compliantmech}
		
			Contrary to the two previous sections where the main goal was to \emph{maximize the mean stiffness} of the structure, the objective now is to design a flexible structure capable of transferring an action (force or displacement) from the \emph{input port} to the \emph{output port}, obtaining a desired force or displacement at that port. The corresponding objective function ${\cal J}^{(III)}\left({\mathbf{u}}(\desvar),\desvar\right)$ can be expressed as 
			\begin{equation} \label{eq_compliant_mechanism_cost_function}
				{\cal J}^{(III)}\left({\mathbf{u}}(\desvar),\desvar\right) \equiv -l_2\left(\mathbf{u}_{\desvar}^{(1)}\right)  \mcolon
			\end{equation}
			where $l_2\left(\mathbf{u}_{\desvar}^{(1)}\right)$ corresponds to the rhs term (\ref{eq_rhs_structural_problem}) of the elastic problem (\ref{eq_weak_problem2}) for ${\mathbf w} = \mathbf{u}_{\desvar}^{(1)}$ with the boundary traction ${\bm{t}_n^{(2)}}\bbx$ being a \emph{dummy constant force value} applied only at the \emph{output port} in the desired direction. 
			
			Mimicking the procedure detailed for the other two topology optimization problems in Sections \ref{sec_theroy_prob_compliance} and \ref{sec_theroy_prob_multicompliance}, the new mathematical problem is given by
			
			\clearpage
			\begin{widetext}
			\begin{equation} \label{eq_minimization_restricted_compliant_mechanism}
				\left[\begin{split}
					&\underset{\desvar\in{\mathscr{U}}_{ad}} {\operatorname{min}}\;{\cal J}^{(III)}\left({\mathbf{u}}(\desvar),\desvar\right)\equiv -\int_{\Omega}{
					\hat{\mathbf{u}}_{\desvar}^{{(1)}\text{T}} \mathbf{B}^{\text{T}}\bbx\ {\mathbb{C}}_{\desvar}\bbx\ \mathbf{B}\bbx \hat{\mathbf{u}}_{\desvar}^{(2)}}\domega = - \mathbf{1}^{\text{T}}\hat{\mathbf{u}}_{\desvar}^{(1)} \hspace{2.5cm} &&(a) \\
					&\text{subject to:} \\
					&\hspace{0.95cm}{\cal C}_0(\desvar)\equiv\int_{\Omega}{c_0(\desvar,\bx)\domega}\le0 &&(b-1) \\
					&\hspace{0.95cm}{\cal C}_k(\desvar)\le0, \quad k:1\dots N &&(b-2) \\
					&\text{governed by:} \\
					&\hspace{0.95cm}\begin{split}
						{\mathbb K}_{\desvar}\hat{\mathbf{u}}_{\desvar}^{(1)}=\mathbf{f}^{(1)}& \text{, with }  \mathbf{f}^{(1)}=\int_{\partial_{\sigma}\Omega^{(1)}}\mathbf{N_{u}}^{\text{T}}\bbx{\bm{t}_n}^{(1)}\bbx\dgamma \\
						{\mathbb K}_{\desvar}\hat{\mathbf{u}}_{\desvar}^{(2)}=\mathbf{f}^{(2)}& \text{, with }  \mathbf{f}^{(2)}= \mathbf{1} = \int_{\partial_{\sigma}\Omega^{(2)}}\mathbf{N_{u}}^{\text{T}}\bbx{\bm{t}_n}^{(2)}\bbx\dgamma
					\end{split}   &&(c) 
				\end{split}\right. \mdot
			\end{equation}
			\end{widetext}
			
			As anticipated in expression (\ref{eq_compliant_mechanism_cost_function}), the \emph{compliant mechanism problem} (\ref{eq_minimization_restricted_compliant_mechanism}) is not self-adjoint, so an \emph{auxiliary state problem} must be solved in addition to the \emph{original state problem}. The additional system presents the same stiffness matrix ${\mathbb K}_{\desvar}$ but a different force vector ${\mathbf f}^{(2)}$, consisting in a \emph{dummy constant force} at the output port, which solution is denoted as $\mathbf{u}_{\desvar}^{(2)}$.  Additional springs, denoted by $K_{in}$ and $K_{out}$, must be considered in the \emph{input and output ports}, respectively, to ensure convergence.
				
	\subsection{General algorithm}
	
		The flowchart of the general algorithm, used to obtain the optimal topology layouts, is illustrated in Figure \ref{fig_algorithm}. Note that each technique will present variations of this optimization algorithm, which will be specified in the corresponding sections (Sections \ref{sec_theroy_SIMP} to \ref{sec_theroy_LevelSet}). Nevertheless, the methods addressed in this paper exhibit a similar updating scheme.  
		
		The main part of the algorithm consists in solving the \emph{state equation} (\ref{eq_equilibrium}) to obtain the \emph{unknown field} $\mathbf{u}(\desvar)$, and computing the corresponding \emph{sensitivities} along with the \emph{objective function value} (equations (\ref{eq_minimization_restricted_compliance})-a, (\ref{eq_minimization_restricted_multi})-a or (\ref{eq_minimization_restricted_compliant_mechanism})-a). After computing the topological derivatives of the objective function, some type of \emph{regularization} must be applied to them (e.g.,~sensitivity filtering and/or temporal regularization) to improve numerical stability and ensure convergence. The topology $\desvar$ is then updated accordingly to each approach following a \emph{optimality criterion} so that the objective function is minimized. For a given volume constraint, topology and objective function convergence is sought, thus obtaining the optimal topology design.
		
		Depending on the topology optimization method, a set of intermediate optimal topologies is obtained when using \emph{time-advancing schemes}, while a single optimal design is only obtained (the last one) for \emph{single-time-step methods} (see Section \ref{sec_Introduction}). For the first family of approaches, the algorithm previously described must be repeated for each time-step until the convergence criteria are met, thus obtaining a set of converged solutions over the \emph{Pareto Frontier} of optimal solutions between the \emph{objective function} and the \emph{volume constraint}.

		\colorlet{lcnorm}{black}
		\begin{figure*}[t]
			\centering
			\begin{tikzpicture}[%
			    >=triangle 60,              
			    start chain=going below,    
			    node distance=5mm and 60mm, 
			    every join/.style={norm},   
			    ]
				\tikzset{
				  base/.style={draw, on chain, on grid, align=center, minimum height=1ex},
				  proc/.style={base, rectangle, text width=15em},
				  test/.style={base, diamond, aspect=2, text width=5em},
				  term/.style={proc, rounded corners},
				  coord/.style={coordinate, on chain, on grid, node distance=4mm and 35mm},
				  norm/.style={->, draw, lcnorm},
				  it/.style={font={\small\itshape}}
				}
				\node [term]       	(term1)		{Start Topology Optimization};
				\node [proc, join,left=of term1,xshift=-2cm] 	(p1)  	{Pre-processing and data initialization}; 
				\node [proc, join] 	(p2)  	{Optimize next time-step (increase reference volume)};
				\node [proc, join] 	(p3)  	{Solve equilibrium equations (FEM), compute sensitivity and compute objective function};
				\node [proc, join] 	(p4)  	{Apply filtering technique}; 
				\node [proc, join] 	(p7)  	{Update topology $\desvar$}; 
				\node [test, join] 	(t1) 	{Convergence?}; 
				\node [test] 	    (t2) 	{Last time-step?};
				\node [term,right=of t2,xshift=2cm] 		(p9)    {Post-processing and Exit}; 
				\node [coord, right=of p3] 	(c1)  {}; 
				\node [coord, right=of t1] 	(c2)  {}; 
				\node [coord, left=of t2] 	(c3)  {}; 
				\node [coord, left=of p2] 	(c4)  {}; 
				\path (t1.south) to node [near start, xshift=6.8em] {$yes$, Optimal topology layout } (t2);
		  		\draw [*->,lcnorm] (t1.south) -- (t2);
				\path (t2.east) to node [near start, xshift=1em,yshift=1em] {$yes$} (p9.west); 
		  		\draw [*->,lcnorm] (t2.east) -- (p9.west);
				\path (t2.west) to node [yshift=1em] {$no$} (c3); 
		  		\draw [*->,lcnorm] (t2.west) -- (c3) |- (p2);
				\path (t1.east) to node [yshift=1em] {$no$} (c2); 
		  		\draw [*->,lcnorm] (t1.east) -- (c2) |- (p3);
			\end{tikzpicture}
			\caption{The general flowchart for topology optimization approaches.}
			\label{fig_algorithm}
		\end{figure*}
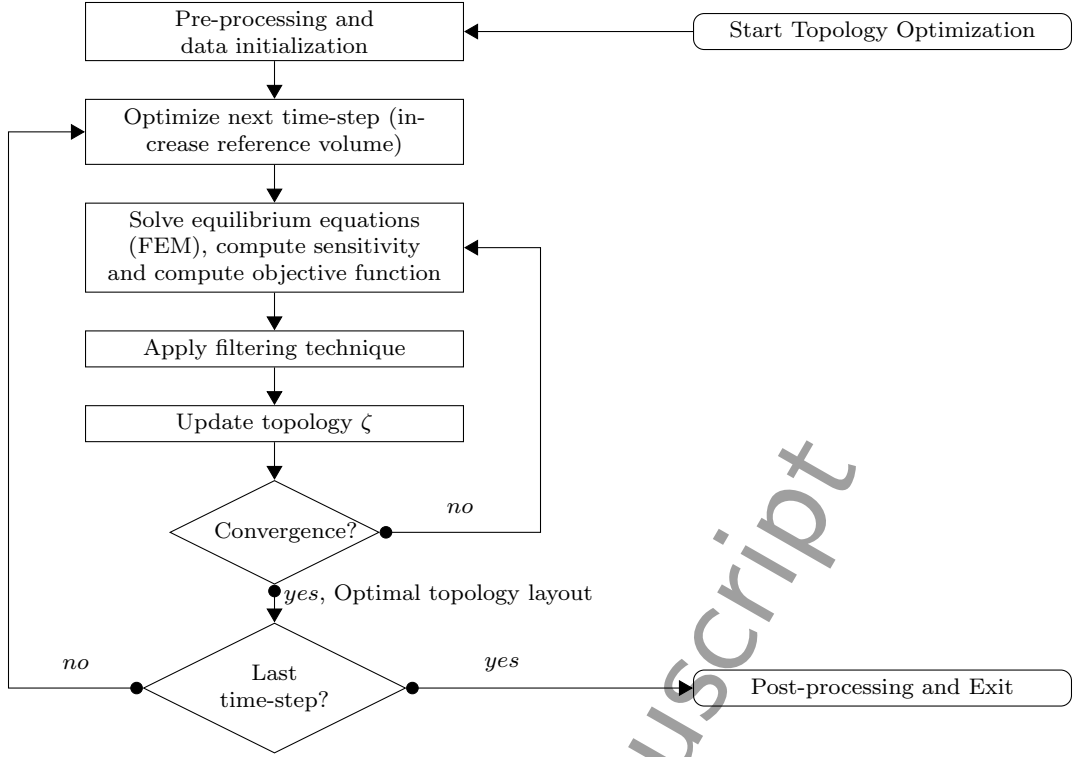
	
\section{Topology Optimization methods} \label{sec_theroy_methods}

	In the following subsections, the specific details of each considered topology optimization method are described, focusing on the differences among them.

	\subsection{SIMP method} \label{sec_theroy_SIMP}
	
		As aforementioned, the SIMP approach employs \emph{element-wise density variables} $\rho_e$ as \emph{design variables} to describe the topology layout. Therefore, the design domain $\Omega$ is discretized into cells or voxels\footnote{Henceforth, both cells and voxels will be referred to as elements.} and each element $e$ is assigned a density $\rho_e$. Although $\rho_e$ would ideally be equal to 1 for material and 0 for void, the design variable is here relaxed by allowing intermediate values $0\le \rho_e \le 1$. As a consequence, additional constraints (\ref{eq_minimization_restricted}-b-2) must be added to the original \emph{topology optimization problem} subject to the volume constraint. Furthermore, the material interpolation of the constitutive tensor $\mathbb{C}_\rho$ is defined as
		\begin{equation} \label{eq_C_SIMP}
			\mathbb{C}_\rho (\rho_e) = \mathbb{C}^- + \rho_e^p \left( \mathbb{C}^+ - \mathbb{C}^- \right), \quad \rho_e \in[0,1] \mcolon
		\end{equation}
		where $\mathbb{C}^+$ and $\mathbb{C}^-$ correspond to the constitutive tensor of the stiff and soft materials, respectively. In addition, the parameter $p$ stands for the penalization factor (typically $p\ge3$). For a constant Poisson ratio, equation (\ref{eq_C_SIMP}) can be written in terms of the Young's modulus $E$ as
		\begin{equation}
			\mathbb{C}_\rho (\rho_e) = \left( E^- +  \rho_e^p \left( E^+ - E^- \right) \right) \mathbb{C}_{(1)}   \mcolon
		\end{equation}
		where $E^+$ and $E^-$ represent the Young's modulus for the stiff and soft materials, respectively, and $\mathbb{C}_{(1)}$ corresponds to the constitutive tensor with unit Young's modulus. Assuming that $E^-$ can be defined proportional to the high Young's modulus, $E^+$, via the \emph{contrast factor} $\alpha$, the resultant constitutive tensor $\mathbb{C}_\rho (\rho_e)$ is defined as $\mathbb{C}^+$ multiplied by a coefficient, which depends on the \emph{design variable of the element}, i.e.,~
		\begin{equation}
			\mathbb{C}_\rho (\rho_e) = \left( \alpha +  \rho_e^p \left( 1 - \alpha \right) \right) \mathbb{C}^+ \mdot
		\end{equation}
		The global stiffness matrix ${\mathbb K}_{\rho}$ is obtained via the assembly of element stiffness matrices defined as
		\begin{align}
			\begin{aligned}
				{\mathbb K}_e (\rho_e) &= \int_{\Omega_e}\mathbf{B}^{\text{T}}\bbx\ {\mathbb{C}}_{\rho}\bbx\ \mathbf{B}\bbx \domega  \\
				&= \left( \alpha +  \rho_e^p \left( 1 - \alpha \right) \right) \int_{\Omega_e}\mathbf{B}^{\text{T}}\bbx\  \mathbb{C}^+\ \mathbf{B}\bbx \domega  \\
				&= \left( \alpha +  \rho_e^p \left( 1 - \alpha \right) \right) {\mathbb K}_e^+ \mcolon
			\end{aligned}
		\end{align}
		with ${\mathbb K}_e^+ $ being the element stiffness matrix considering stiff material for element $e$.
		
		Taking into account these two characteristics, the \emph{topology optimization problem} (\ref{eq_minimization_restricted}) becomes
		\begin{equation} \label{eq_top_prob_SIMP}
			\left[\begin{split}
				&\underset{\rho\in{\mathscr{U}}_{ad}} {\operatorname{min}}\;{\cal J}\left({\mathbf{u}}(\pmb{\rho}),\pmb{\rho}\right)\equiv\int_{\Omega}{j({\mathbf{u}}(\pmb{\rho}),\pmb{\rho},\bx)}\domega  &&(a) \\
				&\text{subject to:} \\
				&\hspace{0.5cm}{\cal C}_0(\pmb{\rho})\equiv  \frac{\vert \Omega ({\pmb{\rho}}) \vert}{\vert \Omega \vert} - f  \le0 &&(b-1) \\
				&\hspace{0.5cm}{\cal C}_e(\rho_e)\le0 \rightarrow 0\le\rho_e\le1, \; e:1\dots N_e &&(b-2) \\
				&\text{governed by:} \\
				&\hspace{0.5cm}{\mathbb K}_{\rho}\hat{\mathbf{u}}_{\rho}^{(i)}=\mathbf{f}^{(i)} &&(c) 
			\end{split}\right. \mcolon
		\end{equation}
		where $\vert \Omega ({\pmb{\rho}}) \vert$ and $\vert \Omega \vert$ are respectively the stiff material volume and the design domain volume, and $f$ is the prescribed volume fraction\footnote{The term $\vert (\cdot) \vert$  denotes the Lebesgue measure of $(\cdot)$.}.
		
		The sensitivity of the objective function (\ref{eq_top_prob_SIMP}-a), ${\partial{\cal J}\left({\mathbf{u}}(\pmb{\rho}),\pmb{\rho}\right)}/{\partial \rho_e}$, is obtained via the adjoint technique, so that the derivative of the unknown field ${\mathbf{u}}_\rho$ with respect to the density $\rho_e$, ${\partial{\mathbf{u}}(\pmb{\rho})}/{\partial \rho_e}$, is not required to be computed. Additionally, the sensitivity of the volume constraint (\ref{eq_top_prob_SIMP}-b-1) with respect to the density of the element $e$ is equal to ${\vert \Omega_e \vert} / {\vert \Omega \vert}$. According to \cite{Bendsoe2004}, the sensitivity of the objective function for the three \emph{topology optimization problems} addressed in this work are given by
		\begin{empheq}[right=\hspace{-0.2cm}]{align}
			& \frac{\partial{\cal J}^{(I)}\left({\mathbf{u}}(\pmb{\rho}),\pmb{\rho}\right)}{\partial \rho_e} =  - \omega_e (\rho_e)\  \hat{\mathbf{u}}_e (\pmb{\rho})  {\mathbb K}_e^+ \hat{\mathbf{u}}_e (\pmb{\rho}) \\
			& \frac{\partial{\cal J}^{(II)}\left({\mathbf{u}}(\pmb{\rho}),\pmb{\rho}\right)}{\partial \rho_e} = -  \omega_e (\rho_e) \sum_{i=1}^{n_l}{ \hat{\mathbf{u}}_e^{(i)} (\pmb{\rho})  {\mathbb K}_e^+  \hat{\mathbf{u}}_e^{(i)} (\pmb{\rho})   } \  \\
			& \frac{\partial{\cal J}^{(III)}\left({\mathbf{u}}(\pmb{\rho}),\pmb{\rho}\right)}{\partial \rho_e} = \omega_e (\rho_e)\ \hat{\mathbf{u}}_e^{(1)} (\pmb{\rho})  {\mathbb K}_e^+ \hat{\mathbf{u}}_e^{(2)} (\pmb{\rho}) 
		\end{empheq}
		where $\omega_e (\rho_e) = p \rho_e^{p-1} \left( 1 - \alpha \right)$ and $\hat{\mathbf{u}}_e^{(i)} (\pmb{\rho})$ denotes the nodal displacements of element $e$ and the state equation $(i)$.
		
		As mentioned in Section \ref{sec_Introduction}, a regularization technique must be applied to the original topology optimization problem to ensure the existence of a solution and avoid the formation of the so-called checkerboard patterns. Different filtering techniques modifying the element sensitivity are studied for the comparison, including a radial sensitivity filter computed using the convolution function (see Section \ref{sec_theroy_SIMP_convn}) and the filter based on \emph{Helmholtz type differential equation} (Section \ref{sec_theroy_SIMP_PDE}). The solution to the filtering technique is denoted by $\widetilde{(\cdot)}$ and replaces the non-filtered sensitivity.
		
		Considering the sensitivities of the objective function and volume constraint, the corresponding \emph{topology optimization problem} can be solved by means of the \emph{Optimality Criteria} (OC) method. The \emph{OC method} seeks the fulfillment of the Karush-Kuhn-Tucker (KKT) conditions
		\begin{equation} \label{eq_KKT_SIMP}
			\frac{\widetilde{\partial{\cal J}}}{\partial \rho_e} + \lambda \frac{\partial{\cal C}_0}{\partial\rho_e} = 0, \; e:1\dots N_e \mcolon
		\end{equation} 
		where $\lambda$ is the \emph{Lagrange multiplier} associated with the volume constraint ${\cal C}_0(\pmb{\rho})$ such that the volume constraint is met, and must be computed via a root-finding algorithm (e.g.,~a bisection method). Note that the element density $\rho_e$ must be in the range of 0 to 1. The \emph{optimality conditions} can be expressed as $B_e=1$, where
		\begin{equation} \label{eq_Be_OC_SIMP}
			B_e = - \frac{\widetilde{\partial{\cal J}}}{\partial \rho_e} \left( \lambda \frac{\partial{\cal C}_0}{\partial\rho_e} \right)^{-1} \mdot
		\end{equation}
		A \emph{heuristic updating scheme}, proposed by \citet{Bendsoe1988}, is used to update the design variables and achieve convergence. For \emph{minimum mean compliance}, the scheme is defined as
		\begin{equation} \label{eq_OC_SIMP}
			\rho_e^{(k+1)} = \left\{
			\begin{split}
				& \max\left(0,\rho_e^{(k)}-m\right)  &&\resizebox{.7\width}{!}{%
				$\text{if } \rho_e^{(k)} B_e^\eta \le \max\left(0,\rho_e^{(k)}-m\right)$}\\
				& \min\left(1,\rho_e^{(k)}+m\right)  &&\resizebox{.7\width}{!}{%
				$\text{if } \rho_e^{(k)} B_e^\eta \ge \min\left(1,\rho_e^{(k)}+m\right)$}\\
				& \rho_e^{(k)} B_e^\eta  &&\resizebox{.7\width}{!}{%
				$\text{otherwise}$}
			\end{split}\right. \mcolon
		\end{equation}
		where $m$ is a positive move limit, $\eta$ is a numerical damping coefficient and $k$ represents the iteration counter. These two numerical parameters are typically set $0.2$ and $0.5$, respectively, for \emph{minimum mean compliance}. Equation (\ref{eq_Be_OC_SIMP}) is modified for \emph{compliant mechanism optimization problems} to just account for positive sensitivities as
		\begin{equation} \label{eq_Be_OC_SIMP_mech}
			B_e = \max\left( \epsilon,- \frac{\widetilde{\partial{\cal J}}}{\partial \rho_e} \right)\left( \lambda \frac{\partial{\cal C}_0}{\partial\rho_e} \right)^{-1} \mcolon
		\end{equation}
		with $\epsilon$ being a small positive value. Equivalently, the updating scheme (\ref{eq_OC_SIMP}) can be expressed as
		\begin{equation}
			\rho_e^{(k+1)} = \left\{
			\begin{split}
				& \max\left(0,\rho_e^{(k)}-m\right) &&\resizebox{.7\width}{!}{%
				$\text{if } \rho_e^{(k)} B_e^\eta \le \max\left(\epsilon,\rho_e^{(k)}-m\right)$}\\
				& \min\left(1,\rho_e^{(k)}+m\right) &&\resizebox{.7\width}{!}{%
				$\text{if } \rho_e^{(k)} B_e^\eta \ge \min\left(1,\rho_e^{(k)}+m\right)$}\\
				& \rho_e^{(k)} B_e^\eta &&\resizebox{.7\width}{!}{%
				$\text{otherwise}$}
			\end{split}\right. \mcolon
		\end{equation}
		with $m=0.1$ and $\eta=0.3$.
		 
		In the following subsections, three variations of the \emph{SIMP approach} are introduced, mainly by changing the filter used to regularize the sensitivity (for instance, using a distance filter computed via a convolution function and a Helmholtz-type filter) or by trying a \emph{time-advancing strategy} (with multiple steps). 
	
		\subsubsection{SIMP method using PDE-like filter: SIMP$^{(I)}$} \label{sec_theroy_SIMP_PDE}
		
			In this case, the sensitivities are regularized via a \emph{Helmholtz-type PDE equation} with homogeneous Neumann boundary conditions, as detailed in \cite{Lazarov2010,Andreassen2010}. The \emph{regularized sensitivities} ${\widetilde{\partial{\cal J}}}/{\partial \rho_e}$ correspond to the solution of
			\begin{equation} \label{eq_Lap_filter_SIMP}
				\left\{
				\begin{split}
					&\widetilde{\zeta}-R_{min}^2\Delta_\bx\widetilde{\zeta}=\zeta& &\quad in \; \Omega\\
					&\nabla_\bx\widetilde{\zeta}\cdot\mathbf{n}={0}& &\quad on \; \partial\Omega
				\end{split}
				\right. \mcolon				
			\end{equation}
			where $\zeta$ and $\widetilde{\zeta}$ are $\rho_e \: {\partial{\cal J}}/{\partial \rho_e}$ and $\rho_e \: {\widetilde{\partial{\cal J}}}/{\partial \rho_e}$, respectively, and $\Delta_\bx(\bx,\cdot)$ and $\nabla_\bx(\bx,\cdot)$ are the Laplacian and gradient operators. The filter radius $R_{min}$ is equal to $r_{min}/(2\sqrt{3})$, with $r_{min}$ being the filter radius of \emph{distance-based filters} (see subsection \ref{sec_theroy_SIMP_convn}). 
			
			As reported by \citet{Lazarov2010}, this type of filtering technique provides computational advantages when regularizing the sensitivities for complex non-uniform meshes in terms of memory storage and computational complexity when compared to classical filtering procedures. Although, for structured meshes, as in the cases addressed in this paper, this performance improvement may not be observed.
			
		\subsubsection{SIMP method using a time-advancing scheme: SIMP$^{(II)}$} \label{sec_theroy_SIMP_incr}
		
			An \emph{incremental-time-advancing scheme} can be implemented on top of SIMP$^{(I)}$. The volume reference of the volume constraint is iteratively updated, thus obtaining a set of intermediate converged solutions. Once the convergence is achieved, the \emph{reference volume fraction} $f$ in equation (\ref{eq_top_prob_SIMP}-(b-1)) is decreased and the topology optimization procedure is repeated for the new volume constraint. At the first iteration of each time-step, the volume constraint must be fulfilled, so that the \emph{Helmholtz-type PDE filter} keeps the volume constant.
		
		\subsubsection{SIMP method using convolution filter: SIMP$^{(III)}$} \label{sec_theroy_SIMP_convn}
		
			The filter in this approach modifies the sensitivities ${\partial{\cal J}}/{\partial \rho_e}$ by means of a standard distance filter as follows
			\begin{equation} 
				\frac{\widetilde{\partial{\cal J}}}{\partial \rho_e} = \frac{1}{\max{(\gamma,\rho_e)} \sum_{i\in N_{ei}}^{}{H_{ei}} } \sum_{i\in N_{ei}}^{}{H_{ei}\rho_i 		\frac{\partial{\cal J}}{\partial \rho_i}} \mcolon
			\end{equation}
			where $\gamma$ is a small positive number to avoid division by zero and $N_{ei}$ denotes the set of elements $i$ for which the center-to-center distance, $\operatorname{dist}(e,i)$, of element $i$ to element $e$ is smaller than a filter radius $r_{min}$, defined by the user, i.e.,~$N_{ei}=\{i\in N_e \;/\; \operatorname{dist}(e,i) \le r_{min} \}$. The function $H_{ei}$ corresponds to the weight factor function (of a linearly decaying filter kernel) given by
			\begin{equation} \label{eq_H_filter}
				H_{ei} = \max{( 0, r_{min} - \operatorname{dist}(e,i) )} \mdot
			\end{equation}
			This sensitivity filter can be mathematically written using a convolution product of the filter function $H(\mathbf{x}-\mathbf{y})$ and the sensitivity of the objective function ${\partial{\cal J}}/{\partial \rho(\mathbf{x})}$ as
			\begin{align}
				\begin{aligned}
					&\frac{\widetilde{\partial{\cal J}}}{\partial \rho} \bbx = \frac{1}{\hat{\rho}\bbx}\left(H*\left(\rho \frac{\partial{\cal 	J}}{\partial \rho}\right)\right)\bbx \\
					&= \frac{1}{\hat{\rho}\bbx \int_{\mathbb{B}_r}{ H(\mathbf{x}-\mathbf{y}) \ d\mathbf{y}}} \int_{\mathbb{B}_r}{ H (\mathbf{x}-\mathbf{y}) \rho (\mathbf{y}) \frac{\partial{\cal J}}{\partial \rho}  (\mathbf{y}) \ d\mathbf{y}}	\mcolon
				\end{aligned}
			\end{align}
			where $\mathbb{B}_r$ is a sphere in 3D and a circle in 2D with center at $\bx$, and radius $r_{min}$ and $\hat{\rho}$ is equal to $\max{(\gamma,\rho_e)}$.
		
	\subsection{SOFTBESO method} \label{sec_theroy_SOFTBESO}
	
		The original \emph{ESO} and \emph{BESO methods} use a \emph{discrete density variable} $\chi_e=\{0,1\}$ as \emph{design variable} in a \emph{hard-kill topology optimization procedure} where elements with low \emph{rejection criterion} are removed from the topology layout. In particular, in BESO, elements can be added in specific areas by using extrapolation techniques. However, as mentioned in Section \ref{sec_Introduction}, this set of \emph{hard-kill approaches} suffers from numerical instabilities failing in some circumstances to obtain convergent solutions. For that reason, this paper focuses on \emph{soft-kill evolutionary techniques}, and in particular, in the \emph{bi-directional evolutionary (BESO) approach} proposed by \citet{Huang2008}.
		
		In this context, the design variable, now termed as the \emph{element-wise density variable} $\rho_e=\{\rho_{min},1\}$, is defined using the original \emph{SIMP material interpolation}, thus relaxing the design variable with a penalization factor $p$. Therefore, the material interpolation of the constitutive tensor is given by
		\begin{equation} \label{eq_C_BESO}
			\mathbb{C}_\rho (\rho_e) = \rho_e^p \mathbb{C}^+ = \rho_e^p E^+ \mathbb{C}_{(1)}, \; \rho_e =\{\rho_{min},1\} \mdot
		\end{equation}
		Assuming that the Young's modulus of the void material, $E^-$, can be expressed, in terms of $\alpha$, as $\alpha$ times the Young's modulus of the stiff material $E^+$, then $\rho_{min}$ must be equal to $\sqrt[\leftroot{4}\uproot{3}p]{\alpha}$. A similar procedure as the one defined for \emph{SIMP-based approaches} is used here to compute the element stiffness matrix, i.e.,
		\begin{align}
			\begin{aligned}
				{\mathbb K}_e (\rho_e) &= \int_{\Omega_e}\mathbf{B}^{\text{T}}\bbx\ {\mathbb{C}}_{\rho}\bbx\ \mathbf{B}\bbx \domega \\ 
								       &= \rho_e^p \int_{\Omega_e}\mathbf{B}^{\text{T}}\bbx\  \mathbb{C}^+\ \mathbf{B}\bbx \domega  = \rho_e^p {\mathbb K}_e^+ \mcolon
			\end{aligned}
		\end{align}
		with ${\mathbb K}_e^+$ being the element stiffness matrix considering stiff material for element $e$. 

		The corresponding \emph{topology optimization problem} (\ref{eq_minimization_restricted}) for \emph{BESO} can be written as
		\begin{equation} \label{eq_top_prob_SOFTBESO}
			\left[\begin{split}
				&\underset{\rho\in{\mathscr{U}}_{ad}} {\operatorname{min}}\;{\cal J}\left({\mathbf{u}}(\pmb{\rho}),\pmb{\rho}\right)\equiv\int_{\Omega}{j({\mathbf{u}}(\pmb{\rho}),\pmb{\rho},\bx)}\domega &&(a) \\
				&\text{subject to:} \\
				&\hspace{0.25cm}{\cal C}_0(\pmb{\rho})\equiv  \frac{\vert \Omega (\pmb{\rho}) \vert}{\vert \Omega \vert} - f  \le0 &&(b-1) \\
				&\hspace{0.25cm}{\cal C}_e(\rho_e)\le0 \rightarrow \rho_e=\{\rho_{min},1\}, \; e:1\dots N_e &&(b-2) \\
				&\text{governed by:} \\
				&\hspace{0.25cm}{\mathbb K}_{\rho}\hat{\mathbf{u}}_{\rho}^{(i)}=\mathbf{f}^{(i)} &&(c) 
			\end{split}\right. \mcolon
		\end{equation}
		where the volume fraction $f$ is decreased at each iteration until the desired final fraction $\overline{f}$, following an exponential expression $f_{k+1}=\max{(\overline{f},(1-ER)f_{k})}$. The evolutionary volume ratio $ER\ll1$ corresponds to the maximum volume fraction decreased at each iteration. Notice that the convergence in topology and objective are not met until the desired final fraction $\overline{f}$ is reached.
		
		As detailed in Section \ref{sec_theroy_SIMP}, the sensitivities of the objective function (\ref{eq_top_prob_SOFTBESO}-a) ${\partial{\cal J}}/{\partial \rho_e}$ for the considered topology optimization problems are defined as
		\begin{empheq}[right=\hspace{-0.2cm}]{align}
			& \frac{\partial{\cal J}^{(I)}\left({\mathbf{u}}(\pmb{\rho}),\pmb{\rho}\right)}{\partial \rho_e} = \resizebox{.9\width}{!}{%
			$-  \omega_e (\rho_e) \  \hat{\mathbf{u}}_e (\pmb{\rho})  {\mathbb K}_e^+ \hat{\mathbf{u}}_e (\pmb{\rho}) $} \mcolon\\
			& \frac{\partial{\cal J}^{(II)}\left({\mathbf{u}}(\pmb{\rho}),\pmb{\rho}\right)}{\partial \rho_e} =  \resizebox{.9\width}{!}{%
			$-  \omega_e (\rho_e) \ \sum_{i=1}^{n_l}{ \hat{\mathbf{u}}_e^{(i)} (\pmb{\rho})  {\mathbb K}_e^+  \hat{\mathbf{u}}_e^{(i)} (\pmb{\rho})   } $} \mcolon\\
			& \frac{\partial{\cal J}^{(III)}\left({\mathbf{u}}(\pmb{\rho}),\pmb{\rho}\right)}{\partial \rho_e} = \resizebox{.9\width}{!}{%
			$ \omega_e (\rho_e) \ \hat{\mathbf{u}}_e^{(1)} (\pmb{\rho})  {\mathbb K}_e^+ \hat{\mathbf{u}}_e^{(2)} (\pmb{\rho}) $} \mcolon
		\end{empheq}
		with $\omega_e(\rho_e)$ being equal to $p \rho_e^{p-1}$.
		
		For the regularization procedure, a \emph{linear distance-based filter}, similar as the one used in \emph{SIMP$^{(III)}$}, is applied to the sensitivities ${\partial{\cal J}}/{\partial \rho_e}$. The filtered sensitivities are obtained as
		\begin{equation} 
			\frac{\widetilde{\partial{\cal J}}}{\partial \rho_e} = \frac{1}{ \sum_{i\in N_{ei}}^{}{H_{ei}} } \sum_{i\in N_{ei}}^{}{H_{ei} \frac{\partial{\cal J}}{\partial \rho_i}} \mcolon
		\end{equation}
		which can also be computed using the convolution function of $H_{ei}$ (\ref{eq_H_filter}) and the non-regularized sensitivities ${\partial{\cal J}}/{\partial \rho_e}$. In addition to the spatial filtering (used to address the mesh-dependency problem), a temporal filtering is also applied by averaging the sensitivity numbers with historical information, thus improving convergence. The temporal filter can be expressed as
		\begin{equation}
			\left.\frac{\widetilde{\partial{\cal J}}}{\partial \rho_e} \right\vert_{e,k} = \frac{1}{2} \left( \left. \frac{\widetilde{\partial{\cal J}}}{\partial \rho_e} \right\vert_{e,k} + \left.\frac{\widetilde{\partial{\cal J}}}{\partial \rho_e} \right\vert_{e,k-1} \right) \mcolon
		\end{equation}
		where $k$ corresponds to the iteration number. Notice that the \emph{temporal-filtered sensitivity} used in the \emph{optimality criteria} (\ref{eq_KKT_SOFTBESO}) replaces the spatial-filtered sensitivity.
		
		The optimality criterion for the \emph{topology optimization problem} (\ref{eq_top_prob_SOFTBESO}) can easily be derived if no restriction is imposed on the design variable, i.e.,
		\begin{equation} \label{eq_KKT_SOFTBESO}
			\frac{\widetilde{\partial{\cal J}}}{\partial \rho_e} + \lambda \frac{\partial{\cal C}_0}{\partial\rho_e} = 0, \; e:1\dots N_e \mcolon
		\end{equation} 
		where the \emph{Lagrange multiplier} $\lambda$ must be computed so that the volume constraint ${\cal C}_0(\pmb{\rho})$ is fulfilled. For \emph{minimum mean compliance problem}, the corresponding updating scheme can be expressed as
		\begin{equation} \label{eq_update_softbeso_compliance}
			\rho_e^{k+1}= \left\{
			\begin{split}
				& 1           && \text{if }  \left( -\tfrac{\widetilde{\partial{\cal J}}}{\partial \rho_e} / \tfrac{\partial{\cal C}_0}{\partial\rho_e} - \lambda \right) \ge 0 \\
				& \rho_{min}  && \text{if }  \left( -\tfrac{\widetilde{\partial{\cal J}}}{\partial \rho_e} / \tfrac{\partial{\cal C}_0}{\partial\rho_e} - \lambda  \right) < 0
			\end{split}
			\right. \mcolon
		\end{equation}
		although the number of elements changing from the void domain to the material domain (or equivalently a volume fraction) is limited by a factor $AR_{max}$. Consequently, if the number of elements changing from the void domain to the material domain is greater than the \emph{maximum volume addition ratio}, only the $AR_{max}$ elements from the void domain with the highest sensitivity are added to the material domain. In the material domain, the $AR_{max}+ER$ elements with the lowest sensitivity are removed from the material domain and replaced with void material, thus satisfying the volume constraint. This procedure ensures that not many elements are added in a single iteration, causing the structure to loose its integrity.
		
		For \emph{compliant mechanism synthesis} \cite{Huang2010,Huang2014}, the updating equation (\ref{eq_update_softbeso_compliance}) is relaxed to 
		\begin{equation}
			\rho_e^{k+1}= \left\{
			\begin{split}
				& \min{\left(\rho_e^{(k)}+m,1\right)}           && \text{if }  \resizebox{.85\width}{!}{%
				$\left( -\tfrac{\widetilde{\partial{\cal J}}}{\partial \rho_e} / \tfrac{\partial{\cal C}_0}{\partial\rho_e} - \lambda \right) \ge 0 $} \\
				& \max{\left(\rho_e^{(k)}-m,\rho_{min}\right)}   && \text{if }  \resizebox{.85\width}{!}{%
				$\left( -\tfrac{\widetilde{\partial{\cal J}}}{\partial \rho_e} / \tfrac{\partial{\cal C}_0}{\partial\rho_e} - \lambda  \right) < 0 $}
			\end{split}
			\right.	
		\end{equation}
		with $m=0.1$, where the design variable $\rho_e$ can now take intermediate values.
				
	\subsection{\methodabb{} method} \label{sec_theroy_UNVARTOP}
	
		As in \emph{Level-set}, the zero-level of the \emph{level-set function} is used to precisely define the boundaries of the material domain, although no updating equation is defined in terms of the \emph{discrimination function} $\psi$. Instead, the \emph{characteristic function} $\overline{\chi}\bbx=\{\beta,1\}$ is employed as the design variable, which is computed from the \emph{discrimination function} at each iteration via the Heaviside function $\cal{H}_\beta(\psi\bbx)$ \footnote{The image set $\{1,0\}$ of the \emph{Heaviside function} is relaxed to $\{1,\beta\}$, this being highlighted with the subscript $(\cdot)_\beta$ in the \emph{Heaviside function} symbol.} (see Figure \ref{fig_design_domain}). As result, the material interpolation of the constitutive tensor $\mathbb{C}_\chi$ is given by
		\begin{equation} \label{eq_C_VARTOP}
			\mathbb{C}_\chi \bbx = \chi^m\bbx  \mathbb{C}^+, \; \chi\bbx\in[\beta,1] \mcolon
		\end{equation} 
		with $\beta$ being the \emph{relaxation factor}. 
		It is important to stress that, once the domain is discretized by finite elements, the \emph{characteristic function} will take 1 or $\beta$ in the majority of the domain, whereas it will only take values ranging 1 to $\beta$ in the elements bisected by the material boundary $\Gamma$.
		The relaxation factor $\beta$ is defined such that the Young's modulus for the void material is $\alpha$ times that of the stiff material, i.e.,
		\begin{align}
			\begin{aligned}
				\mathbb{C}_\chi^- \bbx &= \alpha \mathbb{C}^+ = \beta^m \mathbb{C}^+ \\ 
									   &= \beta^m E^+ \mathbb{C}_{(1)} \quad\quad\quad \text{for } \bx \in \Omega^- \mcolon
			\end{aligned}
		\end{align}
		with $\beta$ being
		\begin{equation}
		    \beta = \sqrt[\leftroot{4}\uproot{3}m]{\alpha} \mcolon
		\end{equation}
		where $m$ is an \emph{exponential factor} defined by the user and $\alpha$ is the \emph{contrast factor} for the Young's modulus. The element stiffness matrix $\mathbb{K}_e(\chi\bbx,\bx)$ is obtained as
		\begin{align} \label{eq_stiffness_vartop}
			\begin{aligned}
				\mathbb{K}_e(\chi\bbx,\bx) &= \int_{\Omega_e}\mathbf{B}^{\text{T}}\bbx\ {\mathbb{C}}_{\chi}\bbx\ \mathbf{B}\bbx \domega \\ 
										   &= \int_{\Omega_e}\mathbf{B}^{\text{T}}\bbx\ \chi^m\bbx  \mathbb{C}^+\ \mathbf{B}\bbx \domega \mdot
			\end{aligned}
		\end{align}
		
		The \emph{topology optimization problem} (\ref{eq_minimization_restricted}) is now written as
		\begin{align} \label{eq_top_prob_VARTOP}
			\left[\begin{aligned}
				&\underset{\chi\in{\mathscr{U}}_{ad}} {\operatorname{min}}\;{\cal J}\left({\mathbf{u}}(\chi),\chi\right)\equiv\int_{\Omega}{j({\mathbf{u}}(\chi),\chi,\bx)}\domega &&(a) \\
				&\text{subject to:} \\
				&\hspace{0.5cm}\begin{aligned}
					{\cal C}_0(\chi) &\equiv \frac{\vert \overline{\Omega^-} \vert}{\vert \Omega \vert} - \frac{1}{\vert \Omega \vert} \int_\Omega{ \frac{1-\chi\bbx}{1-\beta} \domega} \\
									 &= t - \frac{\vert \Omega^- ({\chi}) \vert}{\vert \Omega \vert}  = 0
				\end{aligned} &&(b-1) \\
				&\text{governed by:} \\
				&\hspace{0.5cm}{\mathbb K}_{\chi}\hat{\mathbf{u}}_{\chi}^{(i)}=\mathbf{f}^{(i)} &&(c) 
			\end{aligned}\right. \mcolon
		\end{align}
		where the volume constraint ${\cal C}_0(\chi)$ has been expressed in terms of the soft material fraction in contrast to equations (\ref{eq_top_prob_SIMP}) and (\ref{eq_top_prob_SOFTBESO}). The term $t$ stands for the \emph{pseudo-time variable}, used as \emph{time-advancing parameter}.
		
		The \emph{relaxed topological derivative} (RTD), used as an approximation to the \emph{exact geometric topological derivative}, for a functional ${\cal{F}}\bchi:L^2(\Omega)\rightarrow\mathbb{R}$ is defined as
		\begin{align} \label{eq_RTD_VARTOP}
			\begin{aligned}
				&\dfrac{\delta {\cal F}(\chi)}{\delta \chi}\bbxhat= \left[\dfrac{\partial f({\mathbf{u}}(\chi),{\chi},{\bf x})}{\partial {\chi}}\right]_{{\bf x}=\hat{\bf x}}\Delta \chi\bbxhat \text{, } \\
				&\text{ with } {\Delta\chi}\bbxhat=
						\left\{
						\begin{aligned}
							-(&1-\beta)<0  \;\;\;\;\textit{for} \;\;\bxhat\in\Omega^{+}  \\ 
					 		(&1-\beta)>0 \;\;\;\;\textit{for} \;\;\bxhat\in\Omega^{-}
						\end{aligned} 
						\right.
						\mcolon
			\end{aligned} 
		\end{align}
		where ${\Delta \chi}\bbxhat$ is termed the \emph{exchange function} and stands for the signed variation of $\chi\bbxhat$, due to that material exchange. The sensitivity of the volume constraint (\ref{eq_top_prob_VARTOP}-b-1) with respect to the \emph{characteristic function} is equal to $\text{sgn}(\Delta\chi\bbxhat)/{\vert\Omega\vert}$, where $\text{sgn}(\cdot)$ denotes the sign function of $(\cdot)$. The optimality condition for the \emph{constrained topology optimization problem} can be written as
		\begin{equation} \label{eq_optimal_condition}
			 \left(\dfrac{\partial j \left({\mathbf{u}}(\chi),\chi,\hat{\mathbf x} \right)}{\partial \chi}{\Delta\chi}\bbxhat
					+\lambda\,\frac{\text{sgn}(\Delta\chi\bbxhat)}{\vert\Omega\vert}\right)>{0}\;\; \forall\hat{\mathbf{x}}\in\Omega \mcolon
		\end{equation}
		where $\lambda$ stands for a \emph{Lagrange multiplier} enforcing volume constraint ${\cal C}_0\bchi=0$. Therefore, the \emph{closed-form non-linear solution} for the \emph{topology optimization problem} (\ref{eq_top_prob_VARTOP}), termed as \emph{cutting}\&\emph{bisection} algorithm, can be expressed as
		\begin{equation} \label{eq_closed_form}
			\left\{\begin{split}
				&\psi_{\chi}({\bf x},\lambda)\coloneqq {\xi}\left({\mathbf{u}}(\chi),\chi,{\mathbf x} \right) - {\lambda}/{\vert\Omega\vert}\\
				&\chi({\bf x},\lambda)={\cal H}_{\beta}\left[ \psi_{\chi}({\bf x},\lambda) \right] \\
				&{\cal C}_0(\chi({\bf x},\lambda))=0
			\end{split}\right. \mcolon
		\end{equation}
		where ${\xi}\left({\mathbf{u}}(\chi),\chi,{\mathbf x} \right)$  is termed the \emph{pseudo-energy} and depends on the specific objective function. The Lagrange multiplier $\lambda$ is computed using an efficient bisection algorithm and its value fulfils the volume constraint, as in the previous approaches.
		
		For the three \emph{topology optimization problems}, the \emph{pseudo-energies} at point $\bxhat$ are given by
		\begin{empheq}[right=\hspace{-0.2cm}]{align}
			& {\xi}^{(I)}\left({\mathbf{u}}(\chi),\chi,\bxhat \right) = \resizebox{.85\width}{!}{%
			$\omega\bbxhat \ \hat{\mathbf{u}} (\chi,\bxhat)  {\mathbb K}_e^+ \hat{\mathbf{u}} (\chi,\bxhat)$} \mcolon \label{eq_xi_VARTOP_compliance}\\
			& {\xi}^{(II)}\left({\mathbf{u}}(\chi),\chi,\bxhat \right) = \resizebox{.85\width}{!}{%
			$\omega\bbxhat \ \sum_{i=1}^{n_l}{ \hat{\mathbf{u}}^{(i)} (\chi,\bxhat)  {\mathbb K}_e^+  \hat{\mathbf{u}}^{(i)} (\chi,\bxhat)   }$} \mcolon \label{eq_xi_VARTOP_compliance_multi}\\
			& {\xi}^{(III)}\left({\mathbf{u}}(\chi),\chi,\bxhat \right) = \resizebox{.85\width}{!}{%
			$-\omega\bbxhat \ \hat{\mathbf{u}}^{(1)} (\chi,\bxhat)  {\mathbb K}_e^+ \hat{\mathbf{u}}^{(2)} (\chi,\bxhat)$}  \mcolon \label{eq_xi_VARTOP_mechanism}
		\end{empheq}
		for non-bisected elements, with $\omega\bbxhat$ being equal to $m \chi^{m-1}\bbxhat (1-\beta)$ \footnote{For points in the boundary material, the \emph{characteristic function} and the element stiffness matrix should be replaced with the corresponding ones, as detailed in \cite{Oliver2019}.}.  However, the \emph{pseudo-energy} must be first shifted and normalized, according to 
		\begin{equation} \label{eq_thry_shift_norm}
			\hat{\xi}\left({\mathbf{u}}(\chi),\chi,\bxhat \right)=\dfrac{\xi\left({\mathbf{u}}(\chi),\chi,\bxhat \right)-\chi\bbxhat\Delta_{shift}}{\Delta_{norm}} \mcolon
		\end{equation}
		thus obtaining a \emph{modified energy density} $\hat{\xi}\left({\mathbf{u}}(\chi),\chi,\bxhat \right)$. The terms $\Delta_{shift}$ and $\Delta_{norm}$ correspond, respectively, to the shifting and normalization parameters defined at the first iteration\footnote{The terms $\Delta_{shift}$ and $\Delta_{norm}$ are defined as $\operatorname{min}(\xi_0,0)$ and $\operatorname{max}(\operatorname{range}(\xi_0),\operatorname{max}(\xi_0))$, respectively.}. This variable must be later regularized using a Laplacian regularization (similar as the one described in Section \ref{sec_theroy_SIMP_PDE} for SIMP$^{(I)}$). The \emph{pseudo-energy density} actually used in the \emph{closed-form solution} (\ref{eq_closed_form}) comes from the resolution of
		\begin{equation} \label{eq_regularized_laplacean_smoothing_psi}
			\left\{
			\begin{split}
				&\hat{\xi}_{\tau}-(\tau h_e)^2\Delta_\bx\hat{\xi}_{\tau}=\hat{\xi}& &\quad in \; \Omega\\
				&\nabla_\bx\hat{\xi}_{\tau}\cdot\mathbf{n}={0}& &\quad on \; \partial\Omega
			\end{split}
			\right. \mcolon
		\end{equation}
		where, $\Delta_\bx(\bx,\cdot)$ and $\nabla_\bx(\bx,\cdot)$ are respectively the Laplacian and Gradient operators, and $\mathbf{n}$ is the outward normal to the boundary of the design domain, $\partial\Omega$. $\tau$ and $h_e$ stand for the dimensionless \emph{regularization parameter} and the typical size of the finite element mesh, respectively.
		
		Contrary to the common \emph{SIMP implementations} (SIMP$^{(I)}$ and SIMP$^{(III)}$) or the \emph{BESO method}, the \emph{\methodabb{}} is formulated under a \emph{time-advancing framework}, where the \emph{pseudo-time} $t$ in equation (\ref{eq_top_prob_VARTOP}-b-1) is iteratively increased, thus obtaining intermediate converged solutions, which are local minima and provide a Pareto Frontier in terms of the volume fraction. Notice that, at every time-step, convergence is achieved unlike the algorithm used for \emph{BESO}.

	\subsection{Level-set method via a Hamilton-Jacobi equation} \label{sec_theroy_LevelSet}
	
		As previously mentioned in the introduction, \emph{Level-set methods} use a \emph{level-set function} (LSF) to implicitly represent the optimal material domain via equation (\ref{eq_domain_splitting}), and the material boundary via the 0-level of the \emph{level-set function}. It is important to note that originally \emph{Level-set approaches} only updated the material boundary based on a differential equation, preventing them from creating new voids. This drawback was first overcome by inserting new voids every certain iteration. However, \citet{Yamada2010}, among other researchers, suggested an approach in which a \emph{Level-set method} was used to update not only the boundary of the material domain, but also the material domain itself, thus allowing to nucleate new voids. This specific approach will be taken as the reference in this article in conjunction with the \emph{relaxed topological derivative} (RTD), defined in equation (\ref{eq_RTD_VARTOP}), to obtain the sensitivity function.
				
		In contrast to \emph{\methodabb{}}, the LSF is updated via a \emph{time-dependent updating equation}, typically using a \emph{Hamilton-Jacobi equation}, although other updating schemes can be used. From the \emph{level-set function}, a \emph{characteristic function} $\overline{\chi}$ can be defined as
		\begin{equation} \label{eq_heaviside_function_LSM}
			\overline{\chi}  (\bx,\phi\bbx) = {\cal H}_\beta (\phi\bbx) = \left\{
			\begin{split}
				& 1 &&\quad \text{for }  \phi\bbx\ge0\\
				& \beta &&\quad \text{for }  \phi\bbx<0
			\end{split}
			\right.
		\end{equation} 	
		with $\beta$ being the \emph{relaxation factor}. Based on the preceding definition of the \emph{characteristic function}, the corresponding material interpolations (equations (\ref{eq_C_VARTOP}) to (\ref{eq_stiffness_vartop})) are defined. The \emph{topology optimization problem} (\ref{eq_minimization_restricted}), in terms of the \emph{level-set function}, becomes
		\begin{equation} \label{eq_top_prob_LEVELSET}
			\left[\begin{split}
				&\underset{\phi\in{\mathscr{U}}_{ad}} {\operatorname{min}}{\cal J}\left({\mathbf{u}}(\phi),\overline{\chi} (\phi)\right)\equiv\int_{\Omega}\resizebox{.85\width}{!}{%
				${j({\mathbf{u}}(\phi),\overline{\chi} (\phi),\bx)}\domega $}  &&(a) \\
				&\text{subject to:} &\\
				&\hspace{0.25cm}{\cal C}_0(\phi)\equiv t - \frac{\vert \Omega^- (\overline{\chi} (\phi)) \vert}{\vert \Omega \vert}  =0 &&(b-1) \\
				&\hspace{0.25cm}{\cal C}_n(\hat{\phi}_n)\le0 \rightarrow -1\le\hat{\phi}_n\le1, \; n:1\dots N_n &&(b-2) \\
				&\text{governed by:} &\\
				&\hspace{0.25cm}{\mathbb K}_{\phi}\hat{\mathbf{u}}_{\phi}^{(i)}=\mathbf{f}^{(i)} &&(c) 
			\end{split}\right. \mcolon
		\end{equation}
		where the nodal \emph{level-set function} $\hat{\phi}_n$ must be bounded between $-1$ and $1$, for convergence reasons. From equation (\ref{eq_top_prob_LEVELSET}), the \emph{Lagrangian function} can be expressed as
		\begin{align}
			\begin{aligned}
				&{\cal L}({\mathbf{u}}(\overline{\chi} (\phi)),\overline{\chi} (\phi),\lambda) = {{\cal J}}({\mathbf{u}}(\overline{\chi} (\phi)),\overline{\chi} (\phi)) \\ 
				&\hspace{2.7cm}+ \left(\lambda + \tfrac{1}{2} s \, {\cal C}_0(\overline{\chi} (\phi))\right) {\cal C}_0(\overline{\chi} (\phi)) \mcolon
			\end{aligned}
		\end{align}
		when an \emph{Augmented Lagrangian method} is used to fulfill the volume constraint. Therefore, the \emph{optimality condition} (\ref{eq_optimal_condition}) for the \emph{characteristic function} $\overline{\chi}  (\bx,\phi\bbx)$ reads
		\begin{align} \label{eq_opt_condition_levelset}
			\begin{aligned}
				&\dfrac{\delta {\cal L}({\mathbf{u}}(\overline{\chi} (\phi)),\overline{\chi} (\phi),\lambda)}{\delta\overline{\chi}}\bbxhat= \left(\dfrac{\partial j \left({\mathbf{u}}(\overline{\chi} (\phi)),\overline{\chi} (\phi),\hat{\mathbf x} \right)}{\partial \overline{\chi} }{\Delta\overline{\chi} }\bbxhat \right. \\
									&\hspace{1cm}\left.+\left(\lambda + s \, {\cal C}_0(\overline{\chi} (\phi)) \right)\frac{\text{sgn}(\Delta\overline{\chi} \bbxhat)}{\vert\Omega\vert}\right)>{0}\;\; \forall\hat{\mathbf{x}}\in\Omega \mcolon
			\end{aligned}
		\end{align}
		where $\lambda$ and $s$ stand for the \emph{Lagrange multiplier} and the penalty factor of the \emph{Augmented Lagrangian method}. Notice that the same \emph{pseudo-energy functions} as in \emph{\methodabb{}} (equations (\ref{eq_xi_VARTOP_compliance}) to (\ref{eq_xi_VARTOP_mechanism})) are obtained for the three topology optimization methods.
		
		Instead of the \emph{closed-form non-linear solution} described in Section \ref{sec_theroy_UNVARTOP}, for each time-step the topology in this approach, $\phi$, is updated via a \emph{Hamilton-Jacobi equation}, i.e.,
		\begin{equation}
			\frac{\partial \phi}{\partial t} = \kappa \dfrac{\delta {\cal L}({\mathbf{u}}(\overline{\chi} (\phi)),\overline{\chi} (\phi),\lambda)}{\delta\overline{\chi} } \quad \text{in } \Omega \mcolon
		\end{equation}
		where the \emph{shape derivative} of the Lagrangian has been replaced with the corresponding \emph{relaxed topological derivative} (\ref{eq_RTD_VARTOP}), and $\kappa$ corresponds to a coefficient of proportionality. Substituting equation (\ref{eq_opt_condition_levelset}) to the preceding equation gives
		\begin{align} \label{eq_HJ_levelset}
			\begin{aligned}
				\phi_{k+1} = \phi_{k} + \Delta t \, \kappa \,  &\left(  \dfrac{\partial j \left({\mathbf{u}}(\overline{\chi} (\phi_k)),\overline{\chi} (\phi_k),\hat{\mathbf x} \right)}{\partial \overline{\chi} }{\Delta\overline{\chi} }\bbxhat \right.\\
				&\left.- \frac{\lambda_k + s \, {\cal C}_0(\overline{\chi} (\phi_k))}{\vert\Omega\vert}   \right)
			\end{aligned}
		\end{align}
		with
		\begin{equation}
			\lambda_{k} = \lambda_{k-1} + s \, {\cal C}_0(\overline{\chi} (\phi_{k})) \mdot
		\end{equation}
		The new \emph{level-set function} $\phi_{k+1}$ is then regularized via the \emph{Laplacian regularization} (\ref{eq_regularized_laplacean_smoothing_psi}), and the corresponding \emph{characteristic function} $\overline{\chi}_{\phi,k+1} (\bx,\phi_{\tau,k+1}\bbx)$ is computed based on the regularized LSF, $\phi_{\tau,k+1}\bbx$, using equation (\ref{eq_heaviside_function_LSM}).
		
		As defined in equation (\ref{eq_top_prob_LEVELSET}), a \emph{time-advancing framework} can be formulated, thus obtaining a set of intermediate, converged optimal solutions. However, in addition to the topology and objective function criteria, the volume constraint must be checked to ensure convergence for every time-step, since it is no longer enforced at each iteration.
		
\section{Benchmark cases} \label{sec_benchmark}

	A set of four numerical examples in 3D problems is used to compare the six different implementations with each other. The set of benchmark cases contains two \emph{minimum compliance problems} (Section \ref{sec_theroy_prob_compliance}), one \emph{multi-load compliance problem} (Section \ref{sec_theroy_prob_multicompliance}) and one \emph{compliant mechanism synthesis} (Section \ref{sec_theroy_prob_compliantmech}). In all cases and methods, eight-node hexahedral ($Q_1$) finite elements\footnote{The design domains are assumed to be prismatic domains discretized with hexahedral unit cubic finite elements, i.e.,~a regular finite element mesh. Consequently, some of the advantages of \emph{Laplacian filters} over \emph{distance-based filters} may not be noticed.} are used in the solution of the state equation (\ref{eq_equilibrium}).
	
	This set of 3D problems has been carefully selected to be used as benchmark cases using a variety of topology optimization techniques, which have been widely used for this purpose by different researchers. All of them exhibit a high geometric complexity and represent a significant challenge for the considered methods when solving the optimization problems. They are defined in the following subsections.
	
	\begin{figure*}[t]
		\floatbox[{\capbeside\thisfloatsetup{capbesideposition={left,top},capbesidewidth=5cm}}]{figure}[\FBwidth]
		{\caption{Cantilever beam: topology optimization domain with boundary conditions and dimensions. A distributed vertical load $F$ is applied on the bottom-right edge while the displacements $\mathbf{u}$ are prescribed to $\mathbf{0}$ on the left surface of the domain. The rear surface of the domain, in soft gray, represents the surface of symmetry.}\label{fig_cantilever_domain}}
		{\includegraphics[width=7cm]{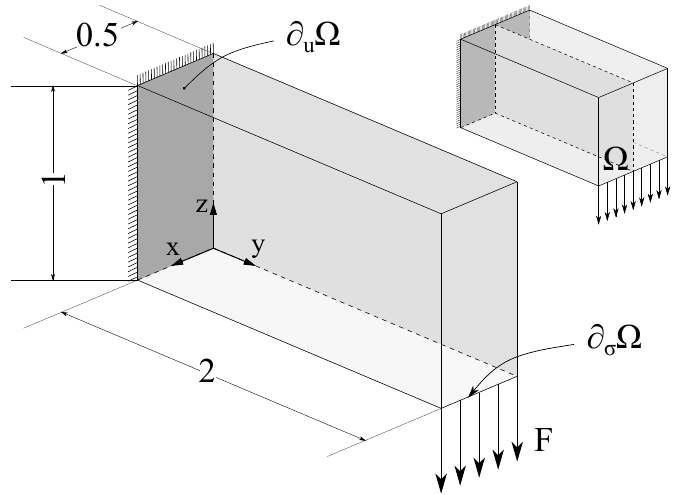}}
	\end{figure*}

	\subsection{Cantilever beam} \label{sec_exmpl_cantilever}
		
		This first numerical example refers to the minimization of the \emph{structural mean compliance} of a cantilever beam in a prismatic domain subjected to specific Dirichlet and Neumann boundary conditions. The displacements are prescribed on the left face of the design domain and a distributed vertical load is applied on the bottom-right edge of it. The analysis domain $\Omega$, displayed in Figure \ref{fig_cantilever_domain}, corresponds to a prism of (relative) dimensions 1x2x1, with the largest dimension oriented in the $y$-axis. Thanks to the symmetry with respect to the $y-z$ plane, half of the domain is discretized with $50$x$200$x$100$ unit cubic hexahedral elements.
		
		This benchmark will be used to check the correctness of the implementations as well as to have a first comparison of the results obtained with each technique in terms of the number of iterations, objective function and topology quality. Some optimal layouts for different volume fractions can be found in \cite{Aage2014,Lazarov2010}.

	\subsection{L-shaped structure} \label{sec_exmpl_Lshape}

		\begin{figure*}[pt]
			\floatbox[{\capbeside\thisfloatsetup{capbesideposition={left,center},capbesidewidth=6cm}}]{figure}[\FBwidth]
			{\caption{L-shaped structure: topology optimization domain with boundary conditions. A point-wise vertical force $F$ is applied on the right-bottom surface while the displacements $\mathbf{u}$ are prescribed on the top-left boundary of the design domain. The rear boundary corresponds to the $y-z$ symmetry surface. A considerable fraction of the domain (on the top-right side of the domain) is prescribed to void material.} \label{fig_Lshape_domain}}
			{\includegraphics[width=7cm]{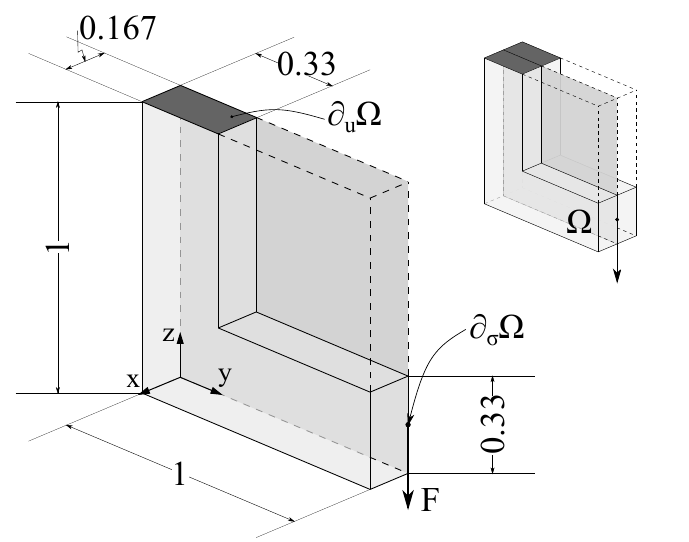}}
		\end{figure*}
		
		This second example tackles the optimization of a simplified version of a hook, as it is shown in Figure \ref{fig_Lshape_domain}. This optimization problem, as the previous one, also corresponds to the minimization of the \emph{mean compliance} (\ref{eq_minimization_restricted_compliance}). The analysis domain is split into two regions: 1) the L-shaped structure, which corresponds to the design domain $\Omega$, and 2) a prismatic volume prescribed as void in the top right area, defined by $y\geq\frac{1}{3}$ and $z\geq\frac{1}{3}$. A single vertical load is applied as illustrated in Figure {\ref{fig_Lshape_domain}} at point $x=0$, $y=1$ and $z=\frac{1}{6}$ and the displacements at the top side, near the left edge, are prescribed (i.e., $y\leq\frac{1}{3}$ and $z=1$). The design domain, with symmetry in the $y-z$ plane, is discretized with a structured mesh of $30$x$180$x$180$ hexahedral elements.
		
		Similarly to the previous problem, this example will provide a comparison between the approaches, but now with a more complex design domain (a rough edge domain with a point-wise load), thus showing the performance of the methods against point loads and non-rectangular design domains. The reader is sent to \cite{Liu2014} for the reference optimal solution.

	\subsection{Multi-load cantilever beam} \label{sec_exmpl_cantilever_multi}
	
\begin{figure*}[b]
			\floatbox[{\capbeside\thisfloatsetup{capbesideposition={right,center},capbesidewidth=4.5cm}}]{figure}[\FBwidth]
			{\caption{Multi-load Cantilever beam: (a) topology optimization domain with boundary conditions. The displacements are prescribed on the left surface of the domain, and a vertical distributed downward force $F_1$ is applied in the first loading case (b), whereas a vertical distributed upward force $F_2$, in the second loading case (c). The rear boundary of the domain corresponds to the symmetry surface.} \label{fig_cantilever_multi_domain}}
			{\includegraphics[width=11cm]{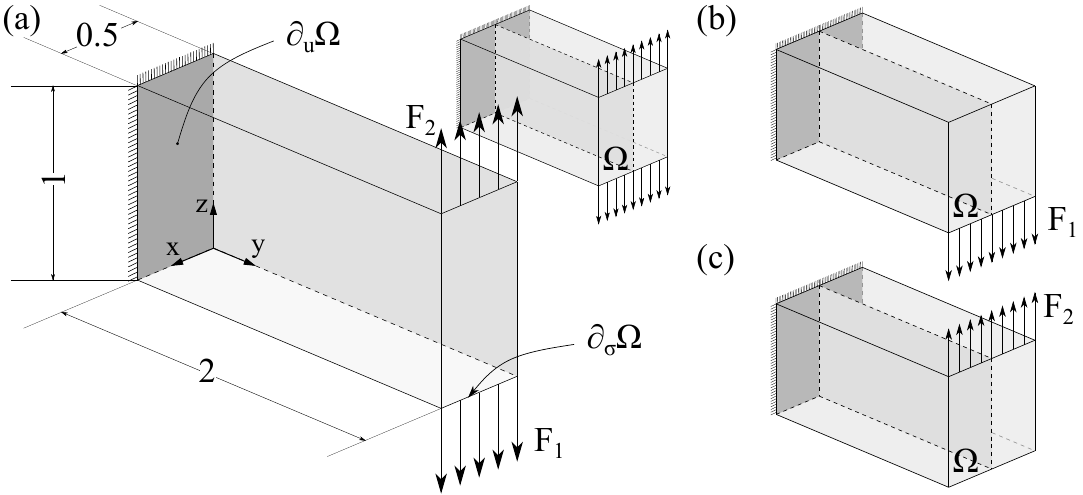}}
		\end{figure*}

		A \emph{multi-load topology optimization problem} with two different loading conditions not applied at once is optimized in this example. Both the dimensions of the prismatic domain, $\Omega$, and the discretization mesh are the same ones as in the first example (see Section \ref{sec_exmpl_cantilever}). However, in this case, the displacements of all the nodes on the left side are imposed and two loading conditions are applied on the top and bottom-right edges. In the first loading condition, a vertical distributed downward force is applied on the bottom-right edge while in the second one, a distributed force with the same magnitude is applied upwards on the upper-right edge, as displayed in Figure {\ref{fig_cantilever_multi_domain}}.
		
		Through this numerical case, it is aimed at determining whether the methods are capable of obtaining symmetric designs when two opposite forces are applied in the design domain, as well as to compare the topology quality and computational cost of the resultant optimal topologies with all the considered techniques.

	\subsection{Gripper compliant mechanism} \label{sec_exmpl_complmech}
		
\begin{figure*}[t]
			\floatbox[{\capbeside\thisfloatsetup{capbesideposition={left,center},capbesidewidth=6.5cm}}]{figure}[\FBwidth]
			{\caption{Gripper (\emph{compliant mechanism}): (a) topology optimization domain with boundary conditions. The displacements are prescribed at the bottom part of the left surface of the domain, and a positive, horizontal distributed load is applied at the top of the left surface for the state equation (b), while a positive, vertical distributed dummy load is applied at the jaws of the gripper for the \emph{additional state equation} (c). The top and rear surfaces correspond to the $x-y$ and $x-z$ symmetries, respectively.} \label{fig_complmech_domain}}
			{\includegraphics[width=8.5cm]{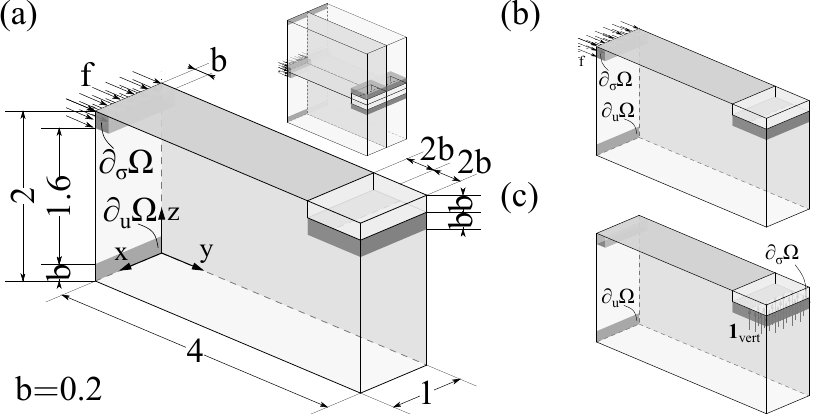}}
		\end{figure*}

		In this last numerical example, a \emph{compliant mechanism} is designed where the vertical displacement at the \emph{output port} is maximized. The displacements are prescribed near the bottom edge at the left side of the domain ($y=0$ and $z\leq0.2$). As illustrated in Figure \ref{fig_complmech_domain}, a positive, horizontal distributed load is applied at the \emph{input port} ($y=0$ and $z\geq1.8$), while a vertical upward \emph{dummy load} is applied at the \emph{output port} ($z=1.8$ and $y\ge3.6$). The analysis domain $\Omega$, whose (relative) dimensions are $2$x$4$x$4$, is discretized with a mesh of $100$x$200$x$200$ hexahedral elements. However, thanks to the two existing symmetries, only a quarter of the domain is analyzed, thus leading to 1.000.000 finite elements. In addition, two regions near the \emph{input} and \emph{output ports} are prescribed to stiff material to guarantee stiff material in those areas ($\Delta z=0.2$). 

		Additionally, surface distributed springs are included in the \emph{input} and \emph{output ports} (in the same direction as the target displacement) to restrict the displacement amplitude at these areas and simulate both the input work of the actuator and the elastic reaction work at the output port. The corresponding numerical values for the springs are $K_{in}=1.5\cdot10^{-1}N/m^3$ and $K_{out}=1.5N/m^3$, while the distributed forces are $f_1=3.81\cdot10^{-3}N/m^2$ and $f_2=3.81\cdot10^{-4}N/m^2$, respectively. Note that the optimal solution will heavily depend on the ratios of these parameters, however not all parameter combinations will ensure a convergent admissible solutions. For that reason and due to the non-semi-definite \emph{topological derivative}, this last example will provide an analysis of the performance of the different techniques with respect to the \emph{design of compliant mechanisms}, produced either with localized hinges or deformable bars (for optimal reference solutions refer to \cite{Takezawa2010,Yamada2010,Huang2014}).
				
\section{Comparison of methods} \label{sec_comparison_results}

	\subsection{Comparison settings}
		
		In the following subsections, the basis of the comparison will be detailed, specifying the platform on which Matlab will be executed, the versions for the Matlab codes of each approach as well as the specific parameters used in each method and numerical example. In addition, it is important to define equivalent convergence criteria for the different techniques in order to guarantee a fair comparison in terms of the computational cost.
		
		\subsubsection{Computing cluster features}

			The benchmark cases are solved on a cluster, in which each node consists of two AMD EPYC 7451 with 24 cores (48 threads) each one at 2.9 GHz and 1 TB DDR4 RAM memory at 2666 MHz. Each example is solved using eight cores and 99 GB of RAM memory to ensure enough memory for each of the numerical benchmarks. In this way, a greater number of cases can be solved at once without affecting the result of each approach. All these cases are computed using modified codes in Matlab 2018b under Scientific Linux 7.2 (based on RedHat Enterprise 7.2).
			
		\subsubsection{Matlab codes}
			
			All the codes used in this paper are 3D extensions of the respective 2D codes, already published by their respective developers, preserving the original algorithmic structure. 
			
			Firstly, \emph{SIMP-based method} codes are based on the 2D implementation initially introduced by \citet{Sigmund2001} in the 99-line program for two-dimensional topology optimization. This program was later improved by \citet{Andreassen2010}, who vectorized the element loops in the assembly and filtering strategies. The code in Matlab was later extended to three-dimensional topology optimization problems by \citet{Liu2014}, who provided the analytical element stiffness matrix for a cubic hexahedral element. Therefore, the \emph{SIMP method with PDE-filtering} (SIMP$^{(I)}$) and \emph{SIMP method with convolution filter} (SIMP$^{(III)}$) take the basic scheme from the 82-line program (using a PDE filter) and the 71-line program (with the \lstinline|conv2| function) from \cite{TOPOPT2018}, respectively, and implement the formulation for the 3D elastic problem from \cite{Liu2014}. Both approaches use the \emph{optimality criteria (OC) method} combined with a \emph{sensitivity filtering} ($ft=1$) to solve the corresponding \emph{topology optimization problem}. It is worth noticing that the $L_\infty$ norm of the design variable has been replaced by a $L_2$ norm normalized with the size of the domain. In addition, some minor changes to the \lstinline|OC| Matlab function have been done to correctly consider active and passive elements. As mentioned, the \emph{SIMP method using time-advancing scheme} (SIMP$^{(II)}$) employs the same scheme as the \emph{SIMP method with PDE-filtering} (SIMP$^{(I)}$), but this time using a \emph{time-advancing scheme}, similar as the one implemented in \emph{\methodabb{}} \cite{Oliver2019,Yago2020}.
			
			Secondly, a \emph{Soft-kill BESO} code, implemented in Matlab, has been adapted from the one presented by \citet{Huang2010} in chapter 4 for \emph{2D topological stiffness optimization}. This code has been extended to three-dimensional problems mimicking the Matlab code for SIMP$^{(III)}$, since they share most of the general scheme of the algorithm. Some modifications have been done to adapt the specific updating scheme, the sensitivity filtering along with the corresponding temporal filtering (the so-called \emph{averaging scheme} in \cite{Huang2010}). Additional minor changes must be implemented in the sensitivity and stiffness computations, since the original SIMP material interpolation is used instead of the one implemented in \cite{Andreassen2010}, with two unique discrete values: $\rho=\{1,\rho_{min}\}$ for the \emph{minimum mean compliance problem} (see Section \ref{sec_theroy_SOFTBESO}). In this particular scenario, the minimum value $\rho_{min}$ is imposed to elements in the void domain to avoid \emph{zero stiffness elements}. Additionally, a similar $L_2$ norm of the topology, implemented for \emph{SIMP-based methods}, is here also used to check if the topology has converged in addition to the existing one in objective function.
			
			Thirdly, the Matlab code for \emph{\methodabb{}} is a 3D extension of the corresponding program for 2D topology optimization problems provided in \cite{Yago2020}. The element stiffness matrices, as a product of the strain-displacement matrix with the nominal constitutive tensor, are precomputed for the non-bisected and bisected elements, here termed \emph{mixed elements}. In this way, once the type of element has been determined, the global stiffness matrix can be quickly computed and assembled. In addition, the \emph{pseudo-energy density} is also easily computed from the matrices calculated with the reference element. In this case, a \emph{Laplacian smoothing}, applied to the \emph{pseudo-energy density} at each iteration, is precomputed at the first iteration as implemented in SIMP$^{(I)}$. Finally, the \emph{Lagrange multiplier} is computed using the \emph{closed-form optimality method} in conjunction with a \emph{modified marching cubes method} to compute the volume, explained in \citet{Oliver2019}. The convergence criteria defined in \cite{Oliver2019} are replaced with the objective function criterion and the topology criterion in terms of a \emph{relaxed design variable}, in addition to the volume constraint.
			
			Finally, the \emph{Level-set} method using the \emph{Relaxed Topological Derivative} (RTD) corresponds to a modification of the previous code for \emph{\methodabb{}}, where the updating scheme is changed. Instead of the original \emph{closed-form optimality criteria}, a \emph{Hamilton-Jacobi} equation is used to update the \emph{level-set function} with a given $\Delta t$, as detailed in \citet{Oliver2019}. An \emph{Augmented Lagrangian method} is used to ensure the volume constraint equation ${\cal C}_0$. The stiffness matrices as well as the different terms required to compute the \emph{pseudo-energy density} are precomputed at the initial iteration. The same \emph{Laplacian smoothing} as in \emph{\methodabb{}} is here applied to the \emph{level-set function} at each iteration. In addition to the two existing convergence criteria, the volume constraint must be also considered in the outer loop. 
		
		\subsubsection{Guidelines for the comparison} \label{sec_convergence_criteria}
		
			The general guidelines for a fair comparison are listed below:
			
			\begin{enumerate}
				\item Benchmark cases: the same numerical benchmark cases and finite element meshes must be used for every topology optimization approach. Four benchmark cases will be carried out using dense meshes (around 1 million finite element) in order to obtain \emph{high-quality designs}.
			
				\item Target volume fraction: the same ratio of the final material domain is imposed by the constraint equation in each approach, which is fulfilled through different techniques. The desired volume fraction corresponds to a small ratio of material with respect to the initial design domain, so that a large material reduction is achieved throughout  the topology optimization. For three-dimensional problems using high dense meshes, this value will commonly be between 80\% and 95\% of the design domain, $\Omega$, depending on user requirements. Nevertheless, it will depend on each specific numerical example and its respective boundary conditions as connections between the different boundary conditions areas must be preserved. This ensures a stiff connection between the nodes in which the loads are applied and those where the displacements are prescribed in the elastic problem.
				
				\item Objective function normalization: since not all the methods start from a full material configuration, an initial iteration with this material layout is computed in all Matlab codes as a reference iteration. The objective function value at this iteration, ${\cal J}_0$, is used to normalize the subsequent iterations in each method, thus obtaining equivalent values for each numerical example, technique and volume fraction. However, the use of different design variables, nodal\footnote{The \emph{characteristic function}, defined through the \emph{discrimination function}, is used as design variable in the \emph{Level-set} and \emph{\methodabb{} methods}. Therefore, the material interface is precisely defined by the \emph{level-set function} or \emph{discrimination function}, respectively.} versus element\footnote{The density is defined for each element either using a continuous material interpolation for the \emph{SIMP-like approaches} or a discrete solution for the \emph{Soft-kill BESO method}.} variables, produces huge discrepancies in the actual objective function value since the stiffness of semi-dense elements is underestimated \cite{Sigmund2007}. For that reason, an additional final iteration is computed with an element \emph{black-and-white} configuration, i.e $\chi,\rho=\{1,10^{-9}\}$, thus obtaining a fully equivalent objective function value.  Nevertheless, it is important to point out that this configuration is not practical from a design standpoint as the smoothness of the design is lost in the projection. The reader is addressed to Appendix \ref{app_postprocess} for further details.
				
				\item Contrast factor: since each compared topology optimization approach defines a different material interpolation for the constitutive tensor $\mathbb{C}_\desvar{}$, it is important to ensure the same contrast factor $\alpha$, so that the same Young's modulus is used for the soft material when using the \emph{ersatz material approach}. This parameter may strongly affect the objective function value and the convergence of the topology optimization. A preliminary study has revealed that topology convergence can be achieved for contrast factors up to $\alpha=10^{-6}$, for \emph{minimum mean compliance problems}. Thus, this value will be used as \emph{contrast factor} for this type of problems.
				
				\item Convergence criteria: in order to guarantee a fair comparison, the convergence criteria of each approach must be replaced with the same equivalent conditions:
				
				\begin{itemize}
					\item Volume constraint: the same tolerance $Tol_{{\cal C}_0}=10^{-3}$ in the volume constraint (\ref{eq_minimization_restricted}-b-1) is assumed for all the topology optimization approaches, except in \emph{Level-set}. In this approach, the tolerance is slightly relaxed to $5\cdot10^{-3}$, in order to improve convergence when using the \emph{Augmented Lagrangian method}. 
				
					\item Objective function criterion: a (weighted) moving mean of the relative objective function $\cal J$ is evaluated along $n$ consecutive iterations as
					\begin{equation}
						\Delta {\cal J}_k = \frac{1}{n} \sum_{i=k-n}^{i=k} {\frac{ \vert {\cal J}_{i} - {\cal J}_{i-1}  \vert }{ {\cal J}_0  }} \mcolon
					\end{equation}
					where the parameter $k$ denotes the $k$-th iteration and ${\cal J}_0$ stands for the objective function at the initial iteration. The appropriate number of iterations $n$ depends on the topology optimization technique, as the number of iterations per time-step will not be the same. Notice that, for \emph{time-advancing schemes}, the $n+1$ iterations correspond to the same time-step, thus avoiding variations in the objective function due to the change of volume constraint. For all the benchmark cases, a tolerance $Tol_{{\cal J}}=10^{-3}$ in the objective function is prescribed.
					
					\item Topology criterion: a $L_2$ norm between two consecutive iterations is evaluated in a \emph{relaxed design variable} $\rho$ as
					\begin{equation}
						\Delta {\rho}_k = \frac{1}{\vert \Omega_0 \vert^{1/2}} \left( \int_\Omega { \left( \rho_k  - \rho_{k-1} \right)^2  \domega} \right)^{1/2} \mcolon
					\end{equation}
					where $k$ represents the iteration number and $\vert \Omega_0 \vert$ stands for the material volume at the first iteration. The design variable $\rho_k$ corresponds to a \emph{relaxed characteristic function} for discrete design variables (for instance, in the \emph{\methodabb{}}, \emph{SOFTBESO}, and \emph{Level-set methods}) or to the density variable for the \emph{SIMP approaches}. In Appendix \ref{app_convergence}, the reader can find the exact definition of this topology criterion for discrete design variables. For this criterion, the convergence tolerance is $Tol_\desvar = 2.5\cdot10^{-3}$.
					
				\end{itemize}
				
				For \emph{incremental time-advancing methods}, such as the \emph{\methodabb{}}, \emph{SIMP$^{(II)}$}, and \emph{Level-set methods}, a linear variation of the tolerances in cost and topology is defined, starting from a higher value for the first time-step (around one order of magnitude higher) to the value established in the last time-step. Consequently, all approaches obtain the convergence with the same criteria for the last increment (i.e.,~for the same stiff material fraction), thus resulting in a fair comparison. 
				
				It is important to stress that the objective function criterion is not a reliable indicator of convergence in \emph{compliant mechanism synthesis problems}. The normalized objective function oscillates significantly more than in \emph{minimum mean compliance problems} in all methods, thus preventing to obtain an optimal solution. This oscillation may be related to this specific type of problem, and, in particular, to the initial value of the objective function which may be null. As a consequence, the normalization of the objective function can not be performed, thus invalidating this convergence criterion. For these reasons, the objective function criterion has not been considered in the last numerical benchmark of this paper. However, it can be used in the other three benchmark cases since a proportionality between the objective and topology criteria is observed, both monotonously converging to the optimal values. In addition, the normalization problem is not detected in \emph{minimum mean compliance problems} as the external work is different from 0 in any case. In case it would also be omitted in these examples, no substantial change would be observed with respect to the obtained results, just resulting in minimal variations in the number of iterations to converge.
				
			\end{enumerate}
		
		\subsubsection{Parameter definition}
		
		    For each topology optimization method and benchmark case, a specific set of parameters must be defined. These parameters define the material behavior (via the \emph{contrast factor} or minimum Young's modulus), the volume fraction, the convergence tolerances, the exact updating parameters for the design variable, and the the ones for the regularization technique.
		    
		    Whenever possible, common values in material properties, volume fraction, and convergence tolerances will be imposed for the different benchmark cases and topology optimization methods. However, the specific updating parameters depend on each approach and numerical example due to convergence issues. In particular, a consistent difference is noticed regarding the definition of the parameters for \emph{minimum mean compliance} and \emph{compliant mechanism synthesis problems}, as already commented.
		
		    Regarding the target volume fraction, a $10\%$ volume fraction of stiff material and a \emph{contrast factor} $\alpha=10^{-6}$ are imposed for \emph{minimum mean compliance} problems. As for \emph{compliant mechanism synthesis problem}, the volume constraint is applied for a $15\%$ volume fraction, while \emph{contrast factor} is increased to $\alpha=10^{-2}$. In both problems, a linear isotropic material with a Young's modulus $E=1$ and Poisson's ratio $\nu=0.3$ is used.
		
		    The updating and regularization parameters depend on each approach, and in certain cases, on the optimization problem. The exact values of these parameters are detailed in Appendix \ref{app_reproducibility}. In general, the parameters of each method are given as follows:
		    \begin{itemize}[noitemsep]
		        \item \textbf{SIMP-based methods}: the penalty value and the minimum radius are prescribed to $p=3$ and $r_{min}=3$, respectively. A sensitivity filtering ($ft=1$) is used for the topology optimization, as aforementioned. The updating parameters $m$ and $\eta$ are defined according to the optimization problem, corresponding to $0.2$ and $0.5$ for \emph{minimum compliance problems}, and 0.1 and 0.3, respectively, for \emph{compliant mechanism design problem}.
		        \item \textbf{SOFTBESO}: the same values for the penalty factor and minimum radius as those in SIMP are used. The \emph{evolutionary ratio} $ER$ and the \emph{maximum volume addition ratio} $AR_{max}$ are prescribed to $0.01$ and $0.1$, respectively.
		        \item \textbf{\methodabb{}}: the number of steps $n_{steps}$, the exponential factor $m$ and the regularization factor $\tau$ depend on each problem. However, the same value of time-steps as SIMP$^{(II)}$ is used for each benchmark case.
		        \item \textbf{Level-set}: in contrast to SIMP$^{(II)}$ and \methodabb{}, the optimizations are carried out with a single time-step. However, the same exponential factors as the ones in \methodabb{} are employed, and the regularization factor $\tau$ is set to 1. In addition, the time-increment $\Delta t$ and the penalty coefficient $s$ of the \emph{Augmented Lagrangian method} also change with the optimization problem.
		    \end{itemize}
		    For \emph{incremental time-advancing techniques} with multiple time-steps, the volume fraction of stiff material at each step is reduced following an exponential evolution
		    \begin{equation}
		        f_j  = f_0 + \frac{\overline{f}-f_0}{1-e^k}\left( 1- e^{k\frac{j}{n_{steps}}} \right), \; j:1\dots n_{steps}
		    \end{equation}
		    with factor $k=-2$.
			
	\subsection{Results}
			
		The results obtained from the six topology optimization approaches are now compared with each other for every of the numerical benchmarks (see Section {\ref{sec_benchmark}}). The comparison is carried out in terms of the optimal topology, objective function value, and the computational cost, discussing the relative objective function values and the relative computation costs. In addition, an analysis of the convergence is also performed. Finally, an overall comparison of the different methods is made according to the results.
	
		\subsubsection{General discussion of results}\label{subsec_topology}
		
			The optimal topology layout for the required material volume is here compared for the six different approaches from a quantitative and qualitative standpoint. In particular, the quality of the optimal solution is discussed together with the minimum filament size of the resulting (linear) pieces of the final design (bars), the computational cost in terms of the iterations, and the normalized value of the objective function for the four addressed benchmark cases.
			
			\paragraph{Cantilever beam} The final solutions of the Cantilever beam problem for each topology optimization technique are displayed in Table \ref{tab_mths_cantilever}. Although the resultant topologies are quite different, all methodologies except for \emph{SIMP$^{(II)}$} obtain a similar overall optimal design consisting of two separated webs. However, these webs present a different internal layout and topology complexity. The designs obtained from \emph{SIMP$^{(I)}$}, \emph{SIMP$^{(III)}$}, \emph{\methodabb{}}, and \emph{Level-set} illustrate a much simpler design based on bars, while \emph{SOFTBESO} produces an optimal design with a thin web (or a high number of close bars), with almost constant thickness. On the other side, \emph{SIMP$^{(II)}$} finds a different optimal layout with a single continuous central web.
			
			The mean bar width value $\overline{h}$, computed as the ratio between the stiff volume and the surface area of the solution, provides feedback on the complexity of the optimal design. For low $\overline{h}$ values, as in \emph{SOFTBESO}, the optimal solution is made of a large number of thin bars, making it more difficult to manufacture and more likely to buckle. As this number increases, the width of the bars tends to increase, thus simplifying the complexity of the design, as in \emph{SIMP$^{(I)}$}, \emph{VARTOP}, or \emph{Level-set}. Furthermore, these topologies are less prone to buckling effects.
			
			The topologies can be also compared in terms of the corresponding value of the objective function. It can be noted that as the number of bars increases and/or the size of these bars decreases (tending to a single continuous web in the limit), the value of the objective function decreases, as it is observed in the \emph{SOFTBESO} and \emph{SIMP$^{(II)}$ approaches}. On the contrary, \emph{Level-set} and \emph{\methodabb{}} optimize the design layout using thicker bars, thus obtaining a higher compliance value\footnote{The bar width could be reduced by modifying the value of the regularization parameter, $\tau$.}. Similar designs and objective function values are obtained via \emph{SIMP$^{(I)}$} and \emph{SIMP$^{(III)}$}.
			
			Finally, the techniques can be compared in terms of the number of iterations (i.e.,~a computational cost-measure). As detailed in Table \ref{tab_mths_cantilever}, \emph{\methodabb{}} requires fewer iterations (116) to achieve the optimal topology layout while not being so far from the optimal topologies obtained by the other approaches. It is closely followed by \emph{SIMP$^{(III)}$} with 124 iterations, and with a few iterations more one can find \emph{SIMP$^{(I)}$}, \emph{SIMP$^{(II)}$}, and \emph{SOFTBESO} with 175, 231 and 272 iterations, respectively. The \emph{Level-set} method takes many more iterations (1266) to converge.
			
			\begin{table*}
			\caption{Comparison of the results of topology optimization methods for the Cantilever beam. The number of iterations, objective function values, and mean bar widths $\overline{h}$ are given for each of the addressed approaches. The optimal topology is also illustrated in the last two columns, via an isometric view and a side view. \label{tab_mths_cantilever}} 
				\begin{tabular}{ ||m{1.6cm} m{1.25cm} m{1.25cm} m{1.25cm} p{9cm}|| }
					\hline
					Method & Total \newline iterations & Objective \newline function & $\overline{h} $ & Optimal solutions \\
					\hline
					\hline
					SIMP$^{(I)}$
					& 175
					& 6.4710
					& 2.1012
					& \begin{minipage}{9cm}
					  	\includegraphics[width=\linewidth,trim={0 0 1250pt 0},clip]{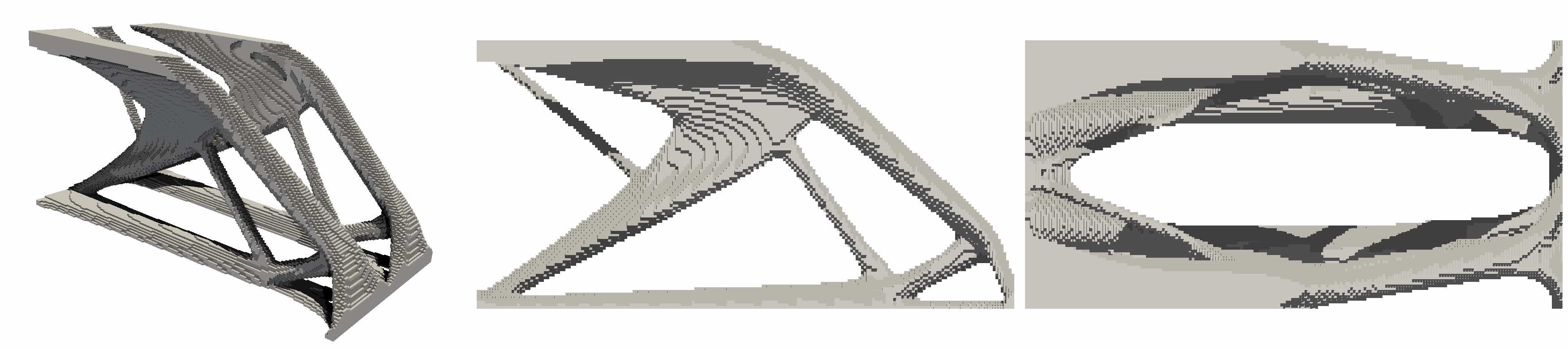}
					  \end{minipage} \\
					\hline
					SIMP$^{(II)}$
					& 231
					& 5.9609
					& 1.7888
					& \begin{minipage}{9cm}
						\includegraphics[width=\linewidth,trim={0 0 1250pt 0},clip]{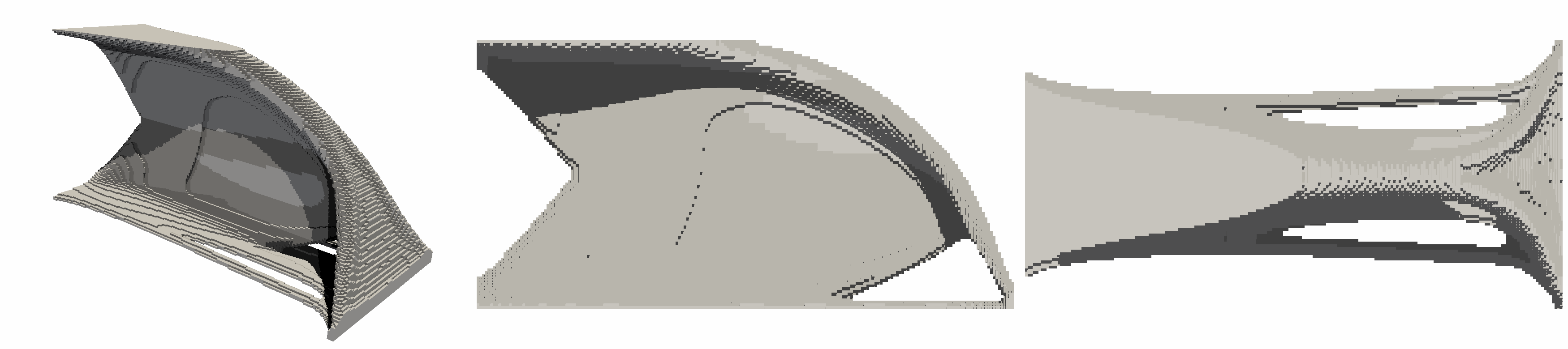}
					\end{minipage} \\
					\hline
					SIMP$^{(III)}$
					& 124
					& 6.7285
					& 1.9700
					& \begin{minipage}{9cm}
						\includegraphics[width=\linewidth,trim={0 0 1250pt 0},clip]{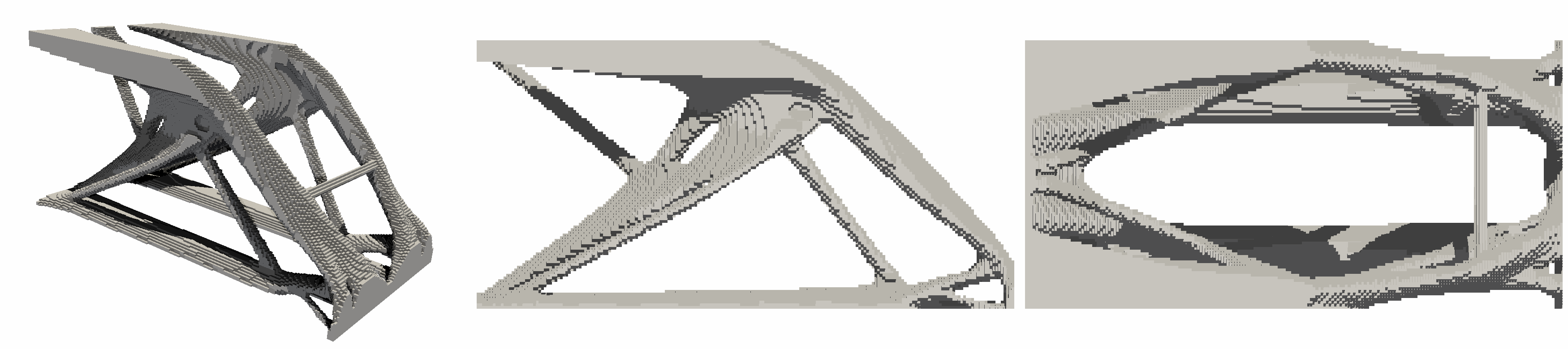}
					\end{minipage} \\
					\hline
					SOFTBESO
					& 272
					& 6.4369
					& 1.4586
					& \begin{minipage}{9cm}
						\includegraphics[width=\linewidth,trim={0 0 1250pt 0},clip]{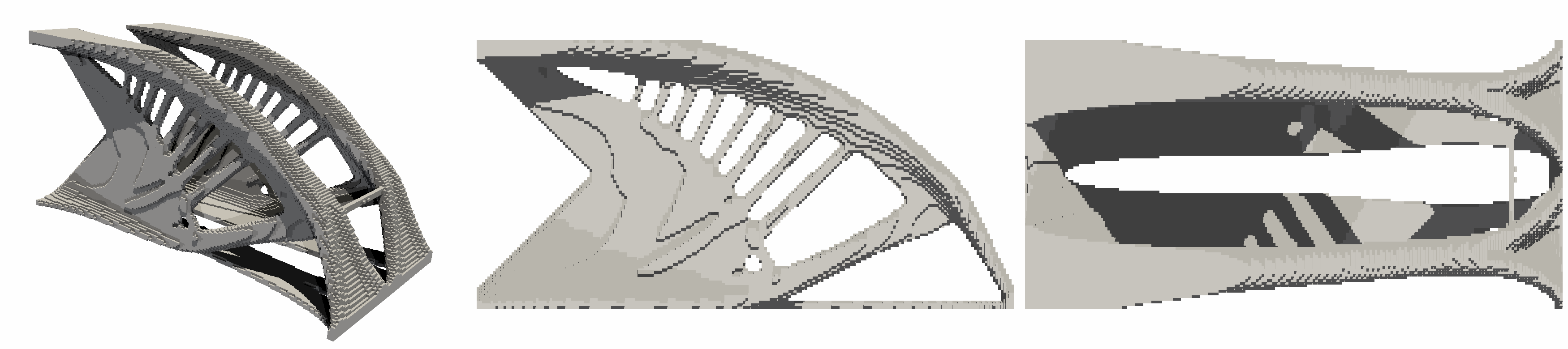}
					\end{minipage} \\
					\hline
					\methodabb{}
					& 116
					& 6.3981
					& 2.2783
					& \begin{minipage}{9cm}
						\includegraphics[width=\linewidth,trim={0 0 1250pt 0},clip]{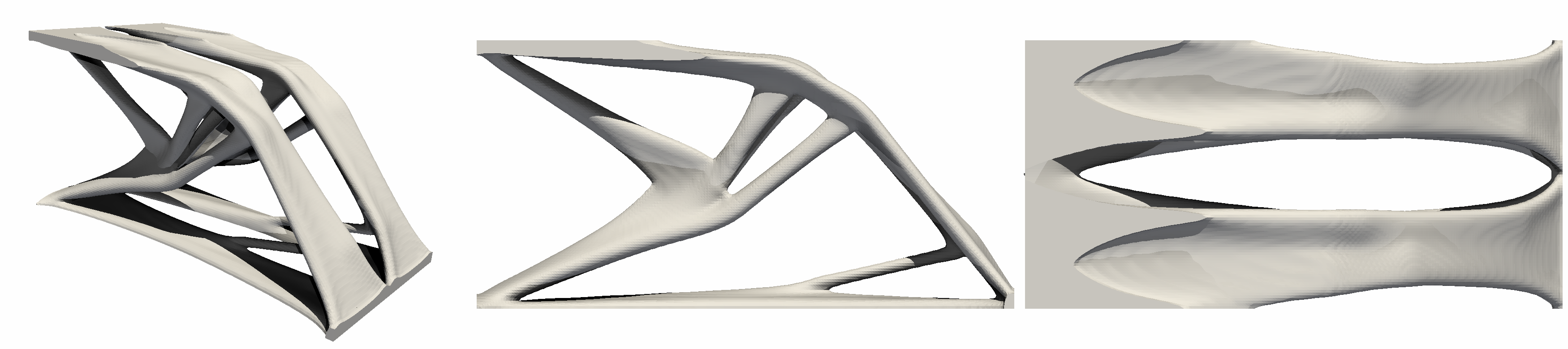}
					\end{minipage} \\
					\hline
					Level-set
					& 1266
					& 6.9494
					& 2.8121
					& \begin{minipage}{9cm}
						\includegraphics[width=\linewidth,trim={0 0 1250pt 0},clip]{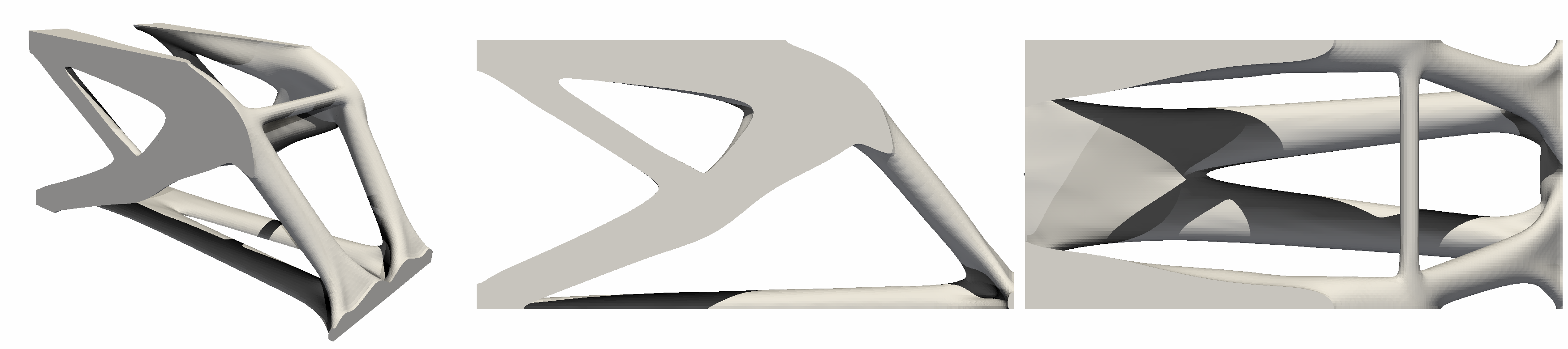}
					\end{minipage} \\
					\hline
				\end{tabular}
				\end{table*}
					
			\paragraph{L-shaped structure} The obtained results for the L-shaped structure are presented in Table {\ref{tab_mths_Lshape}}. As it can be noticed, the overall design of the structure is similar in all the methods. The vertical part, at the left, is almost identical in all the solutions. The most significant differences are found in the lower part of the structure, in which the topological complexity changes with the method. In this case, designs with 2 or 3 webs are obtained according to the approach, which connects the vertical part of the structure with the load application point. In particular, all the methods present a 2 web design except for the solutions of \emph{SIMP$^{(II)}$}, which displays an internal central structure. Additionally, all the designs are mainly constituted by bars, being thinner in \emph{SIMP$^{(II)}$} and \emph{SOFTBESO}, as displayed by the mean bar width value $\overline{h}$.
			
			Regarding the objective function value, the values for all the solutions are around a compliance value of 2.40-2.60 with respect to the initial reference compliance\footnote{An initial iteration with a full stiff material is computed. The passive elements/nodes are accounted for in this initial design.}. \emph{SIMP$^{(II)}$} achieves the topology design with the lowest objective function value, while other approaches produce solutions with a value closer to 2.5. \emph{SOFTBESO} obtains the solution with the highest objective function value (2.6).
			
			Finally, a comparison of the number of iterations shows that, again, \emph{\methodabb{}} requires fewer iterations than the other considered topology optimization techniques. There is not a large difference in the number of iterations with \emph{SIMP$^{(I)}$} or \emph{SIMP$^{(III)}$} (66 and 76 versus 62 of the \emph{\methodabb{}}), although the difference in iterations increases when compared to \emph{SIMP$^{(II)}$}, \emph{SOFTBESO} or \emph{Level-set} techniques, as observed in the previous example.
			
			\begin{table*}
				\caption{Comparison of the results of topology optimization methods for the L-shaped structure. The number of iterations, objective function values, and mean bar widths $\overline{h}$ are given for each of the addressed approaches. The optimal topology is also illustrated in the last two columns. \label{tab_mths_Lshape}} 
				\begin{tabular}{ ||m{1.6cm} m{1.25cm} m{1.25cm} m{1.25cm} p{7cm}|| }
					\hline
					Method & Total \newline iterations & Objective \newline function & $\overline{h} $ & Optimal solutions \\
					\hline
					\hline
					SIMP$^{(I)}$
					& 66
					& 2.4823
					& 2.1194
					& \begin{minipage}{7cm}
					  	\includegraphics[width=\linewidth,trim={0 0 660pt 0},clip]{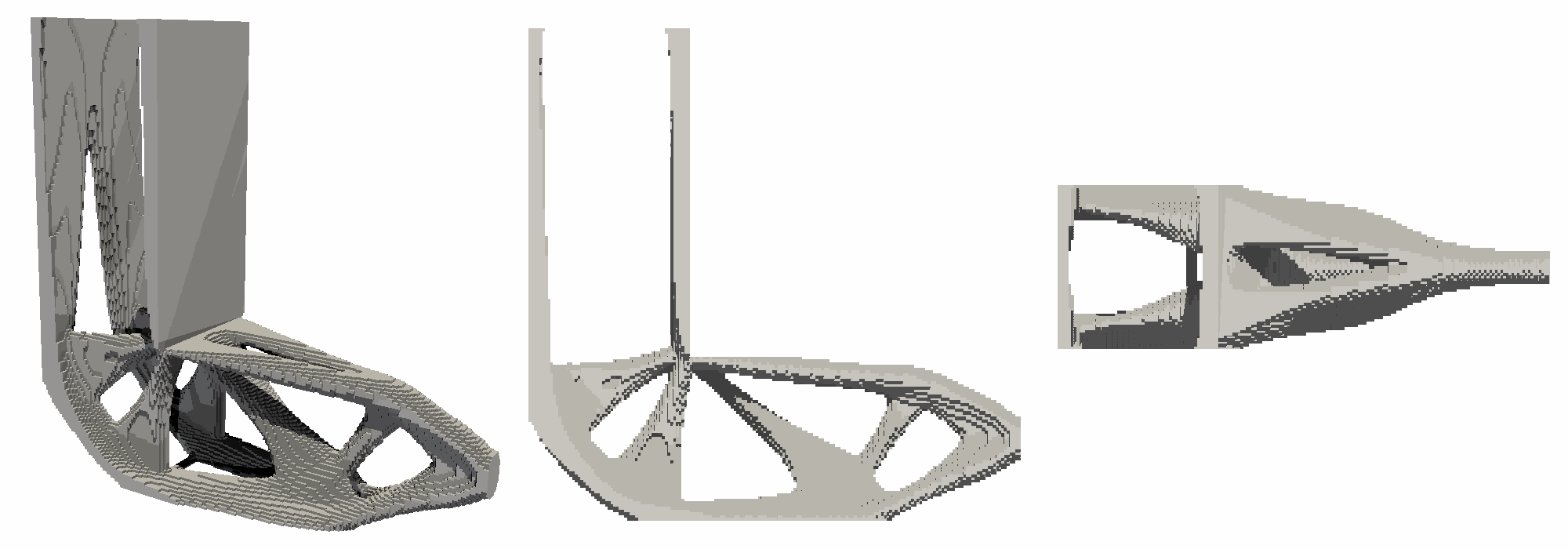}
 					\end{minipage} \\
					\hline
					SIMP$^{(II)}$
					& 140
					& 2.4229
					& 1.7722
					& \begin{minipage}{7cm}
						\includegraphics[width=\linewidth,trim={0 0 660pt 0},clip]{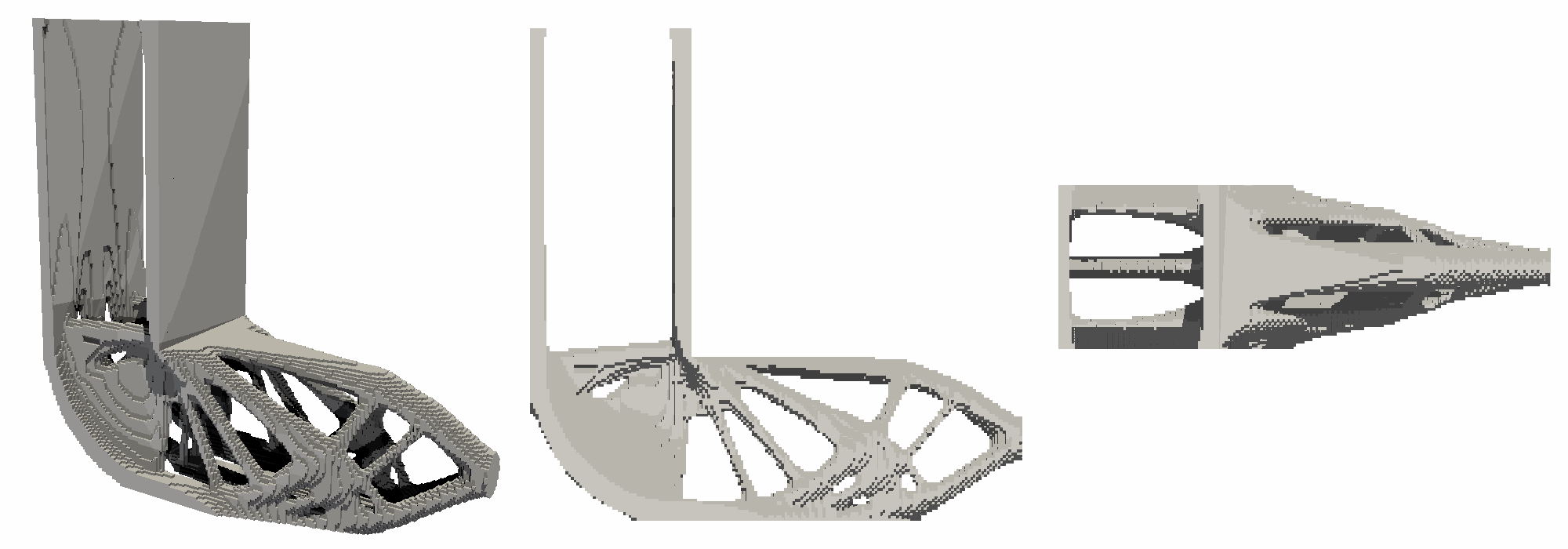}
					\end{minipage} \\
					\hline
					SIMP$^{(III)}$
					& 76
					& 2.4882
					& 2.1108
					& \begin{minipage}{7cm}
						\includegraphics[width=\linewidth,trim={0 0 660pt 0},clip]{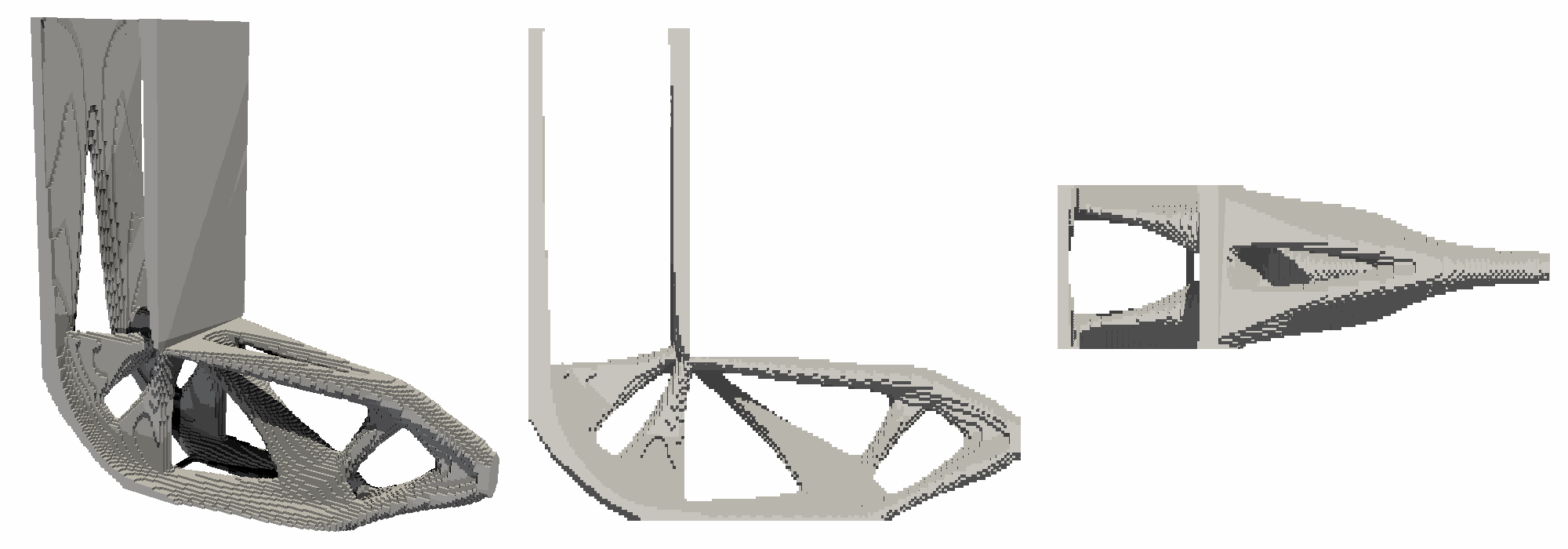}
					\end{minipage} \\
					\hline
					SOFTBESO
					& 190
					& 2.6171
					& 1.5100
					& \begin{minipage}{7cm}
						\includegraphics[width=\linewidth,trim={0 0 660pt 0},clip]{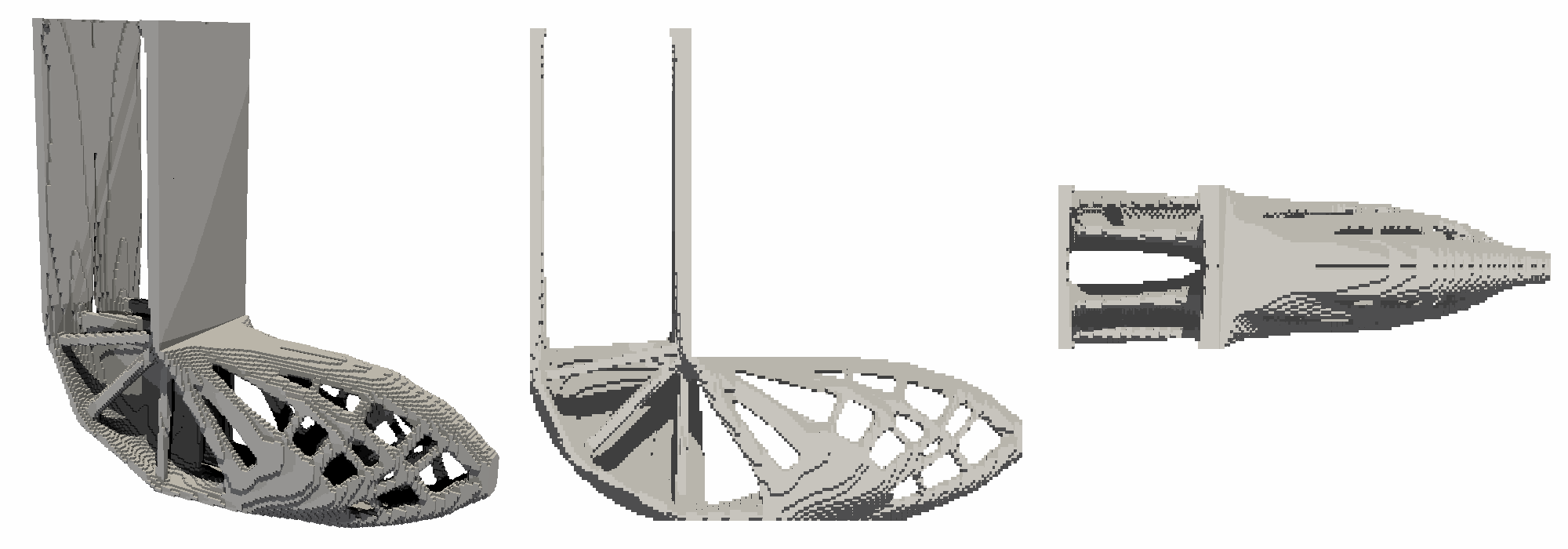}
					\end{minipage} \\
					\hline
					\methodabb{}
					& 62
					& 2.4811
					& 2.4903
					& \begin{minipage}{7cm}
						\includegraphics[width=\linewidth,trim={0 0 660pt 0},clip]{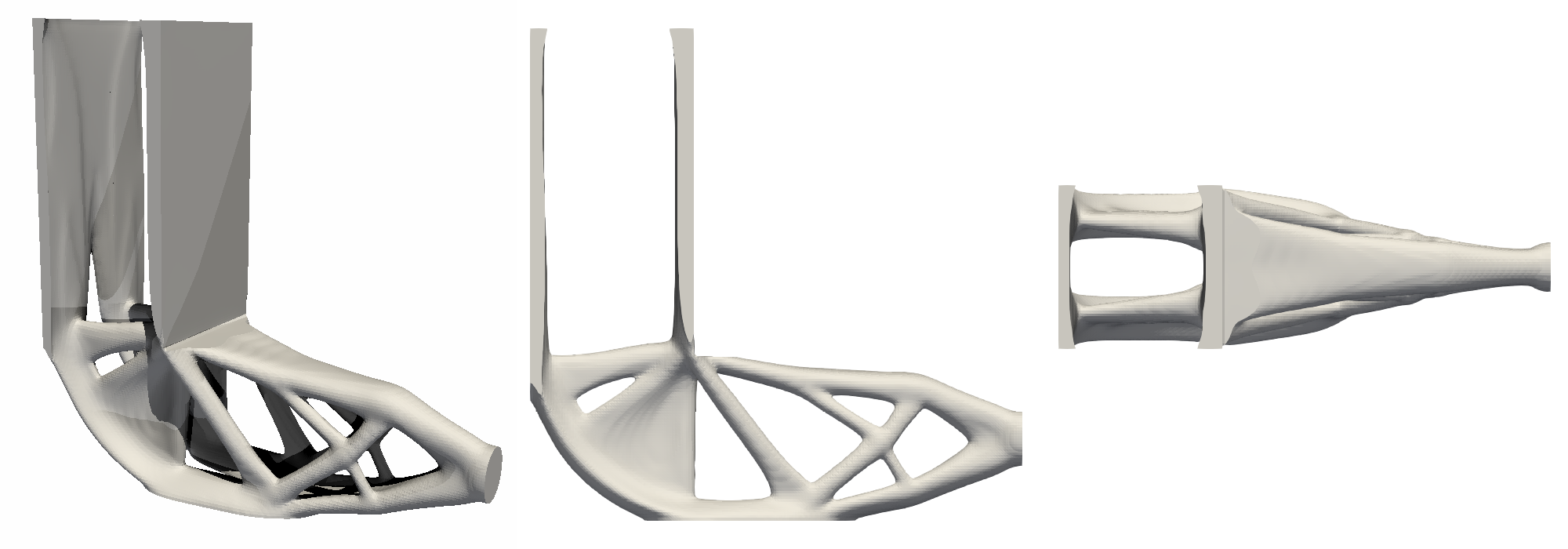}
					\end{minipage} \\
					\hline
					Level-set
					& 1071
					& 2.5163
					& 2.4532
					& \begin{minipage}{7cm}
						\includegraphics[width=\linewidth,trim={0 0 660pt 0},clip]{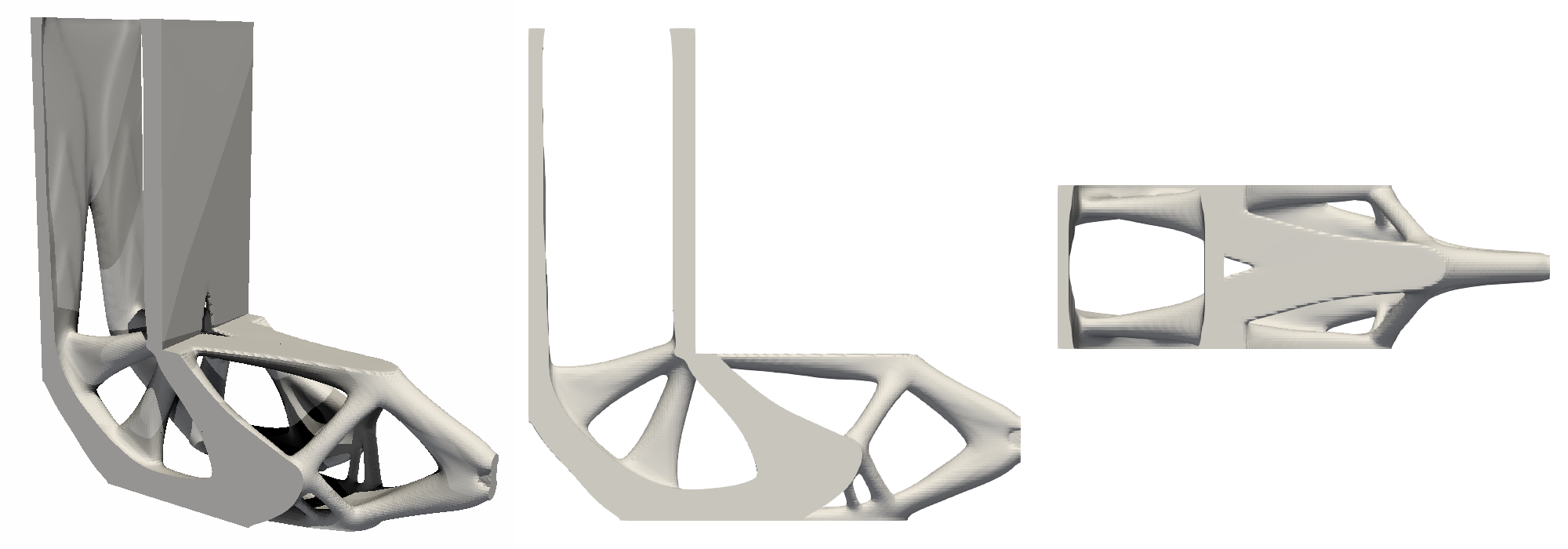}
					\end{minipage} \\
					\hline
				\end{tabular}
			\end{table*}
			
			\paragraph{Multi-load Cantilever beam} Table \ref{tab_mths_Multicantilever} presents the results obtained regarding the \emph{multi-load problem} for the corresponding approaches. In contrast to the previous cases, the resultant optimal solutions are quite different from each other. Although all six methodologies find symmetrical optimal solutions (with respect to the horizontal midplane of the domain), the resulting topologies do not correspond to the symmetrical solutions obtained for the first example (see table {\ref{tab_mths_cantilever}}), which could be intuitively presumed.
			
			Most of the solutions are based on bar designs except for the \emph{SOFTBESO} approach, which consists of a continuous core. For this reason, the mean bar size $\overline{h}$ is the lowest of all techniques. Nevertheless, the solutions obtained with \emph{\methodabb{}}, \emph{SIMP$^{(I)}$}, and \emph{SIMP$^{(III)}$} have many similarities, being the design of these last two techniques practically the same. Furthermore, the optimal layout achieved with \emph{SIMP$^{(II)}$} is made of 3 webs with thinner bars, having a similar overall design. It is important to stress that the solutions obtained using \emph{\methodabb{}} and \emph{Level-set} present the highest mean bar width, thus achieving the best designs from a manufacturing standpoint and buckling resistance. However, the solution of the \emph{Level-set} method corresponds to a different local minimum than the previous ones.
			
			The lowest objective function values are obtained by \emph{SOFTBESO} and \emph{SIMP$^{(II)}$} even though the designs are quite complex and can not be easily manufactured. Conversely, \emph{Level-set} finds the topology layout with the highest value. The other approaches (\emph{\methodabb{}}, \emph{SIMP$^{(III)}$}, and \emph{SIMP$^{(I)}$}) provide sufficiently manufacturable (low complexity) solutions with intermediate values. Similar to the previous cases, methods \emph{SIMP$^{(I)}$}, \emph{SIMP$^{(III)}$} and \emph{\methodabb{}} are the ones with the lowest computational cost, method \emph{SIMP$^{(I)}$} being 30\% faster than the other two methods.
			
			\begin{table*}	
				\caption{Comparison of the results of topology optimization methods for the Multi-load cantilever beam. The number of iterations, objective function values, and mean bar widths $\overline{h}$ are given for each of the addressed approaches. The optimal topology is also illustrated in the last two columns. \label{tab_mths_Multicantilever}}
				\begin{tabular}{ ||m{1.6cm} m{1.25cm} m{1.25cm} m{1.25cm} p{9cm}|| }
					\hline
					Method & Total \newline iterations & Objective \newline function & $\overline{h} $  & Optimal solutions \\
					\hline
					\hline
					SIMP$^{(I)}$
					& 81
					& 7.4271
					& 1.8343
					& \begin{minipage}{9cm}
					  	\includegraphics[width=\linewidth,trim={0 0 1250pt 0},clip]{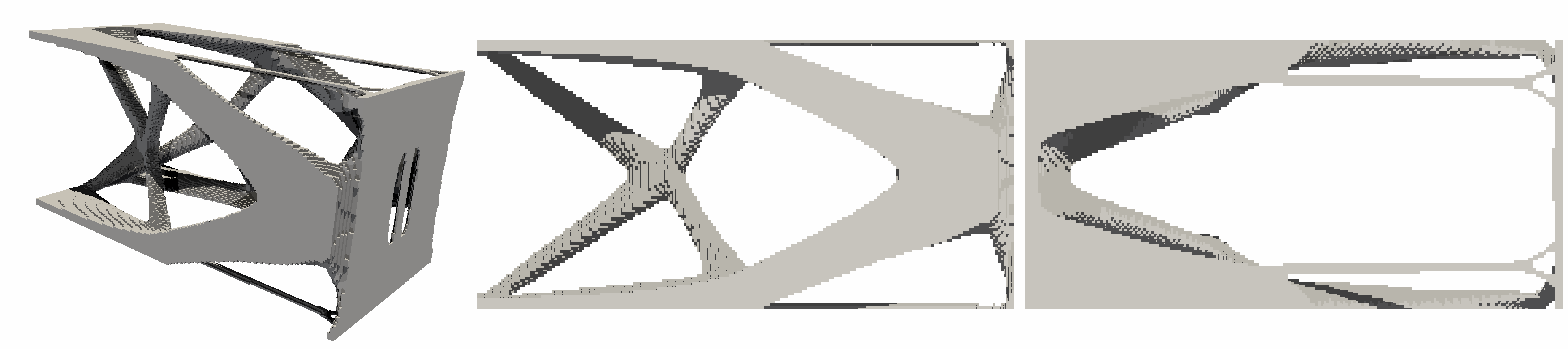}
 					\end{minipage} \\
					\hline
					SIMP$^{(II)}$
					& 208
					& 6.7774
					& 1.4312
					& \begin{minipage}{9cm}
						\includegraphics[width=\linewidth,trim={0 0 1250pt 0},clip]{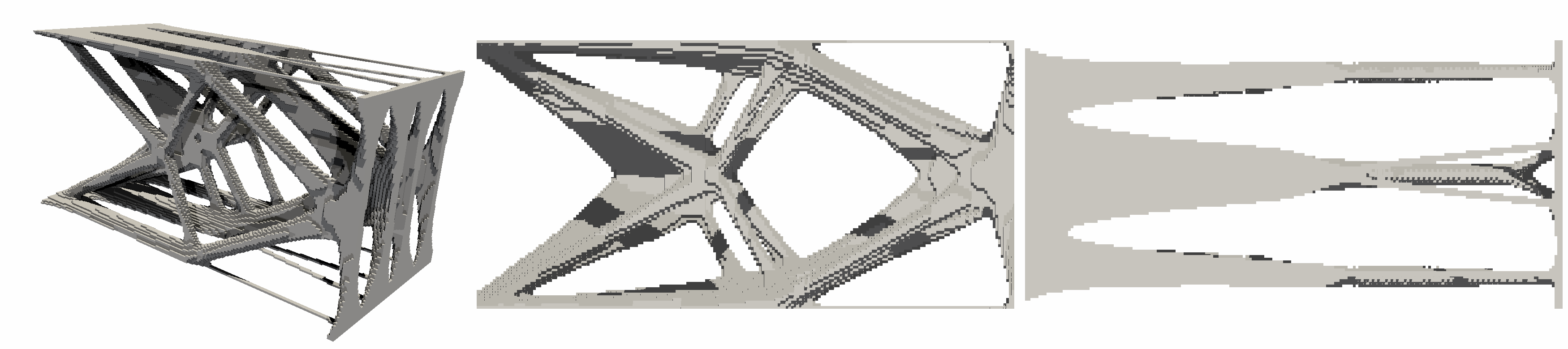}
					\end{minipage} \\
					\hline
					SIMP$^{(III)}$
					& 111
					& 7.3156
					& 1.8300
					& \begin{minipage}{9cm}
						\includegraphics[width=\linewidth,trim={0 0 1250pt 0},clip]{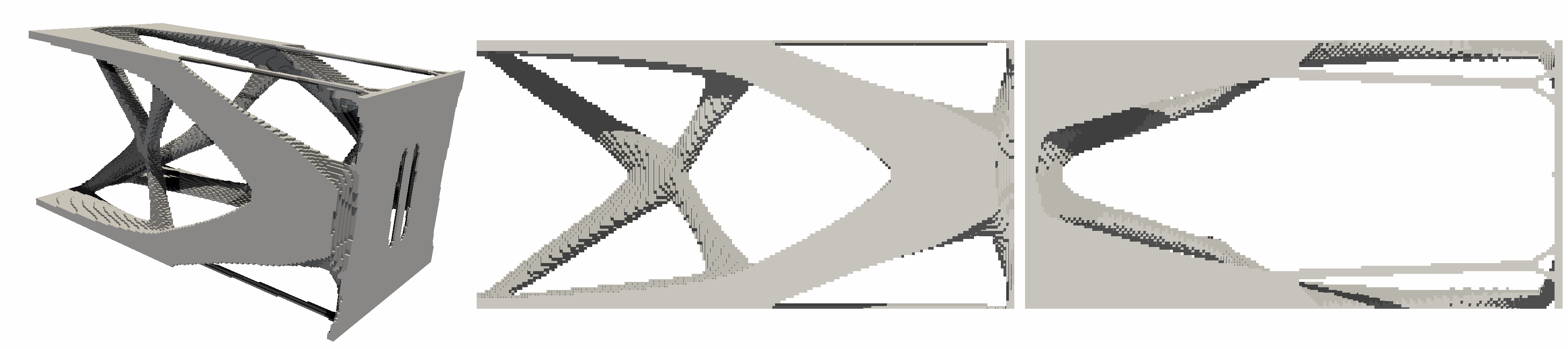}
					\end{minipage} \\
					\hline
					SOFTBESO
					& 249
					& 6.6406
					& 1.3108
					& \begin{minipage}{9cm}
						\includegraphics[width=\linewidth,trim={0 0 1250pt 0},clip]{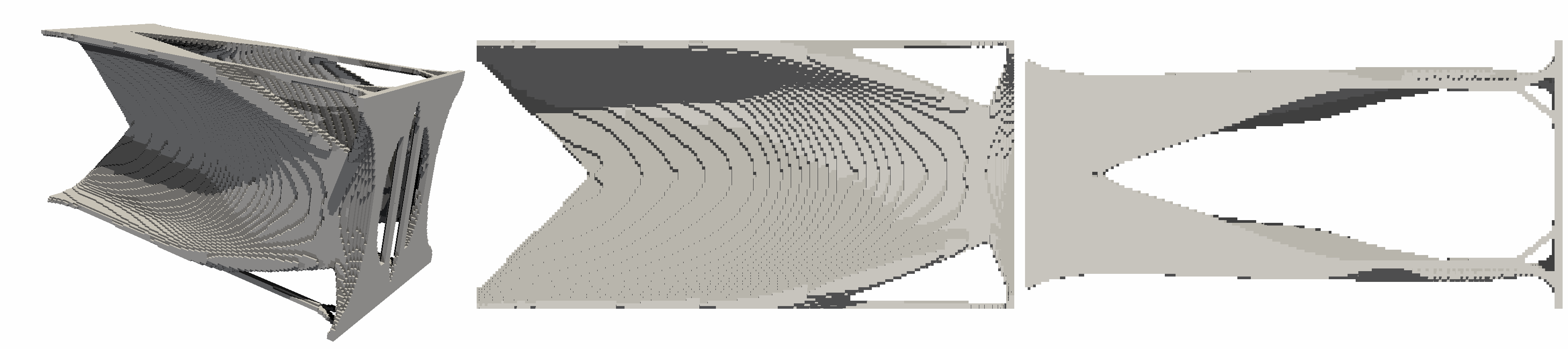}
					\end{minipage} \\
					\hline
					\methodabb{} 
					& 115
					& 7.0459
					& 2.5573
					& \begin{minipage}{9cm}
						\includegraphics[width=\linewidth,trim={0 0 1250pt 0},clip]{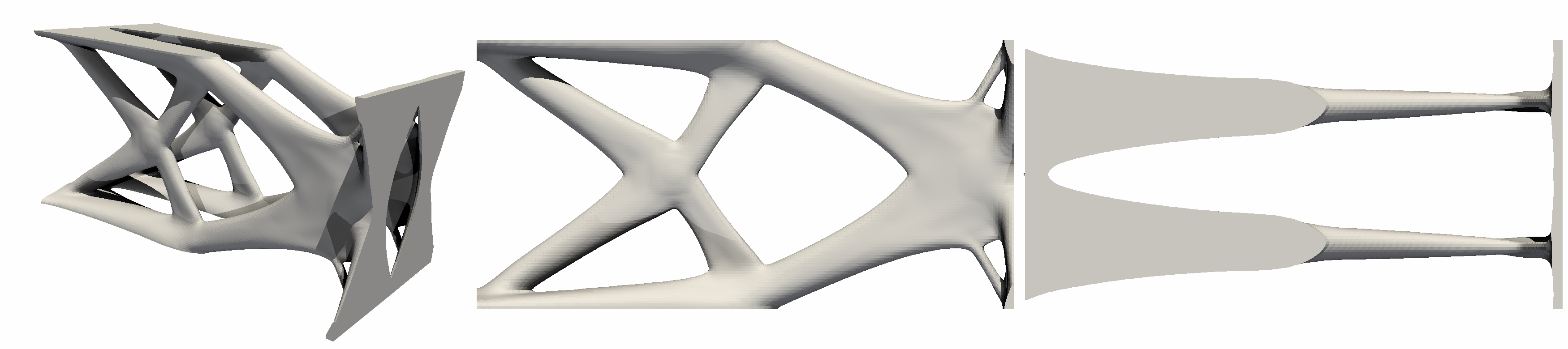}
					\end{minipage} \\
					\hline
					Level-set
					& 694
					& 8.2108
					& 2.5669
					& \begin{minipage}{9cm}
						\includegraphics[width=\linewidth,trim={0 0 1250pt 0},clip]{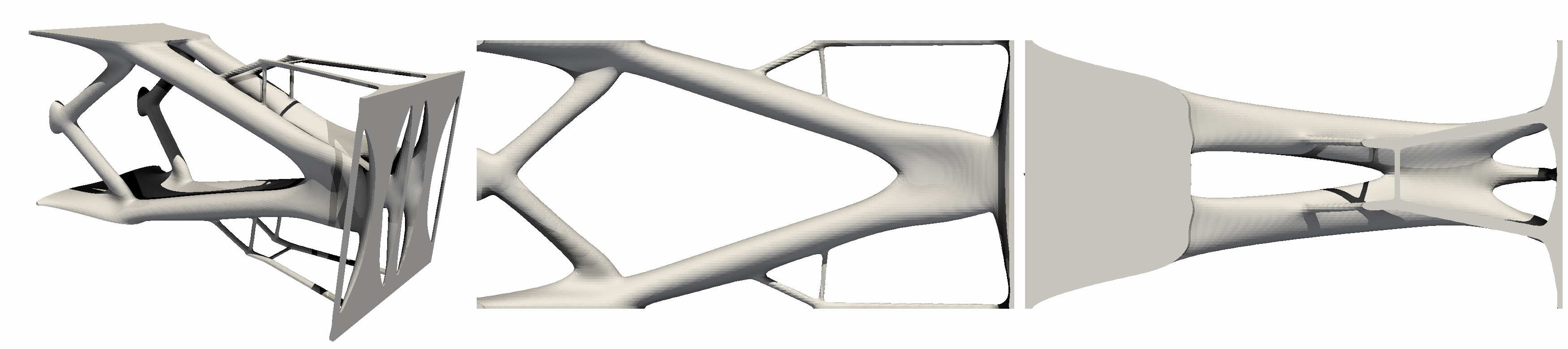}
					\end{minipage}  \\
					\hline
				\end{tabular}
			\end{table*}
						
			\begin{table*}
				\caption{Comparison of the results of topology optimization methods for the Gripper compliant mechanism. The number of iterations, objective function values, and mean bar widths $\overline{h}$ are given for each of the addressed approaches. The optimal topology is also illustrated in the last two columns. \label{tab_mths_gripper}}
				\begin{tabular}{ ||m{1.6cm} m{1.25cm} m{1.5cm} m{1.25cm} p{8cm}|| }
					\hline
					Method & Total \newline iterations & Objective \newline function & $\overline{h} $  & Optimal solutions \\
					\hline
					\hline
					SIMP$^{(I)}$
					& 325
					& -261.6980
					& 2.7041
					& \begin{minipage}{8cm}
					  	\includegraphics[width=\linewidth,trim={0 0 660pt 0},clip]{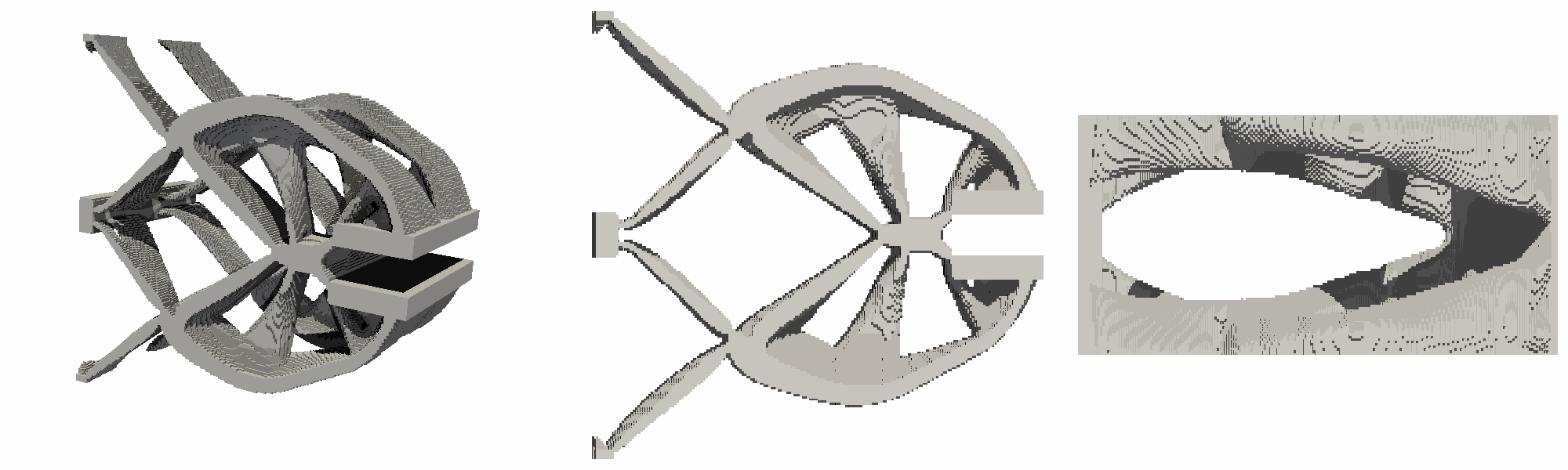}
 					\end{minipage} \\
					\hline
					SIMP$^{(II)}$
					& 297
					& -305.3930
					& 2.5020
					& \begin{minipage}{8cm}
						\includegraphics[width=\linewidth,trim={0 0 660pt 0},clip]{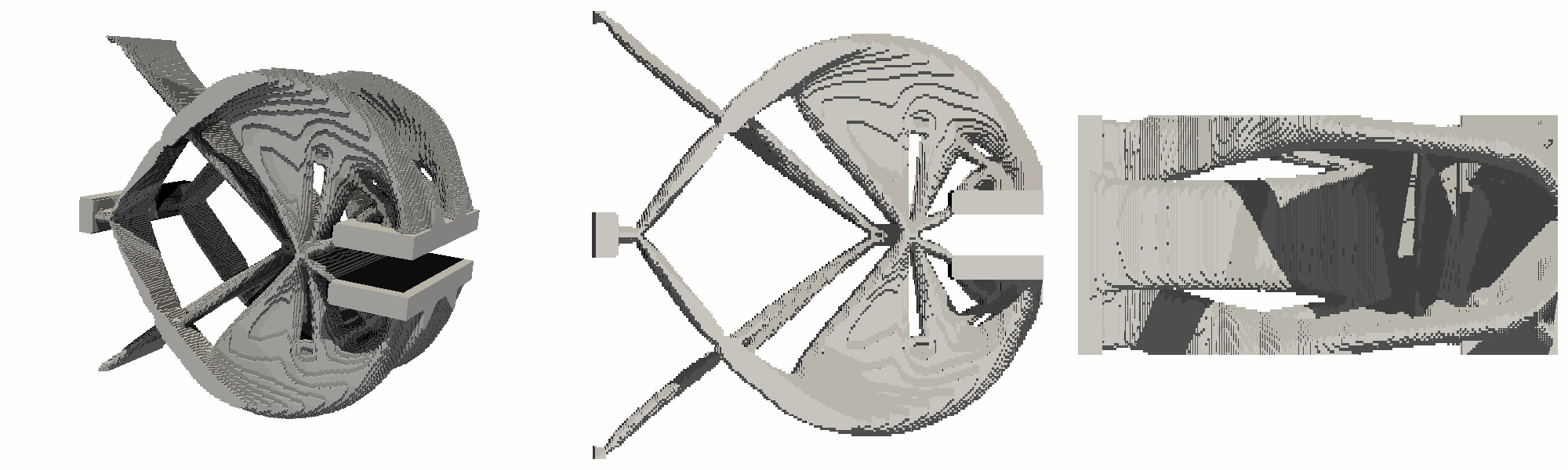}
					\end{minipage} \\
					\hline
					SIMP$^{(III)}$
					& 264
					& -297.0620
					& 2.4805
					& \begin{minipage}{8cm}
						\includegraphics[width=\linewidth,trim={0 0 660pt 0},clip]{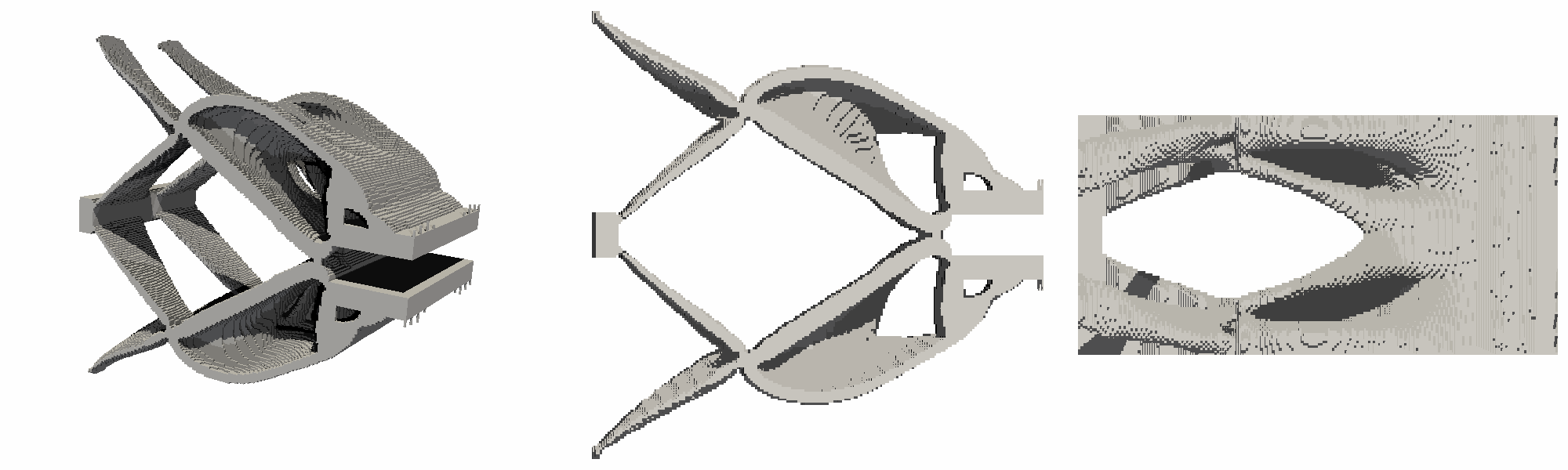}
					\end{minipage} \\
					\hline
					SOFTBESO
					& 207
					& -135.7380
					& 2.5441
					& \begin{minipage}{8cm}
						\includegraphics[width=\linewidth,trim={0 0 660pt 0},clip]{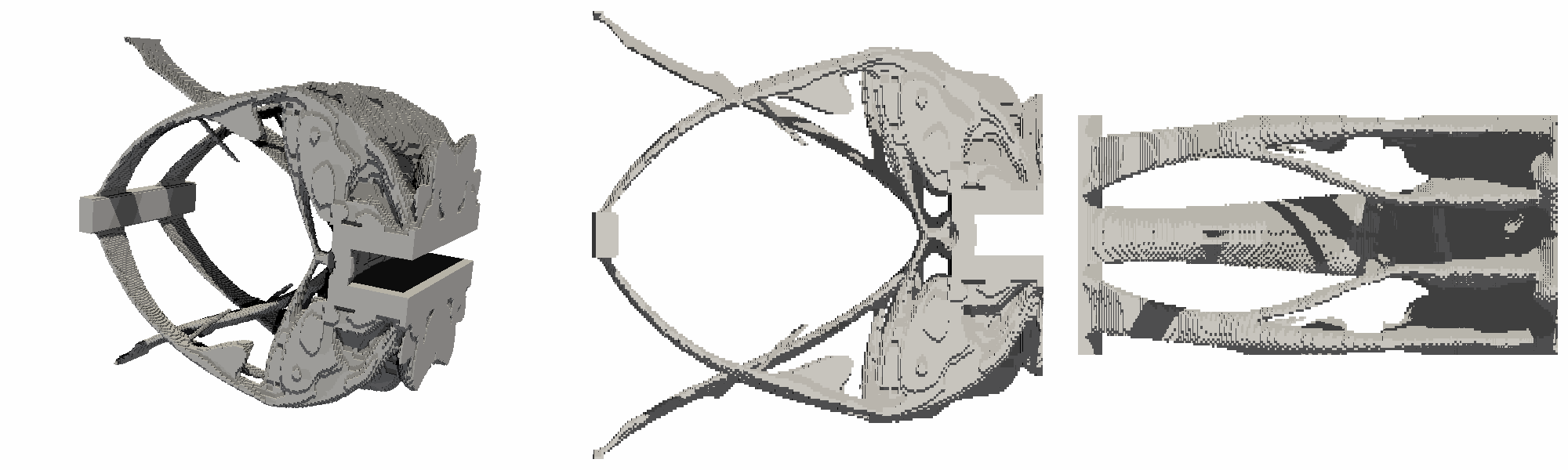}
					\end{minipage} \\
					\hline
					\methodabb{}
					& 32
					& -269.4409
					& 2.4553
					& \begin{minipage}{8cm}
						\includegraphics[width=\linewidth,trim={0 0 660pt 0},clip]{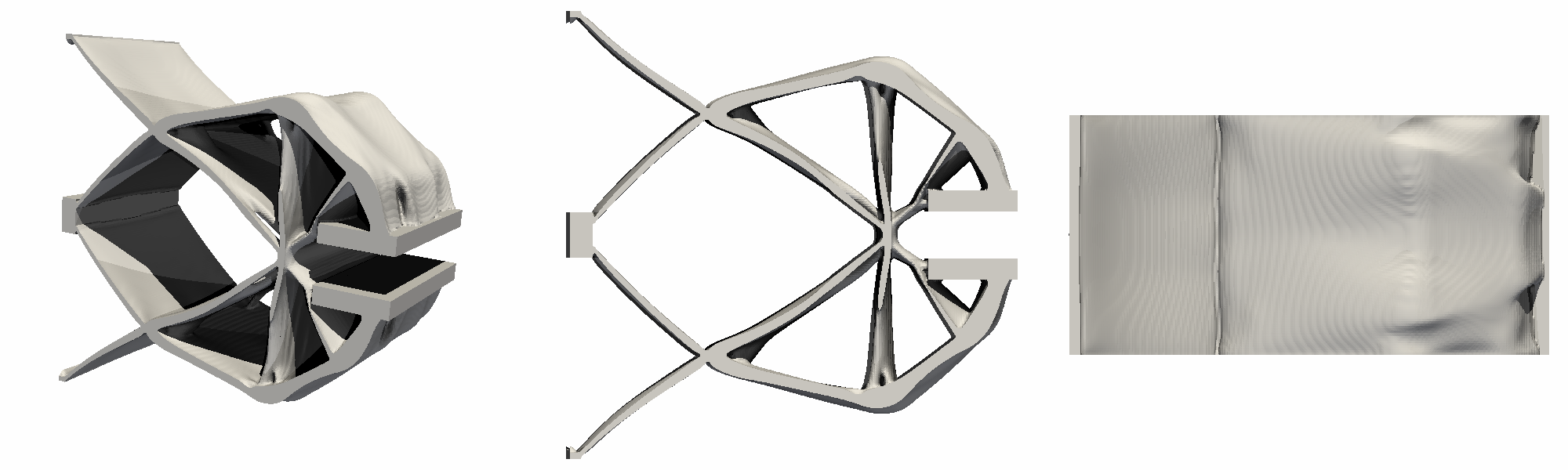}
					\end{minipage} \\
					\hline
					Level-set
					& 176
					& -331.0203
					& 2.5128
					& \begin{minipage}{8cm}
						\includegraphics[width=\linewidth,trim={0 0 660pt 0},clip]{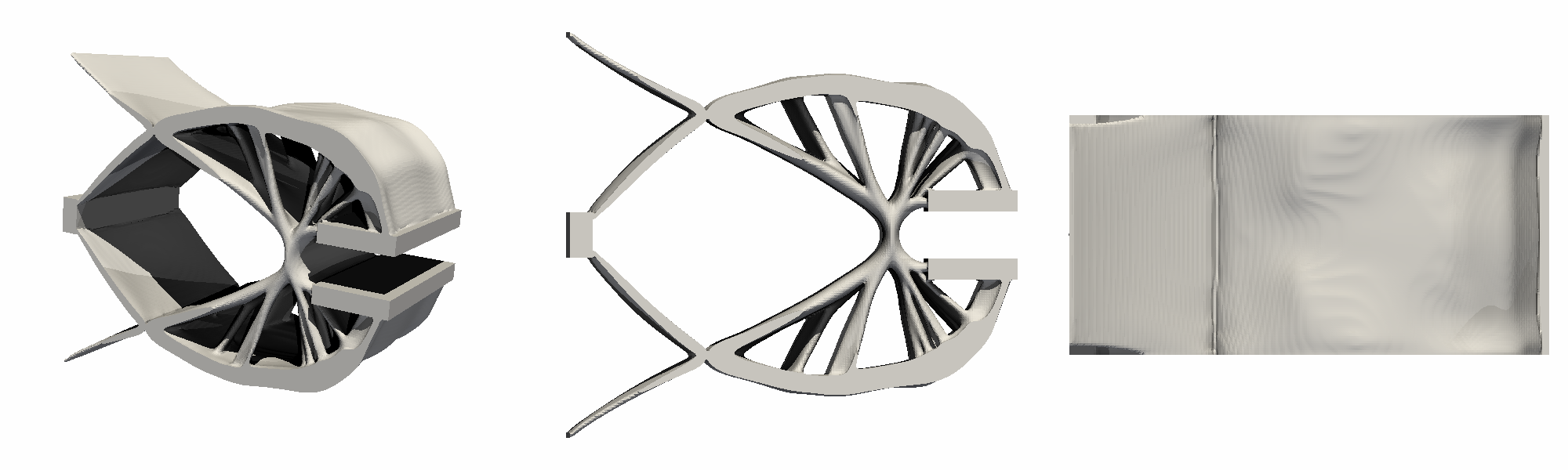}
					\end{minipage} \\
					\hline
				\end{tabular}
			\end{table*}	
			
			\paragraph{Gripper compliant mechanism} The results of this last numerical example are summarized in Table \ref{tab_mths_gripper}. The optimal solutions exhibit resemblance to each other, obtaining the desired mechanism. However, the topology layouts can be grouped into two groups: obtaining 3D-like designs for \emph{SIMP} and \emph{SOFTBESO} methods, while almost 2D-extruded designs are obtained for \emph{\methodabb{}} and \emph{Level-set}. The considered approaches can also be split into two main groups depending on their capability to generate the mechanism either by creating localized hinges or deformable bars. All methods except \emph{SOFTBESO} achieve a design based on localized hinges, thus significantly increasing the value of the objective function. In other words, the same force applied at the \emph{input port} results in a smaller displacement in the target direction at the \emph{output port}. For this reason, it is concluded that \emph{SOFTBESO} has not fully converged to the same local minimum under the given parameters. Regarding the mean bar width, the values for all approaches range between $2.4$ and $2.5$, being equal to $2.7$ for \emph{SIMP$^{(I)}$} as the design is based on a smaller number of thicker bars.
			
			Except for the \emph{SOFTBESO}, the objective function values of the other approaches range between $-260$ and $-331$, showing a larger discrepancy than in the previous benchmark cases. On the other hand, unlike the previous benchmark cases, the best topology design is obtained using the \emph{Level-set method}, even though the layout almost resembles a 2D-extruded design.
			
			Finally, the comparison of the number of iterations reveals that \emph{\methodabb{}} requires much fewer iterations than the other methods. The other topology optimization techniques require between $200$ and $300$ iterations. Therefore, the considered methods can be sorted according to the number of iterations, in increasing order, as follows: \emph{\methodabb{}}, \emph{Level-set}, \emph{SOFTBESO}, \emph{SIMP$^{(III)}$}, \emph{SIMP$^{(II)}$} and \emph{SIMP$^{(I)}$}.
			
			\begin{figure*}[pb]
				\centering
				\includegraphics[width=14cm]{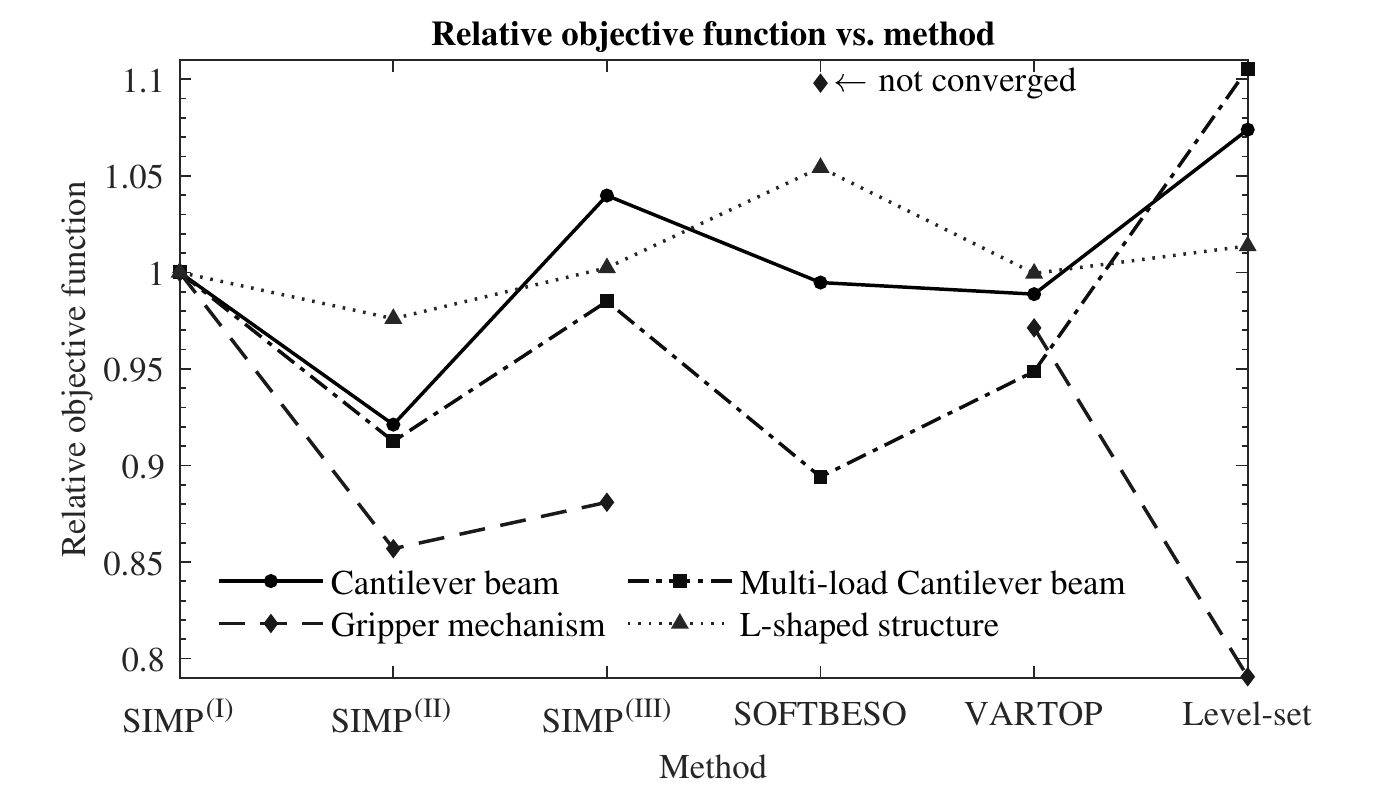}
				\caption{Objective function value for each example and topology optimization approach normalized with the results obtained with the \emph{SIMP$^{(I)}$  method}. Legend: (1) the Cantilever case is represented with a  solid black line, (2) the L-shaped case, with a dotted black line, (3) the Cantilever multi-load case, with a dash-dotted line and (4) the Gripper mechanism with a dashed line, which is interrupted for the \emph{SOFTBESO method} due to lack of convergence.}
				\label{fig_rel_cost}
			\end{figure*}
			
			After analyzing all the results, it can be stated that topologies resulting from \emph{Level-set} and \emph{\methodabb{}} have smooth and accurate interfaces since the solution is defined via a \emph{level-set} or a \emph{discrimination function}. Thus, low complexity topology designs are obtained. On the contrary, \emph{SIMP-based} and \emph{SOFTBESO methods} produce element-wise discontinuous designs. In addition, \emph{SIMP-based approaches} require special post-processing as the design has \emph{semi-dense elements}, thus requiring an extra projection procedure to determine the density value that defines the material interface. In this procedure, bars might be disconnected or broken up, giving as solution non-optimal topologies. Additionally, a smoothing post-processing should be done to achieve crisp and smooth edges from these two family of approaches. 
			
		\subsubsection{Objective function value} \label{subsec_cost_fucntion}
			
			In Figure {\ref{fig_rel_cost}}, the objective function values for each example and topology optimization method are illustrated. The values are normalized with respect to \emph{SIMP$^{(I)}$}. As aforementioned, the objective function for each of the numerical benchmarks does not differ much from one approach to another. The values are between a range of $\pm 15\%$ of the ones obtained using \emph{SIMP$^{(I)}$}.
						
			As observed in the graphic, \emph{SIMP$^{(II)}$} achieves consistently the optimal solutions with the lowest objective function value as a consequence of the larger number of thin straight bars (high topology complexity), as detailed in Section {\ref{subsec_topology}}. Nevertheless, two exceptions are observed, the first one for the \emph{multi-load cantilever} problem where \emph{SOFTBESO} achieves a solution with a lower objective function, and the second one for the Gripper case and the \emph{Level-set}.
						
			It is important to emphasize that a greater variance is only observed in the \emph{Gripper} due to the fact that there is a greater difference in topology among the different approaches. Each technique achieves a characteristic compliant design with the exception of SOFTBESO. This approach obtains a topology layout with an objective function value that is almost two times higher than the one obtained using \emph{SIMP$^{(I)}$}.
			
			\begin{table*}[t]
				\centering
				\caption{Comparison of computational cost in terms of iterations of the considered topology optimization methods.}\label{tab_CPU_iterations}
				
				\begin{tabular}{m{3.8cm}  |C{1.4cm}|C{1.4cm}|C{1.4cm}|C{1.4cm}|C{1.4cm}|C{1.4cm} }
					\toprule
					Numerical Example          & SIMP$^{(I)}$ & SIMP$^{(II)}$ & SIMP$^{(III)}$ & BESO & \methodabb{} & Level-set                 \\ \midrule
					Cantilever beam            & 175          &      231      &       124      & 272  &     116      & 1266     \\ \hline
					L-shaped structure         & 66           &      140      &       76       & 190  &      62      & 1071     \\ \hline
					Multi-load cantilever beam & 81           &      208      &      111       & 249  &     115      & 694      \\ \hline
					Gripper                    & 325          &      297      &      264       &  207 &      32      & 176      \\ \bottomrule 
				\end{tabular}
			\end{table*}
			
		\subsubsection{CPU computation cost: iterations} \label{subsec_comp_cost}
		
			\begin{figure*}[b]
				\centering
				\includegraphics[width=14cm]{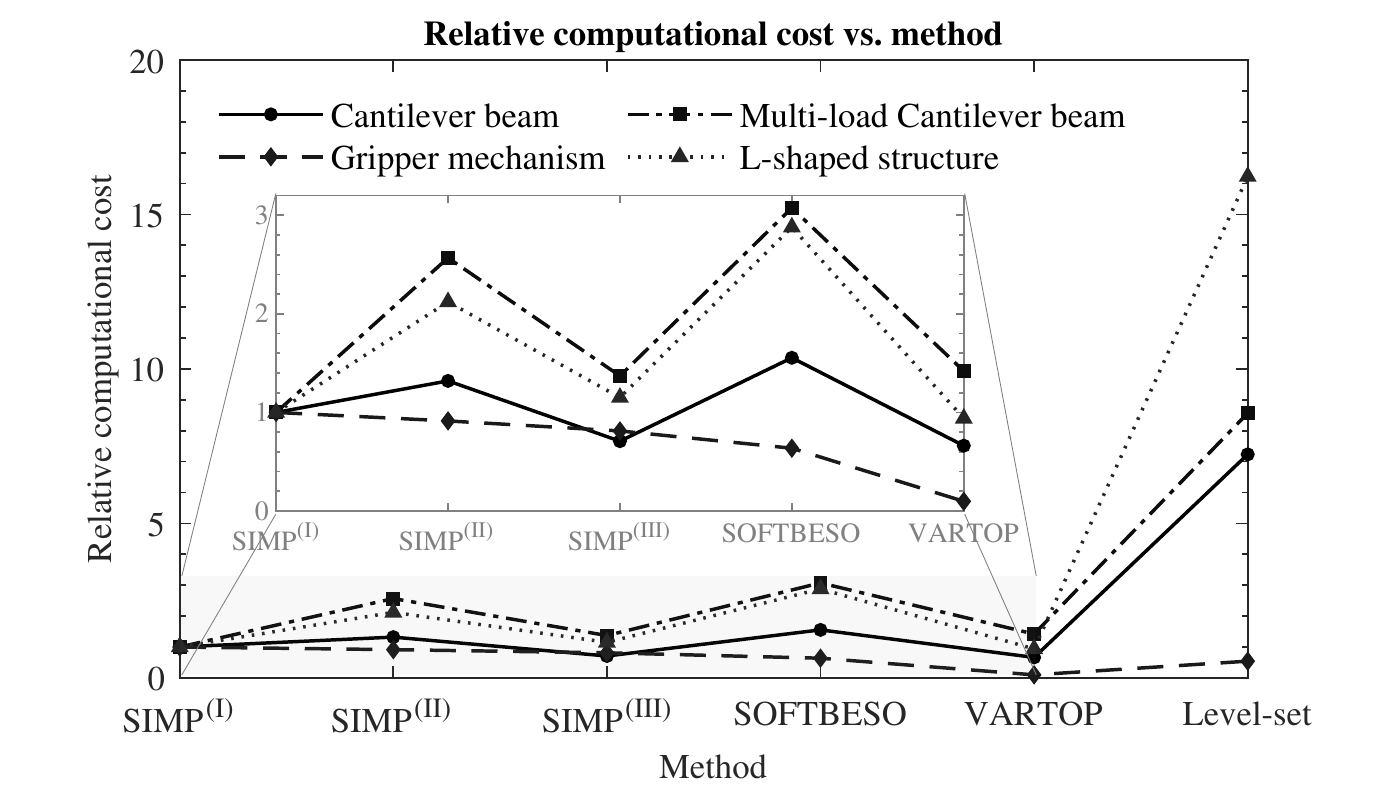}
				\caption{Relative computational cost in terms of the number of iterations. Each numerical example is normalized with the number of iterations of the \emph{SIMP$^{(I)}$}. Legend: (1) the Cantilever case is represented with a  solid black line, (2) the L-shaped case, with a dotted black line, (3) the Cantilever multi-load case, with a dash-dotted line and (4) the Gripper mechanism with a dashed line.}
				\label{fig_cpu_iters}
			\end{figure*}
		
			The computational cost is assessed in this paper according to the number of iterations instead of the computational time. In this way, it is possible to decouple the solution from the platform used to run the topology optimization technique (i.e.,~OS, programming language, and hardware, among others) as well as from the solver used to solve the state problem. It has been observed that the selection of a specific iterative solver may significantly increase the computational time of some approaches with respect to others. Therefore, to remain as unbiased as possible, and in the hypothetical case that all methods would use a direct solver with equivalent computational cost per iteration, the computational cost could be evaluated with the number of iterations, thus obtaining a fair comparison.
					
			The comparison of the computation cost in terms of the number of iterations is shown in Table \ref{tab_CPU_iterations}. The values of the computational cost, normalized with respect to \emph{SIMP$^{(I)}$}, are illustrated in Figure \ref{fig_cpu_iters}. As can be seen, the relative computational cost depends on each numerical example. However, it keeps a certain tendency along the considered approaches for \emph{minimum mean compliance} problems.
			
			Regarding the Cantilever beam, \emph{\methodabb{}} and \emph{SIMP$^{(III)}$} are up to 1.4 times faster than \emph{SIMP$^{(I)}$}, and up to 2 times faster than \emph{SIMP$^{(II)}$} or \emph{SOFTBESO}. \emph{Level-set} turns out to be 7 times more computationally expensive than \emph{SIMP$^{(I)}$}. In addition, it is important to stress that \emph{\methodabb{}} is 7\% faster than \emph{SIMP$^{(III)}$}, even though it provides not only the optimal solution but a set of solutions for different volume fractions (Pareto Frontier).
			
			For the L-shaped structure and the multi-load cantilever beam optimizations, the relative computation costs increase from the previous example, except in \emph{\methodabb{}}. Its relative computational cost becomes almost 1 for the L-shaped structure and even 1.4 for the multi-load cantilever case. \emph{SIMP$^{(I)}$} results in the the fastest approach for this latter benchmark. The advantage over the \emph{SOFTBESO}, \emph{Level-set}, and \emph{SIMP$^{(II)}$ methods} is still present, although no significant improvement in computational cost is obtained with respect to \emph{SIMP$^{(III)}$} and \emph{\methodabb{}}.
			
			As for the \emph{compliant mechanism example}, the previously observed trend does not apply any more. In this case, \emph{\methodabb{}} is the fastest approach by far (almost an order of magnitude faster), followed by the \emph{Level-set} and \emph{SOFTBESO} approaches. Both methods require approximately half as many iterations as \emph{SIMP$^{(I)}$}. \emph{SIMP$^{(III)}$} and \emph{SIMP$^{(II)}$} techniques are respectively $20\%$ and $10\%$ faster than the reference method. This trend change in the computational cost may be caused by the change in the topology optimization problem (i.e.,~the non self-adjoint character of the \emph{problem}).
			
		\subsubsection{Robustness: monotonic convergence degree} \label{sec_robustness}
		
		    The convergence robustness is analyzed through the evolution of the objective function and volume fraction, and the criteria in topology and objective function throughout the optimization. For each technique, the analysis of these variables determine the monotonic convergence degree of every method. The discussion is performed only through the first two examples, since they are representative enough to provide a complete overview of the issue of robustness.
			
\begin{figure*}
				\centering
				\makebox[\textwidth][c]{\includegraphics[width=17.4cm]{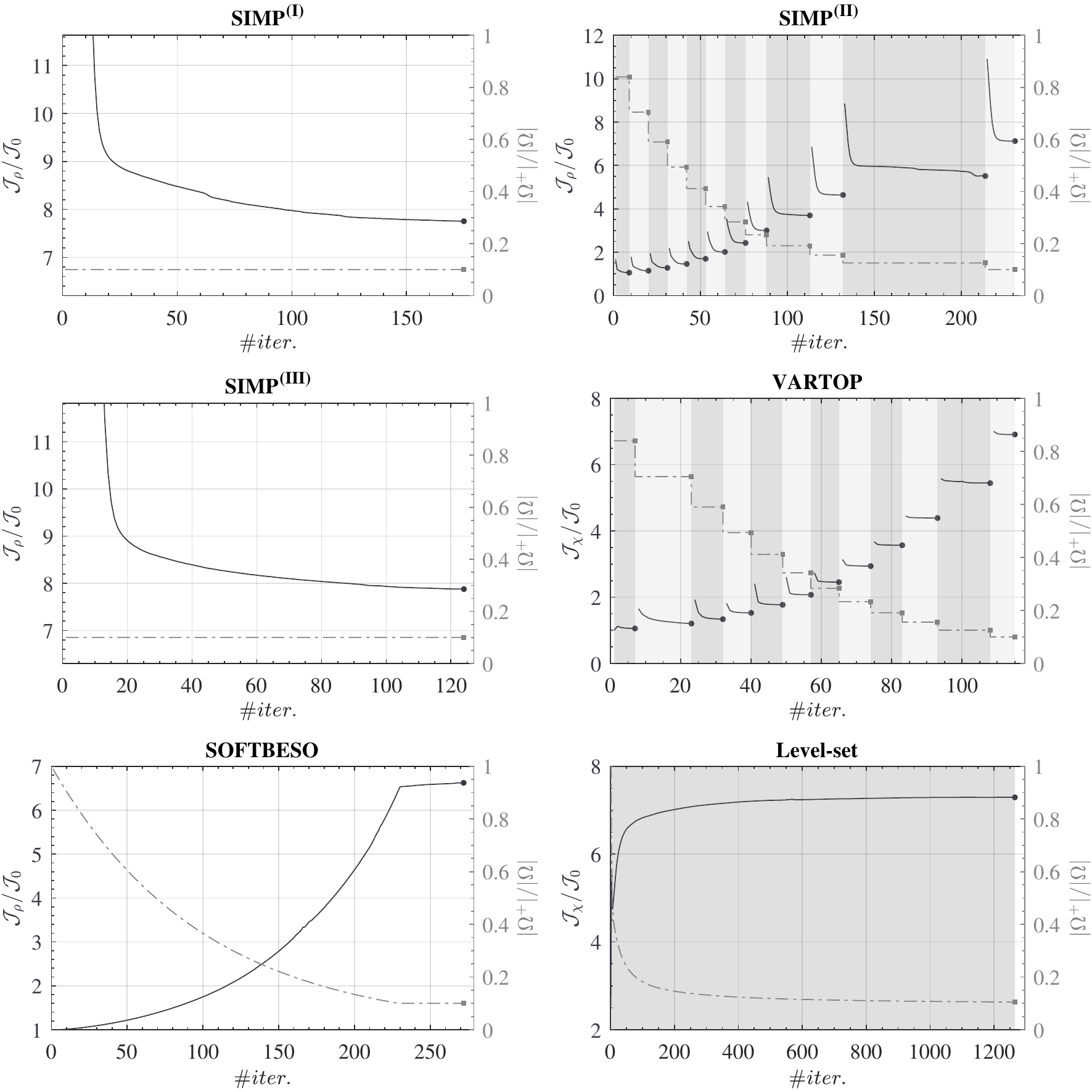}}
				\caption{Evolution histories of the values of the objective function and volume fraction throughout the iterations of the Cantilever beam topology optimization for the six considered methods. \emph{Single-time-step approaches} are illustrated in the first column, while \emph{incremental time-advancing techniques} are depicted in the second column, each time-step being shaded with a different color. The normalized objective function ${\cal J}_\rho$ or ${\cal J}_\chi$ is associated with the left y-axis and represented with a solid black line. On the other side, the volume fraction (i.e.,~the stiff material fraction) $\frac{\vert\Omega^+\vert}{\vert\Omega\vert}$ is represented by a dash-dotted gray line in the right y-axis of each graphic. }
				\label{fig_robustness_value_cantilever}
			\end{figure*}

			The evolution of the objective function for the Cantilever beam is illustrated in Figure \ref{fig_robustness_value_cantilever}. \emph{Single-time-step methods} are represented in the first column while \emph{incremental time-advancing techniques} (i.e.,~\emph{SIMP$^{(II)}$}, \emph{\methodabb{}}, and \emph{Level-set}) are depicted in the second column. Each time-step is shaded with a different color to improve its visualization. The normalized objective function value ${\cal J}/{\cal J}_0$ (solid line colored in black) is illustrated in the left y-axis, while the stiff material fraction (dash-dotted line, colored in gray) is associated with the right y-axis. 
			
			Based on the convergence, the following features can be highlighted: (1) SIMP$^{(I)}$ and SIMP$^{(III)}$ prescribe a constant stiff material fraction (i.e., $f=0.1$) from the initial iteration, and the objective function converges monotonically to a value close to 7.8 \footnote{Note that the objective function value in the graph does differ from Table \ref{tab_mths_cantilever}, since different contrast factor $\alpha$ are used in the optimization and in the post-processing iteration.}, (2) in \emph{SOFTBESO}, the stiff material fraction is gradually reduced from the initial value 1 to the target value 0.1, consequently, the objective function increases until the target volume is achieved, (3) in \emph{SIMP$^{(II)}$} and \emph{\methodabb{}}, the target stiff material fraction is reduced from 1 to 0.1 in 12 time-steps, thereby the objective function is minimized at each time-step, and (4) \emph{Level-set}, which even though it can also be an \emph{incremental time-advancing method}, it has a particular response since the volume constraint is not strictly enforced on each iteration, but it oscillates ruled by an \emph{Augmented Lagrangian method}.
			
			As illustrated in Appendix \ref{app_order_convergence}, the \emph{order of convergence} of the objective function is close to 1 for all the techniques. Therefore, all topology optimization methods have a linear convergence in the objective function.

			\begin{figure*}
				\centering
				\makebox[\textwidth][c]{\includegraphics[width=17.4cm]{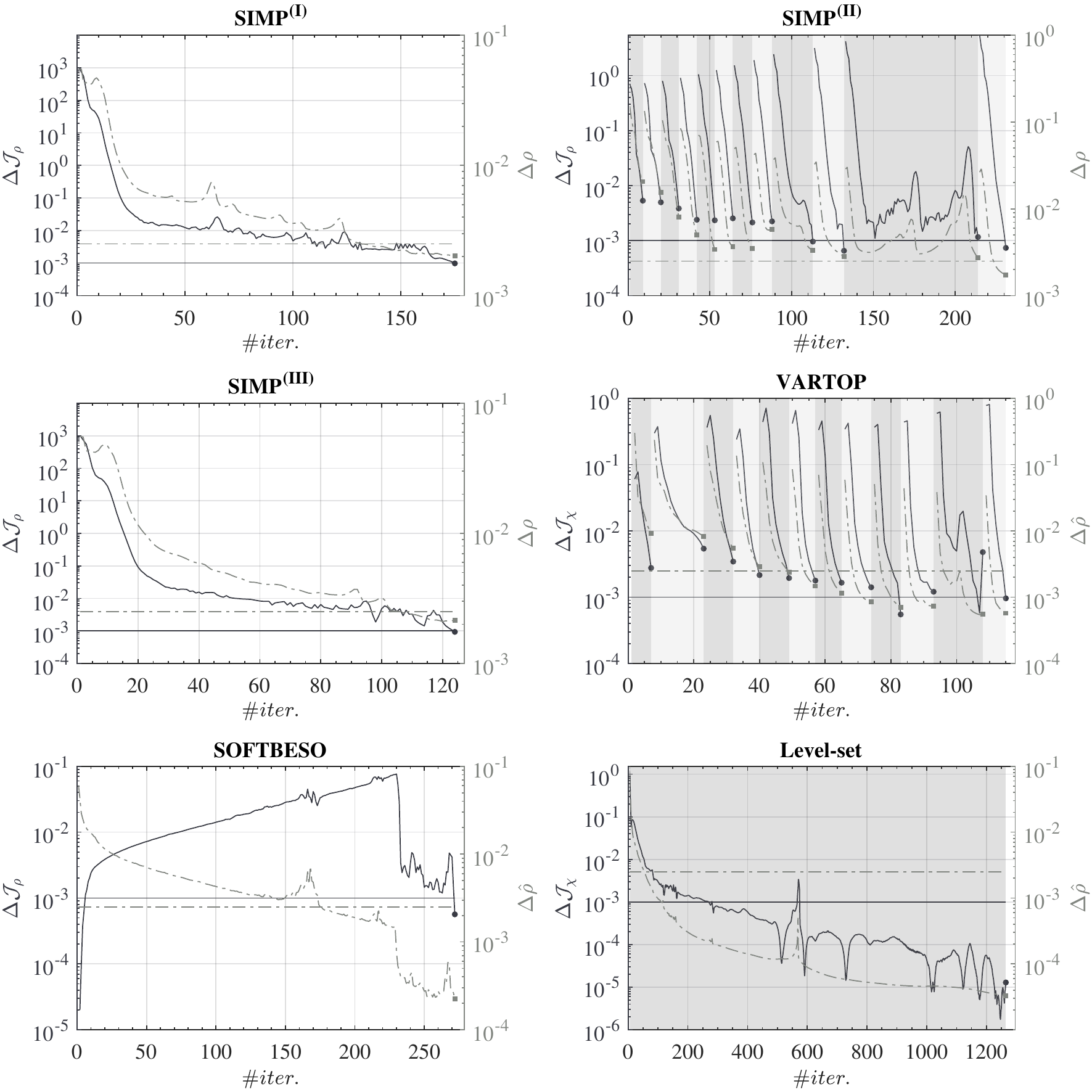}}
				\caption{Evolution histories of the criteria values in the objective function and in the topology throughout iterations of the Cantilever beam topology optimization for the six considered methods. \emph{Single-time-step approaches} are illustrated in the first column, while \emph{incremental time-advancing techniques} are depicted on the second column, each time-step being shaded with a different color. The criterion in objective function, associated with the left y-axis, is represented with a solid black line, while the criterion in the topology is represented by a dash-dotted gray line in the right y-axis of each graphic. In addition, the corresponding maximum tolerances $Tol_{\cal J}$ and $Tol_\desvar$ allowed in the last time-step (or in the entire optimization for \emph{single-time-step methods}) are also displayed in every graphic as horizontal lines with the same properties.}
				\label{fig_robustness_tolerance_cantilever}
			\end{figure*}

			The convergence curves of the objective function and topology criteria are depicted in Figure \ref{fig_robustness_tolerance_cantilever}. The objective function criterion (solid black line) is represented in the left y-axis, while the topology criterion (gray dash-dotted line) in the right y-axis. As in Figure \ref{fig_robustness_value_cantilever}, the previous four different groups can be distinguished,  but now in terms of the convergence criteria. Both \emph{SIMP$^{(II)}$} and \emph{\methodabb{}} show a strictly monotonous convergence within each time-step, only noticing some small oscillations in the second last time-step where a change in topology has taken place. As for the other methods, \emph{SIMP$^{(I)}$} and \emph{SIMP$^{(III)}$} present monotonous convergence with small amplitude oscillations, while some important variations are noticed in \emph{SOFTBESO} once the final stiff material fraction is achieved. Finally, the convergence criterion in the \emph{Level-set method} mimics the trend detected in the objective function and volume fraction with small amplitude oscillations. As a global comment, it can be stated that the objective function criterion corresponds to the most restrictive criterion in all topology optimization approaches, except in the \emph{Level-set method}, in which the volume constraint is the limiting one.
			
			The convergences corresponding to the other examples have been also analyzed in detail. The graphics do not present any significant difference with respect to those already analyzed for the Cantilever beam. However, for completeness reasons the corresponding graphics of the second example are depicted in Appendix \ref{app_Lshape_robustness}.
		
		\subsubsection{Overall performance}	
		
			In this last subsection, instead of comparing the different methods in a quantitative and analytical way, a more qualitative comparison is presented according to the following  aspects: (1) Surface smoothness, (2) Topology complexity, (3) Objective function, and (4) Computational cost. 
									
			The first aspect refers to the surface smoothness required by several manufacturing techniques. In these technologies, sharp edges and noise shells (i.e., abrupt continuous changes) must be avoided in the boundary of the optimal solution. The second criterion takes into account the complexity of the optimal design obtained by each technique, considering other mechanical properties not directly included in the optimization. For instance, designs based on thick bars will have a better structural behavior in buckling or fatigue compared to designs with a greater number of thin bars. These two aspects will also have an impact on the manufacturing challenges, which will decrease as the design becomes smoother and less complex. The third one considers the value of the objective function, or equivalently the efficiency of each method of finding a better local minimum. This criterion gathers the information shown in Figure \ref{fig_rel_cost} regarding the relative objective function values. The last point of comparison globally assesses the computational cost of each method to perform the optimization. Analogous to the last aspect, this criterion gathers the information represented in Figure \ref{fig_cpu_iters} with respect to the relative computational cost.
			
			\definecolor{mycolor1}{RGB}{239,243,255}
			\definecolor{mycolor2}{RGB}{189,215,231}
			\definecolor{mycolor3}{RGB}{107,174,214}
			\definecolor{mycolor4}{RGB}{33,113,181}
			
			\begin{figure*}[b]
				\centering
				\pgfplotsset{width=17cm,height=6cm,compat=1.17}
				\makebox[\textwidth][c]{\begin{tikzpicture}
				\begin{axis}[
				    ybar,
				    bar width = 15pt,
				    legend style={at={(0.5,-0.20)},
				      anchor=north,legend columns=-1},
				    ylabel={\textbf{Comparative scores}},
				    symbolic x coords={SIMP$^{(I)}$,SIMP$^{(II)}$,SIMP$^{(III)}$,BESO,VARTOP,Level-set},
				    xtick=data,
				    ymin = 0,
				    ymax = 6.5,
				    ytick={6,4,2,0},
				    yticklabels={A,B,C,D},
				    minor ytick = {5,3,1},
				    enlarge x limits=0.15,
				    enlarge y limits=0.0,
				    restrict y to domain=0:6,
				    ybar=0pt,
				    xminorgrids=true,
				    ]
				\addplot[fill=mycolor1,draw=black!70,postaction={pattern=dots}] coordinates {(SIMP$^{(I)}$,2) (SIMP$^{(II)}$,2) (SIMP$^{(III)}$,2) (BESO,2) (VARTOP,6) (Level-set,6)};
				\addplot[fill=mycolor2,postaction={pattern=north east lines}] coordinates {(SIMP$^{(I)}$,6) (SIMP$^{(II)}$,4) (SIMP$^{(III)}$,6) (BESO,4) (VARTOP,6) (Level-set,6)};
				\addplot[fill=mycolor3,postaction={pattern=north west lines}] coordinates {(SIMP$^{(I)}$,4) (SIMP$^{(II)}$,6) (SIMP$^{(III)}$,4) (BESO,4) (VARTOP,4) (Level-set,4)};
				\addplot[fill=mycolor4,postaction={pattern=crosshatch}] coordinates {(SIMP$^{(I)}$,6) (SIMP$^{(II)}$,4) (SIMP$^{(III)}$,6) (BESO,4) (VARTOP,6) (Level-set,2)};
				\legend{Surface smoothness,Topology complexity,Objective function, Computational cost} 
				\end{axis}
				\end{tikzpicture}}
				\caption{Qualitative comparison of the studied methods regarding the smoothness of the design (dotted column), the topology complexity (right-inclined lines pattern), the value of the objective function (left-inclined lines pattern) and the computational cost in terms of iterations (column with crosshatch pattern). Each one of the areas is rated qualitatively with the levels A, B, C, or D, being A the best qualification and D the worst one.} \label{fig_overall_performance}
			\end{figure*}
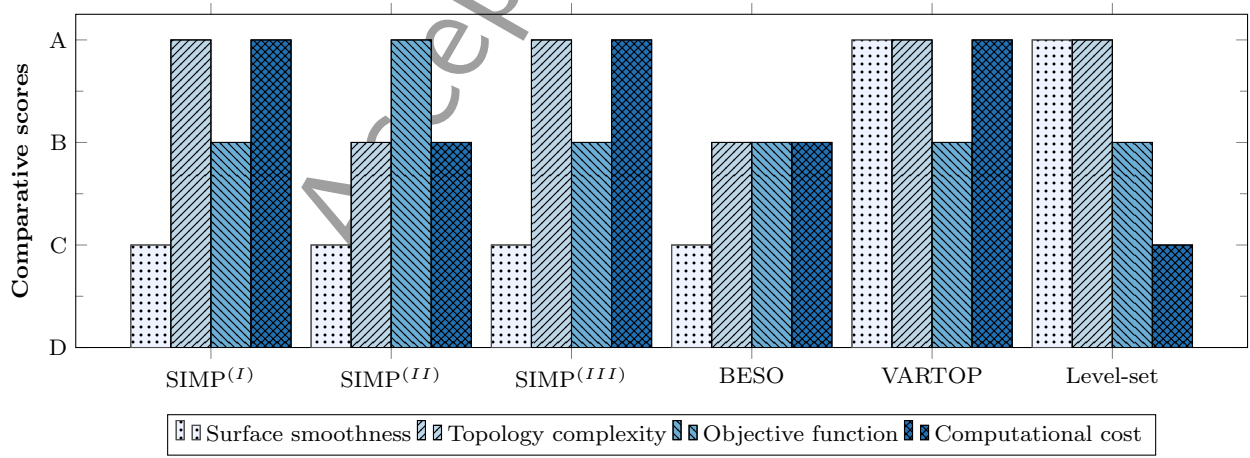

			In Figure \ref{fig_overall_performance}, each one of the aspects is represented with a column bar rated between A and D, with A being the best qualification in that section and D being the worst one. For each approach, four bars of different colors and patterns are represented, each one corresponding to an analyzed aspect.
			
	        \begin{figure*}[t!]
				\centering
				\pgfplotsset{width=17cm,height=6cm,compat=1.17}
				\makebox[\textwidth][c]{\begin{tikzpicture}
				\begin{axis}[
					ybar,
					bar width = 15pt,
					legend style={at={(0.5,-0.20)},
					 anchor=north,legend columns=-1},
					ylabel={\textbf{Comparative scores}},
					symbolic x coords={SIMP$^{(I)}$,SIMP$^{(II)}$,SIMP$^{(III)}$,BESO,VARTOP,Level-set},
					xtick=data,
					ymin = 0,
					ymax = 7,
					ytick={6,4,2,0},
					minor ytick = {5,3,1},
					yticklabels={A ,B ,C ,D },
					enlarge x limits=0.15,
					enlarge y limits=0.0,
					restrict y to domain=0:6,
					ybar=0pt,
					xminorgrids=true,
					]
				\addplot[fill=mycolor3,postaction={pattern=north east lines}] coordinates {(SIMP$^{(I)}$,4) (SIMP$^{(II)}$,3) (SIMP$^{(III)}$,4) (BESO,3)  (Level-set,6) (VARTOP,6)};
			   	\addplot[fill=mycolor4,postaction={pattern=north west lines}] coordinates {(SIMP$^{(I)}$,6) (SIMP$^{(II)}$,5) (SIMP$^{(III)}$,6) (BESO,4)  (Level-set,3) (VARTOP,5)};
				\legend{Topology quality,Computational efficiency}
				\end{axis}
				\end{tikzpicture}}	
				\caption{Qualitative comparison of the studied methods, combining the topology related properties and the computational ones in a single bar. The topology quality is represented with a light blue colored bar and a right tilted line pattern, while the computational efficiency is represented using a dark blue bar with a left tilted line pattern. Each of the criteria is rated qualitatively with the levels A, B$^+$ B, C$^+$, C, D$^+$ or D, being A the best qualification and D the worst one.} \label{fig_overall_performance_global}
			\end{figure*}
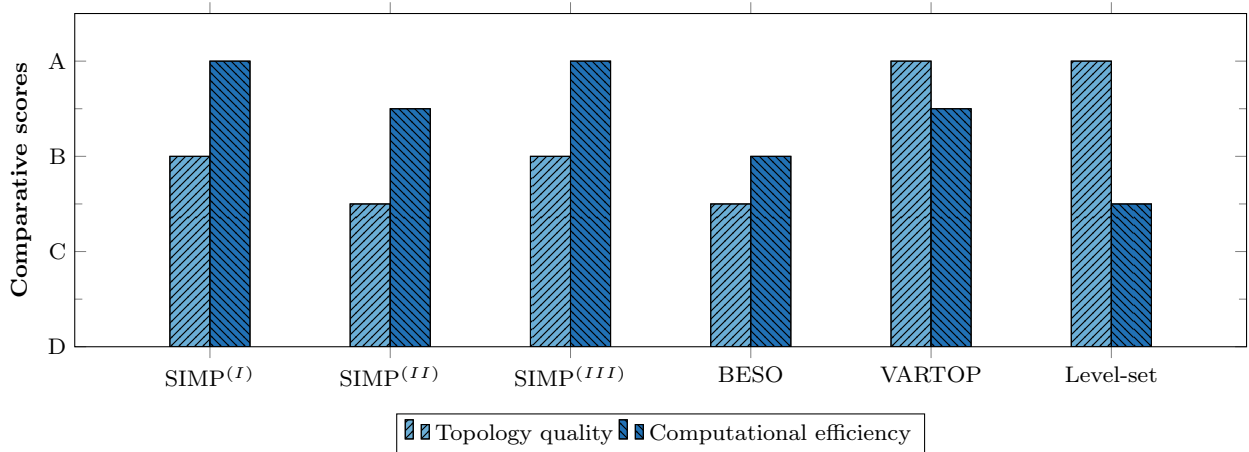			
			
			Regarding the surface smoothness, \emph{\methodabb{}} and \emph{Level-set} provide designs whose surfaces are smooth. On the contrary, all other approaches only achieve element-wise optimal designs, thus the boundary of the solution is defined through abrupt continuous changes. Consequently, additional post-processing procedures are required to manufacture these solutions with smooth boundaries. For this reason, \emph{\methodabb{}} and \emph{Level-set} are evaluated with an A, while the others are rated with a C grade. 
			
			Concerning the topology complexity, it has been noticed in Section \ref{subsec_topology} that the quality of the solutions is reasonably high in almost all the techniques, being this slightly lower for \emph{SIMP$^{(II)}$} and \emph{BESO methods}. In this two techniques, the complexity of the optimal topology increases, obtaining designs based on thinner bars (i.e.~lower bar width) with lower buckling prevention. Accordingly, this two approaches are rated with a B while the others, with an A.
			
			In terms of the objective function, all approaches obtain a similar optimal value, although \emph{SIMP$^{(II)}$} consistently obtains marginally lower values than the other techniques, as detailed in section \ref{subsec_cost_fucntion}. For this reason, \emph{SIMP$^{(II)}$} obtains an A qualification, while the other methods are left with a B. Finally, the comparison of the computational cost, discussed in section \ref{subsec_comp_cost}, is represented in the last column. The computational cost is lower and of similar magnitude for \emph{SIMP$^{(I)}$}, \emph{SIMP$^{(III)}$} and \emph{\methodabb{}}, followed by the \emph{SIMP$^{(II)}$} and \emph{BESO} approaches, and finally by \emph{Level-set}. These three groups are respectively rated with an A, B, and C.

			Figure \ref{fig_overall_performance} can be further simplified by combining the two topology-related features in a single criterion referred to as \emph{topology quality}, and the two criteria related to the objective function and the computational cost in a single criterion called \emph{computational efficiency}. These two criteria are equivalently represented by a bar chart in Figure \ref{fig_overall_performance_global}. From this figure, it can be concluded that \emph{\methodabb{}}, although not being the best approach in all considered aspects in Figure \ref{fig_overall_performance}, is presented as a competitive technique to more conventional \emph{topology optimization approaches}, such as \emph{SIMP$^{(I)}$} and \emph{SIMP$^{(III)}$}. On the other hand, \emph{SOFTBESO} and \emph{Level-set} do not provide any significant advantages, exhibiting mostly deficiencies in topology complexity or computational cost, respectively, for the cases studied in this paper.

\section{Concluding remarks} \label{sec_conclusions}

	This contribution presents a thorough comparison among most of the well-established \emph{topology optimization approaches}, i.e.~the \emph{SIMP}, \emph{Level-set}, and \emph{SOFTBESO methods}, and the \emph{\methodabb{} approach}. A set of well-known 3D numerical benchmarks in the field of structural topology optimization has been addressed to analyze their performance. The corresponding results have been assessed in terms of the optimal topology, the robustness in convergence, the objective function, and the computational cost.
	
	Regarding the topology, a quality dependence has been observed among the assessed methods, being slightly lower for \emph{SIMP$^{(II)}$} and \emph{BESO methods}. The quality and complexity of the topologies can depend on the type of design variable: continuous vs discrete and nodal vs element, as well as the approach to impose the volume constraint.
	
	Regarding the design variable, the methods can be split into three groups: 
	\begin{itemize}[noitemsep]
	    \item \emph{Level-set} and \emph{\methodabb{}} use \emph{nodal scalar functions} to precisely describe the interface, as well as a \emph{discrete characteristic function} to define sharp \emph{white-and-black} configurations.
	    \item \emph{SOFTBESO} uses an \emph{element-wise discrete functions} to define the topology layout\footnote{As aforementioned, a \emph{relaxed characteristic function} (or density variable) is used for \emph{compliant mechanism synthesis}.}. Although obtaining \emph{white-and-black} designs, the material boundary is only defined through elements.
	    \item \emph{SIMP-based methods} use an \emph{element-wise continuous variable}. Consequently, these techniques can not precisely define the boundary, but instead get a blurred interface with \emph{gray semi-dense elements}. By using a projection technique, \emph{white-and-black} designs can be obtained, which interface is defined through elements.
	\end{itemize}
	As a consequence, \emph{SOFTBESO} and \emph{SIMP} techniques may not obtain the best possible optimal solution and would require post-processing techniques to obtain smooth designs which could be easily manufactured. However, there is no guarantee that the resultant topologies are actually optimal layouts.
	
	As aforementioned, the volume constraint methodology may also affect the resultant topology. In those techniques where the volume constraint is gradually imposed using \emph{element-wise variables} (e.g.~\emph{SOFTBESO} and \emph{SIMP$^{(II)}$}), the final topologies tend to be more complex and consist of a larger number of thin bars. As a result, these topologies have worse mechanical behavior and are more challenging and expensive to manufacture.
	
	In terms of the topology, it can be concluded that both a combination of nodal scalar design variables with a gradual incremental volume constraint, and a combination of a continuous element-wise variable with a constant volume constraint provide optimal topologies with high quality. In the comparison, \emph{SIMP$^{(I)}$}, \emph{SIMP$^{(III)}$}, \emph{\methodabb{}}, and \emph{Level-set} achieve optimal designs that are based on thicker bars (with higher mean bar size), thus improving manufacturability with high-quality optimal designs and reducing the buckling proneness.
	
	Concerning the robustness of each method, it has been confirmed that all the techniques have a linear convergence in the objective function regardless of the methodology used to impose the volume constraint. This fact supports the selection of the techniques for the comparison, and the independence with respect to the filtering technique (i.e.~spatial or Helmholtz-type filtering) and the updating scheme of the design variable (i.e. incremental or absolute). These two differences may have an effect on the optimal solution, but not on the \emph{order of convergence}.
	
	In regard to the objective function value, a small variation of $\pm 15\%$ is observed in the four numerical benchmarks among the studied methods with the exception of two tests in the Gripper mechanism. However, \emph{SIMP$^{(II)}$} achieves systematically the lowest objective function values, as the majority of its optimal designs are based on smaller, thinner bars (i.e.,~high complex designs), as mentioned before. Due to this characteristic, these designs are proclive to buckling.
	
	For the first three examples, all studied techniques can be sorted according to a descending number of required iterations as follows: \emph{Level-set}, \emph{SOFTBESO}, \emph{SIMP$^{(II)}$}, \emph{\methodabb{}}, \emph{SIMP$^{(III)}$} and \emph{SIMP$^{(I)}$}. However, the relative computational cost depends on each example and technique, but the same trend is observed. It is important to emphasize that \emph{incremental time-advancing techniques} such as \emph{SIMP$^{(II)}$} and \emph{\methodabb{}} obtain not only the final optimal solution but also a set of intermediate converged solutions at almost the same computational cost (Pareto frontier for the volume fraction). In this scenario, \emph{\methodabb{}} is up to 1.5 times faster than the corresponding \emph{SIMP$^{(II)}$ implementation}. As for the Gripper compliant mechanism, the tendency in computational cost completely changes from the previous examples, observing a very significant reduction with \emph{\methodabb{}} compared to \emph{SIMP$^{(I)}$}. Contrary to the other numerical examples, \emph{SOFTBESO} and \emph{Level-set} also require a lower number of iterations than \emph{SIMP}-based implementations.
		
	In conclusion, the \emph{\methodabb{}}, \emph{SIMP$^{(I)}$}, and \emph{SIMP$^{(III)}$ approaches} present topology layouts with a higher topology quality than the other methods at a lower computational cost, even though their objective function is not minimized as much as in other approaches. 
	
	The authors are aware that, in spite of the efforts done for a fair comparison, a certain degree of subjectivity can still remain in this kind of studies, but they also think that those studies should be presented to the community of structural topology optimization even to be argued and discussed with the aim of the progress of computational topology optimization.

\begin{acknowledgements}

    This research has been funded by the European Research Council (ERC) under the European Union’s Horizon 2020 research and innovation programme (Proof of Concept Grant agreement n 874481) through the project “Computational design and prototyping of acoustic metamaterials for target ambient noise reduction” (METACOUSTIC).
    The authors also acknowledge financial support from the Spanish Ministry of Economy and Competitiveness, through the research grant DPI2017-85521-P for the project “Computational design of Acoustic and Mechanical Metamaterials” (METAMAT) and through the “Severo Ochoa Programme for Centres of Excellence in R\&D” (CEX2018-000797-S).
    D. Yago acknowledges the support received from the Spanish Ministry of Education through the FPU program for PhD grants.
    
\end{acknowledgements}

\section*{Conflict of interest}
	The authors declare that they have no conflict of interest as regards this work.

\begin{appendices}

	\section{Convergence criteria} \label{app_convergence}
	
		Convergence is evaluated in terms of the volume constraint, the objective function, and the topology design, as mentioned in Section \ref{sec_convergence_criteria}. In particular, the topology criterion must be analyzed in detail, since this criterion must be standardized in all methods, each one using a different design variable.   
		
		For \emph{density-based methods}, such as \emph{SIMP}, the topology criterion can be written as a $L_2$ norm of the \emph{element density variable} $\rho_e$ between two consecutive iterations as
		\begin{align}
			\begin{aligned}
				\Delta {\rho}_k &= \frac{1}{\vert \Omega_0 \vert^{1/2}} \left( \int_\Omega { \left( \rho_k  - \rho_{k-1} \right)^2  \domega} \right)^{1/2} \\
								&= \frac{1}{\vert \Omega_0 \vert^{1/2}} \left( \sum_{e=1}^{N_e} { \left( \rho_{e,k}  - \rho_{e,k-1} \right)^2  \vert \Omega_e \vert} \right)^{1/2} \mcolon
			\end{aligned}
		\end{align}
		where $e$ corresponds to the element number and $k$ to the iteration number, and $\vert \Omega_e \vert$ is the volume of element $e$. 
	
		However, for the other approaches, a relaxed \emph{characteristic function} must be used to compute the topology criterion. The element \emph{density variable} or the corresponding element \emph{characteristic function} are regularized via a \emph{Laplacian regularization} (\ref{eq_Lap_filter_SIMP}). Therefore, the topology criterion is computed as
		\begin{align}
			\begin{aligned}
				\Delta \hat{\rho}_k &= \frac{1}{\vert \Omega_0 \vert^{1/2}} \left( \int_\Omega { \left( \mathbf{N} \hat{\pmb{\rho}}_k  - \mathbf{N} \hat{\pmb{\rho}}_{k-1} \right)^2  \domega} \right)^{1/2} \\
									&= \frac{\sqrt{ \left(\hat{\pmb{\rho}}_k - \hat{\pmb{\rho}}_{k-1}\right)^{\text{T}} \mathbb{M}  \left(\hat{\pmb{\rho}}_k - \hat{\pmb{\rho}}_{k-1}\right) }}{\vert \Omega_0 \vert^{1/2}} \mcolon
			\end{aligned}	
		\end{align}		
		with $\hat{\pmb{\rho}}_k$ being the solution to
		\begin{equation}
			\left(\mathbb{M} + (\tau h)^2 \mathbb{K}\right) \hat{\pmb{\rho}}_k = \int_\Omega { \mathbf{N}^{\text{T}} \varphi_e \domega} \mdot
		\end{equation}
		The element variable $\varphi_e$ corresponds to element \emph{density variable} $\rho_e$, the element \emph{characteristic function} $\overline{\chi}_{e,\psi}$ or the element \emph{characteristic function} $\overline{\chi}_{e,\phi}$ for the \emph{BESO},  \emph{\methodabb{}}, or \emph{Level-set methods}, respectively. The matrices $\mathbf{N}$, $\mathbb{M}$ and $\mathbb{K}$ stand for the shape function matrix, the mass matrix and the stiffness matrix, and $\tau$ corresponds to the \emph{regularization parameter} of the topology criterion. It is important to stress that the regularization parameter $\tau$ must be chosen thoroughly so that the two topology criteria are equivalent. Based on the authors' experience, it has been prescribed to $\tau=8$.
		
	\section{Post-processing iteration} \label{app_postprocess}
	
		Due to the discrepancies in the design variables (nodal vs element and continuous vs discrete) and the existence of \emph{semi-dense elements}, the objective function value $\cal J$ can not be directly compared between topology optimization approaches. For that reason, once the optimal topology has converged, an additional iteration must be computed using a \emph{black-and-white} element-wise design with a uniform small \emph{contrast factor} $\alpha=10^{-9}$ for all the studied methods. In this scenario, the topology design is expressed via the \emph{characteristic function} $\overline{\chi}_e=\{1,\beta\}$, as defined in equation (\ref{eq_characteristic_function_general}), with $\beta$ depending on each method so that a constant soft Young's modulus is used throughout the methods, as detailed in Section \ref{sec_theroy_methods}. 
		
		Bear in mind that a projection technique on the \emph{density} or on the \emph{characteristic function} (based on its element definition) is required to obtain an optimal topology layout represented only by elements completely contained in the stiff material domain or in the soft material domain. In this projection technique, the volume must be kept unmodified so that the objective function is computed with the same stiff material fraction. Depending on the topology optimization approach, the element-wise \emph{characteristic function} $\overline{\chi}_e$ is computed as
		\begin{itemize}
			\item For \emph{\methodabb{}} and \emph{Level-set}, a \emph{Heaviside function} with the \emph{actual characteristic function} $\hat{\chi}_{e,\psi}$ (or $\hat{\chi}_{e,\phi}$ for the \emph{Level-set method}) and a reference value computed via a bisection algorithm, i.e.,
			\begin{equation}
				\overline{\chi}_e = {\cal H}_\beta\left( \hat{\chi}_{e,\psi} - \gamma \right) \quad \text{for }\forall e\in N_e \mcolon
			\end{equation}
			with $\beta<\gamma<1$ being computed such that the volume constraint ${\cal C}_0(\overline{\chi}_e)$ in the entire domain is enforced (equations (\ref{eq_top_prob_VARTOP}-b-1) and (\ref{eq_top_prob_LEVELSET}-b-1)).
			
			\item For \emph{density-based approaches} (including \emph{Soft-kill BESO}), a \emph{Heaviside function} with the element density variable $\rho_e$ and a reference value $\overline{\rho}$ computed via a bisection algorithm, i.e.,
			\begin{equation}
				\overline{\chi}_e = {\cal H}_\beta\left( \rho_{e} - \overline{\rho} \right) \quad \text{for }\forall e\in N_e \mcolon
			\end{equation}
			with $\beta$ being 0 for \emph{SIMP methods} or $\sqrt[\leftroot{4}\uproot{3}p]{\alpha}$ for \emph{BESO}.
		\end{itemize}
		
	\section{Parameter definition} \label{app_reproducibility}
	
	    In order to ensure replicability, all the relevant parameters are provided in Tables \ref{tab_parameters_cases_global} and \ref{tab_parameters_cases}. Table \ref{tab_parameters_cases_global} details the values related to the tolerances as well as the values for the contrast factor and the volume fraction for each topology optimization. On the other hand, Table \ref{tab_parameters_cases} provides the specific parameters for updating and regularizing the design variable.
	    	    
		\begin{table*}
		\centering 
		\caption{Global parameters and tolerances used for each benchmark case and topology optimization method. \emph{Volume fraction}, \emph{contrast factor}, and \emph{objective function tolerance} are detailed for each benchmark, while tolerances in \emph{volume fraction} and \emph{topology} are defined for each method. \label{tab_parameters_cases_global}} 
	    {\renewcommand{\arraystretch}{1.25}
	    \begin{tabular}{ m{2.1cm} | C{2.8cm} | C{2.8cm} | C{2.8cm} | C{3.5cm}  }
    		\toprule
	        & \multicolumn{4}{c}{\textbf{Benchmark}} \\ \cline{2-5}
    		\textbf{Method} & \textbf{Cantilever} & \textbf{L-shaped structure}  & \textbf{Multiload Cantilever} & \textbf{Gripper} \\
    		\midrule
    		\multirow[c]{2}{=}{\textbf{SIMP$^{(I)}$}, \textbf{SIMP$^{(II)}$}, \textbf{SIMP$^{(III)}$}, \textbf{BESO}}  & \multicolumn{4}{C{11.9cm}}{ $Tol_{{\cal C}_0}=10^{-3}$, \hspace{0.4cm} $Tol_\desvar = 2.5\cdot10^{-3}$, \hspace{0.4cm} $n = 3$}  \\ \cline{2-5}
    		 & \multicolumn{3}{C{8.4cm} |}{$\alpha=10^{-6}$, \hspace{0.4cm} $\vert \Omega^+ \vert =0.1 \vert \Omega \vert$, \hspace{0.4cm} $Tol_{{\cal J}}=10^{-3}$} & $\alpha=10^{-2}$, \hspace{0.4cm} $\vert \Omega^+ \vert =0.15 \vert \Omega \vert$, \hspace{0.4cm} $Tol_{{\cal J}}=10^{0}$ \\
    		\hline
    		\multirow[b]{2}{*}{\textbf{\methodabb{}}}  &  \multicolumn{4}{C{11.9cm}}{ $Tol_{{\cal C}_0}=10^{-3}$, \hspace{0.4cm} $Tol_\desvar = 2.5\cdot10^{-3}$, \hspace{0.4cm} $n = 2$} \\ \cline{2-5}
    		 & \multicolumn{3}{C{8.4cm} |}{$\alpha=10^{-6}$, \hspace{0.4cm} $\vert \Omega^+ \vert =0.1 \vert \Omega \vert$, \hspace{0.4cm} $Tol_{{\cal J}}=10^{-3}$} & $\alpha=10^{-2}$, \hspace{0.4cm} $\vert \Omega^+ \vert =0.15 \vert \Omega \vert$, \hspace{0.4cm} $Tol_{{\cal J}}=10^{0}$\\
    		\hline
    		\multirow[b]{2}{*}{\textbf{Level-set}}  &  \multicolumn{4}{C{11.9cm}}{ $Tol_{{\cal C}_0}=5\cdot10^{-3}$, \hspace{0.4cm} $Tol_\desvar = 2.5\cdot10^{-3}$, \hspace{0.4cm} $n = 5$} \\ \cline{2-5}
    		 & \multicolumn{3}{C{8.41cm} |}{$\alpha=10^{-6}$, \hspace{0.4cm} $\vert \Omega^+ \vert =0.1 \vert \Omega \vert$, \hspace{0.4cm} $Tol_{{\cal J}}=10^{-3}$} & $\alpha=10^{-2}$, \hspace{0.4cm} $\vert \Omega^+ \vert =0.15 \vert \Omega \vert$, \hspace{0.4cm} $Tol_{{\cal J}}=10^{0}$\\
    		\bottomrule
	    \end{tabular}}
	    \end{table*}	  

		\begin{table*}
		\centering 
		{\renewcommand{\arraystretch}{1.25}
		\caption{Parameters used for each benchmark case and topology optimization method. \label{tab_parameters_cases}}
		\begin{tabular}{ m{2.1cm}  *{4}{|C{3cm}} }
    		\toprule
    		& \multicolumn{4}{c}{\textbf{Benchmark}} \\ \cline{2-5}
    		\textbf{Method} & \textbf{Cantilever} & \textbf{L-shaped structure}  & \textbf{Multiload Cantilever} & \textbf{Gripper} \\
    		\midrule
    		\multirow[c]{2}{*}{\textbf{SIMP$^{(I)}$}}  &  \multicolumn{4}{C{11.9cm} }{$p=3$,  \hspace{0.4cm} $r_{min}=3$,  \hspace{0.4cm} $ft=1$} \\ \cline{2-5}
    		 & \multicolumn{3}{C{9cm} |}{$m=0.2$,  \hspace{0.4cm} $\eta=0.5$} & $m=0.1$,  \hspace{0.4cm} $\eta=0.3$\\
    		\hline
    		\multirow[b]{2}{*}{\textbf{SIMP$^{(II)}$}}  &  \multicolumn{4}{C{11.9cm} }{$p=3$,  \hspace{0.4cm} $r_{min}=3$,  \hspace{0.4cm} $ft=1$,  \hspace{0.4cm} $k=-2$} \\ \cline{2-5}
    		& $m=0.2$,  \hspace{0.4cm} $\eta=0.5$,  \hspace{0.4cm} $n_{steps}=12$ & $m=0.2$,  \hspace{0.4cm} $\eta=0.5$,  \hspace{0.4cm} $n_{steps}=8$ & $m=0.2$,  \hspace{0.4cm} $\eta=0.5$,  \hspace{0.4cm} $n_{steps}=12$ & $m=0.1$,  \hspace{0.4cm} $\eta=0.3$,  \hspace{0.4cm} $n_{steps}=8$\\
    		\hline
    		\multirow[c]{2}{*}{\textbf{SIMP$^{(III)}$}}  &  \multicolumn{4}{C{11.9cm} }{$p=3$,  \hspace{0.4cm} $r_{min}=3$,  \hspace{0.4cm} $ft=1$} \\ \cline{2-5}
    		 & \multicolumn{3}{C{9cm} |}{$m=0.2$,  \hspace{0.4cm} $\eta=0.5$} & $m=0.1$,  \hspace{0.4cm} $\eta=0.3$\\ 
    		\hline
    		\multirow[c]{2}{*}{\textbf{BESO}}  &  \multicolumn{4}{C{11.9cm} }{$r_{min}=3$,  \hspace{0.4cm} $ER=0.01$,  \hspace{0.4cm} $AR_{max}=0.1$} \\  \cline{2-5}
    		 & \multicolumn{3}{C{9cm} |}{$p=3$} & $p=2$,  \hspace{0.4cm} $m=0.1$\\ 
    		\hline
    		\textbf{\methodabb{}} & $m=3$, \hspace{0.1cm} $\tau=1$,  \hspace{0.1cm} $n_{steps}=12$,  \hspace{0.1cm} $k=-2$ & $m=5$,  \hspace{0.1cm} $\tau=1.5$,  \hspace{0.1cm} $n_{steps}=8$,  \hspace{0.1cm} $k=-2$ &  $m=3$,  \hspace{0.1cm} $\tau=1.5$,  \hspace{0.1cm} $n_{steps}=12$,  \hspace{0.1cm} $k=-2$ &  $m=100$,  \hspace{0.1cm} $\tau=0.5$,  \hspace{0.1cm} $n_{steps}=8$,  \hspace{0.1cm} $k=-2$ \\[0.3cm]
    		\hline
    		\textbf{Level-set} & $m=3$,  \hspace{0.1cm} $\tau=1$,  \hspace{0.1cm} $n_{steps}=1$,  \hspace{0.1cm} $\Delta t = 0.1$,  \hspace{0.1cm} $s=10^{-4}$ & $m=5$,  \hspace{0.1cm} $\tau=1$,  \hspace{0.1cm} $n_{steps}=1$,  \hspace{0.1cm} $\Delta t = 0.1$,  \hspace{0.1cm} $s=5\cdot10^{-7}$ & $m=3$,  \hspace{0.1cm} $\tau=1$,  \hspace{0.1cm} $n_{steps}=1$,  \hspace{0.1cm} $\Delta t = 0.1$,  \hspace{0.1cm} $s=10^{-3}$ & $m=100$,  \hspace{0.1cm} $\tau=0.5$,  \hspace{0.1cm} $n_{steps}=1$,    $\Delta t = 0.05$,  \hspace{0.1cm} $s=10^{-2}$ \\
    		\bottomrule
		\end{tabular}}
		\end{table*}	
				
	\section{Order of convergence} \label{app_order_convergence}
	
	    In this appendix, the \emph{order of convergence} of the objective function for the different methods will be evaluated to define an additional parameter regarding the computational robustness (Section \ref{sec_robustness}). As a result, it will be possible to verify whether one method stands out from the others in terms of the \emph{order of convergence}.
	    
	    The \emph{order of convergence}, $p$, for the objective function can be computed from the sequence of iterative values ${\cal{J}}_n/{\cal{J}}_0$ (from $n=0$ to $n=\infty$) that converges to ${\cal{J}}^*/{\cal{J}}_0$, when
	    \begin{equation}
	        \lim_{n\rightarrow\infty} \frac{\vert e_{n+1} \vert}{\vert e_{n} \vert^p} = \mu \mcolon
	    \end{equation}
	    with $p>0$ and $\mu\neq0$ corresponding to the \emph{order of convergence} and \emph{rate of convergence}. The $e_{n+1}$ and $e_{n}$ denote the errors of the objective function at $n$-th and $(n+1)$-th iterations, respectively, with respect to the converged one, ${\cal{J}}^*/{\cal{J}}_0$. The error at each iteration is evaluated as
	    \begin{equation}
	        e_{n} = \frac{{\cal{J}}_n}{{\cal{J}}_0} - \frac{{\cal{J}}^*}{{\cal{J}}_0} \mcolon
	    \end{equation}
	    with ${\cal{J}}^*/{\cal{J}}_0$ being approximated to the normalized objective function value for the last converged optimal solution. For \emph{incremental time-advancing techniques}, the \emph{order of convergence} can be evaluated for each time-step using the corresponding converged objective function value.
	    
\begin{figure*}[h]
			\centering
			\makebox[\textwidth][c]{\includegraphics[width=17.4cm]{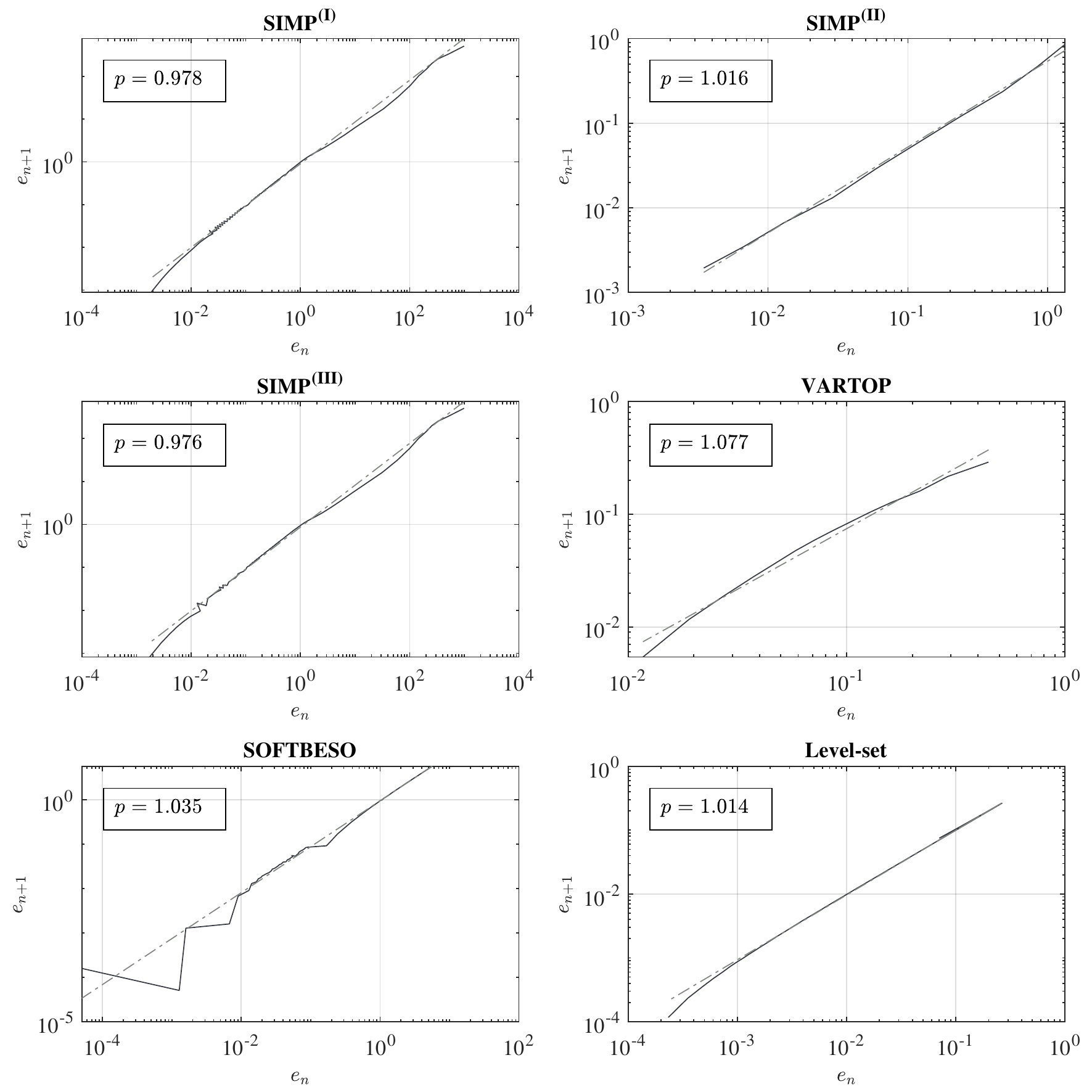}}
			\caption{Iterative sequence of the objective function errors throughout the iterations of the Cantilever beam topology optimization for the six considered methods. \emph{Single-time-step approaches} are illustrated in the first column, while \emph{incremental time-advancing techniques} are depicted in the second column. The objective function error $e_{n}$ is represented with a solid black line, while the corresponding linear regression is represented with a dashed gray line. The order of convergence is included in the top-left corner.}
			\label{fig_convergence_order_Cantilever}
		\end{figure*}

	    The iterative sequence of the error in the objective function $e_{n}$ is illustrated in Figure \ref{fig_convergence_order_Cantilever} for the Cantilever beam benchmark case. As in Figures \ref{fig_robustness_value_cantilever} to \ref{fig_robustness_tolerance_cantilever}, \emph{single-time-step methods} are displayed in the first column while \emph{incremental time-advancing techniques} are depicted in the second column, the \emph{order of convergence} being computed for an intermediate time-step. The corresponding linear regression, used to compute the \emph{order of convergence}, is represented in all the graphics with a dashed line. The exact value for the \emph{order of convergence} is displayed at the top-left corner. As can be observed, the \emph{order of convergence} for all the approaches is close to 1, thus all the addressed methods have a linear convergence in the objective function.
	    
	\section{Robustness of L-shaped structure} \label{app_Lshape_robustness}
	
\begin{figure*}
			\centering
			\makebox[\textwidth][c]{\includegraphics[width=17.4cm]{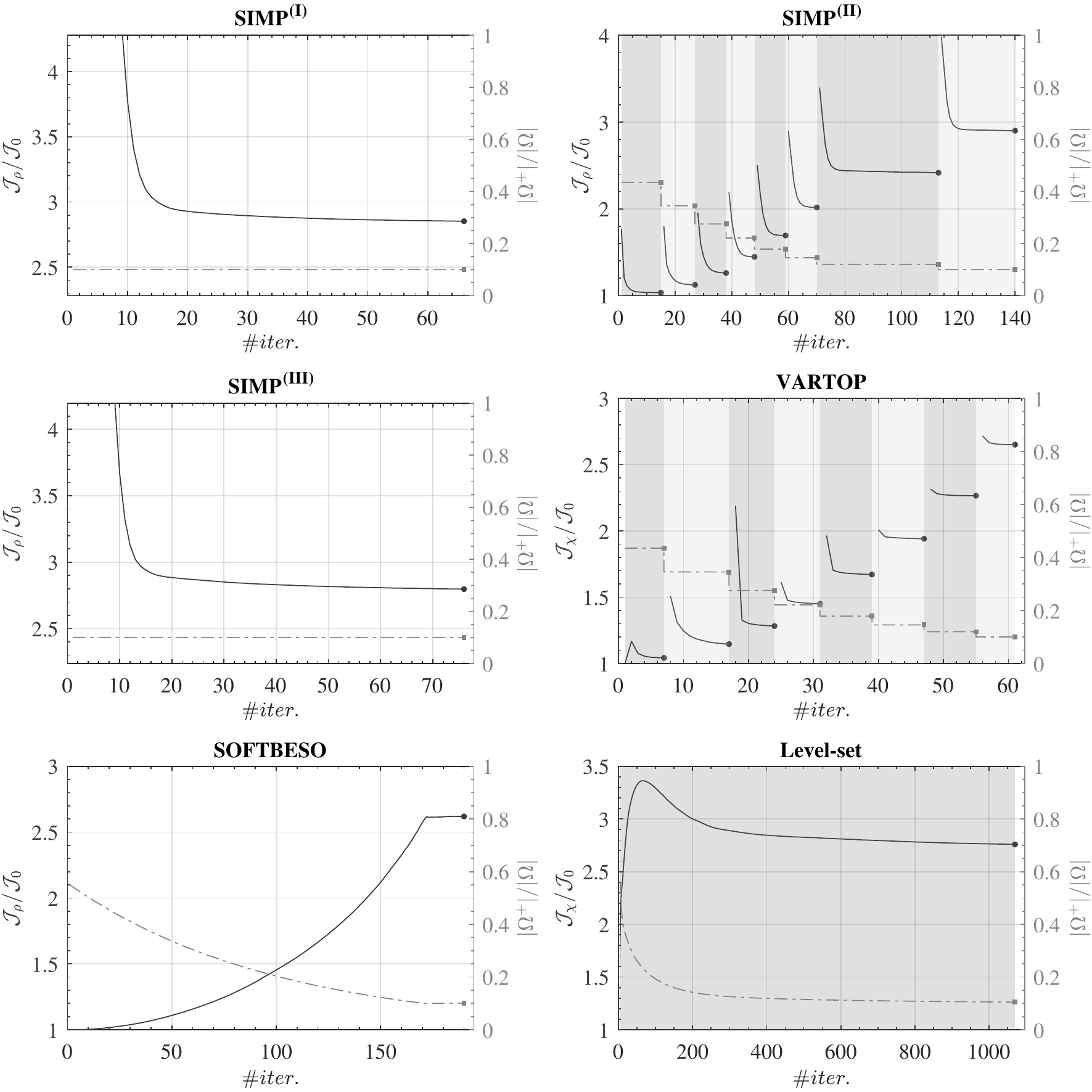}}
			\caption{Evolution histories of the values of the objective function and volume fraction throughout the iterations of the L-shaped structure topology optimization for the six considered methods. \emph{Single-time-step approaches} are illustrated in the first column, while \emph{incremental time-advancing techniques} are depicted in the second column, each time-step being shaded with a different color. The normalized objective function ${\cal J}_\rho$ or ${\cal J}_\chi$ is associated with the left y-axis and represented with a solid black line. On the other side, the volume fraction (i.e.,~the stiff material fraction) $\frac{\vert\Omega^+\vert}{\vert\Omega\vert}$ is represented by a dash-dotted gray line in the right y-axis of each graphic.}
			\label{fig_robustness_value_Lshape}
		\end{figure*}

\begin{figure*}
			\centering
			\makebox[\textwidth][c]{\includegraphics[width=17.4cm]{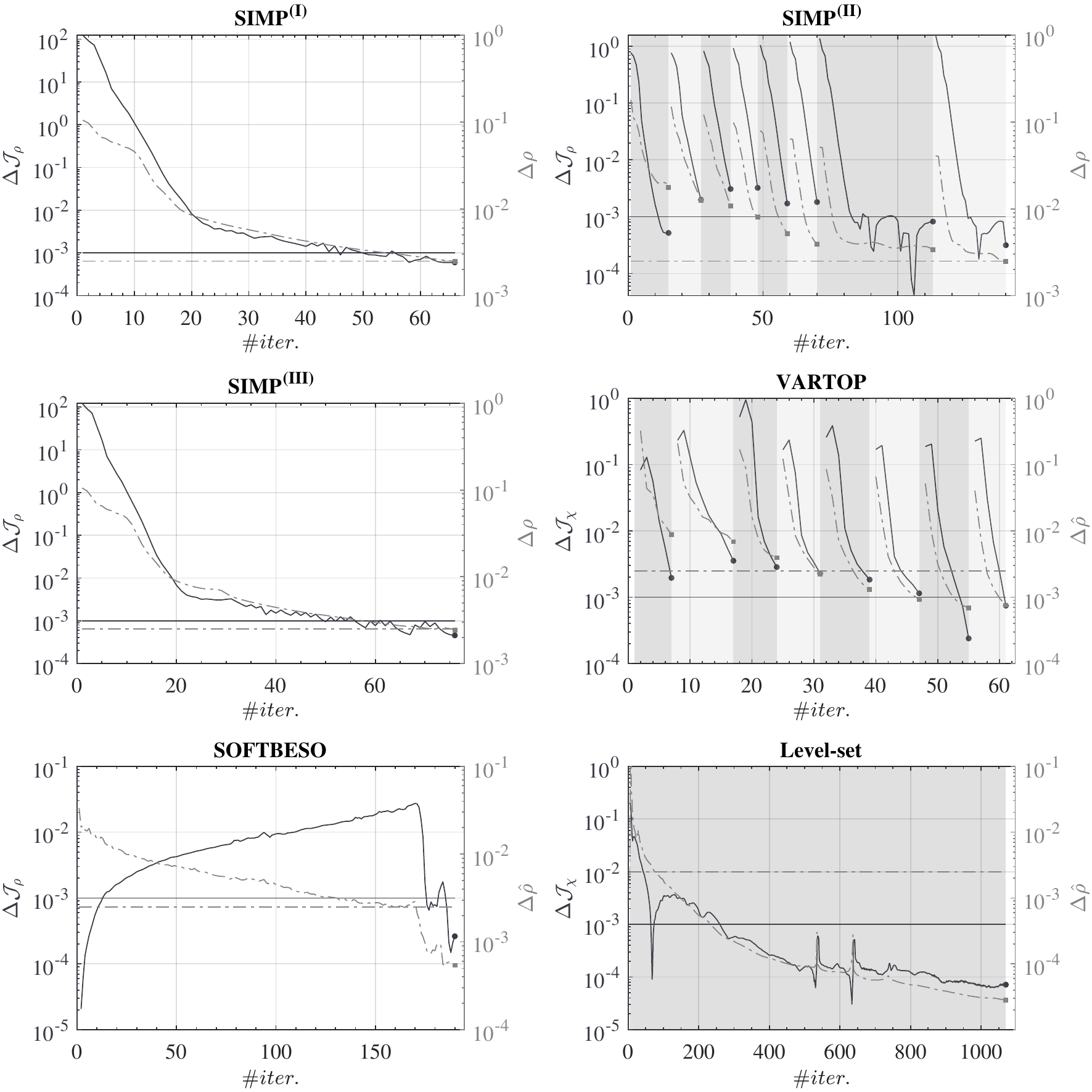}}
			\caption{Evolution histories of the criteria values in the objective function and in the topology throughout iterations of the L-shaped structure topology optimization for the six considered methods. \emph{Single-time-step approaches} are illustrated in the first column, while \emph{incremental time-advancing techniques} are depicted on the second column, each time-step being shaded with a different color. The criterion in objective function, associated with the left y-axis, is represented with a solid black line, while the criterion in the topology is represented by a dash-dotted gray line in the right y-axis of each graphic. In addition, the corresponding maximum tolerances $Tol_{\cal J}$ and $Tol_\desvar$ allowed in the last time-step (or in the entire optimization for single-time-step methods) are also displayed in every graphic as horizontal lines with the same properties.}
			\label{fig_robustness_tolerance_Lshape}
		\end{figure*}

	    Mimicking Figures \ref{fig_robustness_value_cantilever} and \ref{fig_robustness_tolerance_cantilever}, the evolution of the objective function and the stiff material fraction is illustrated in Figure {\ref{fig_robustness_value_Lshape}}, while the evolution of the criteria is depicted in Figure {\ref{fig_robustness_tolerance_Lshape}}.
	
\end{appendices}

\clearpage
\clearpage

\bibliographystyle{elsarticle-num-names}
\bibliography{Comparison_bib}

\begin{thebibliography}{149}
\expandafter\ifx\csname natexlab\endcsname\relax\def\natexlab#1{#1}\fi
\providecommand{\url}[1]{\texttt{#1}}
\providecommand{\href}[2]{#2}
\providecommand{\path}[1]{#1}
\providecommand{\DOIprefix}{doi:}
\providecommand{\ArXivprefix}{arXiv:}
\providecommand{\URLprefix}{URL: }
\providecommand{\Pubmedprefix}{pmid:}
\providecommand{\doi}[1]{\href{http://dx.doi.org/#1}{\path{#1}}}
\providecommand{\Pubmed}[1]{\href{pmid:#1}{\path{#1}}}
\providecommand{\bibinfo}[2]{#2}
\ifx\xfnm\relax \def\xfnm[#1]{\unskip,\space#1}\fi
\bibitem[{Yulin and Xiaoming(2004)}]{Yulin2004}
\bibinfo{author}{M.~Yulin}, \bibinfo{author}{W.~Xiaoming},
\newblock \bibinfo{title}{A level set method for structural topology
  optimization and its applications},
\newblock \bibinfo{journal}{Advances in Engineering Software}
  \bibinfo{volume}{35} (\bibinfo{year}{2004}) \bibinfo{pages}{415--441}.
  \DOIprefix\doi{10.1016/j.advengsoft.2004.06.004}.
\bibitem[{Wang et~al.(2003)Wang, Wang, and Guo}]{Wang2003}
\bibinfo{author}{M.~Y. Wang}, \bibinfo{author}{X.~Wang},
  \bibinfo{author}{D.~Guo},
\newblock \bibinfo{title}{A level set method for structural topology
  optimization},
\newblock \bibinfo{journal}{Computer Methods in Applied Mechanics and
  Engineering} \bibinfo{volume}{192} (\bibinfo{year}{2003})
  \bibinfo{pages}{227--246}. \DOIprefix\doi{10.1016/s0045-7825(02)00559-5}.
\bibitem[{Allaire et~al.(2004)Allaire, Jouve, and Toader}]{Allaire2004}
\bibinfo{author}{G.~Allaire}, \bibinfo{author}{F.~Jouve},
  \bibinfo{author}{A.-M. Toader},
\newblock \bibinfo{title}{Structural optimization using sensitivity analysis
  and a level-set method},
\newblock \bibinfo{journal}{Journal of Computational Physics}
  \bibinfo{volume}{194} (\bibinfo{year}{2004}) \bibinfo{pages}{363--393}.
  \DOIprefix\doi{10.1016/j.jcp.2003.09.032}.
\bibitem[{Bruns and Tortorelli(2001)}]{Bruns2001}
\bibinfo{author}{T.~E. Bruns}, \bibinfo{author}{D.~A. Tortorelli},
\newblock \bibinfo{title}{Topology optimization of non-linear elastic
  structures and compliant mechanisms},
\newblock \bibinfo{journal}{Computer Methods in Applied Mechanics and
  Engineering} \bibinfo{volume}{190} (\bibinfo{year}{2001})
  \bibinfo{pages}{3443--3459}. \DOIprefix\doi{10.1016/s0045-7825(00)00278-4}.
\bibitem[{Wang and Wang(2004)}]{Wang2004b}
\bibinfo{author}{M.~Y. Wang}, \bibinfo{author}{X.~Wang},
\newblock \bibinfo{title}{{\textquotedblleft}color{\textquotedblright} level
  sets: a multi-phase method for structural topology optimization with multiple
  materials},
\newblock \bibinfo{journal}{Computer Methods in Applied Mechanics and
  Engineering} \bibinfo{volume}{193} (\bibinfo{year}{2004})
  \bibinfo{pages}{469--496}. \DOIprefix\doi{10.1016/j.cma.2003.10.008}.
\bibitem[{Coelho et~al.(2007)Coelho, Fernandes, Guedes, and
  Rodrigues}]{Coelho2007}
\bibinfo{author}{P.~G. Coelho}, \bibinfo{author}{P.~R. Fernandes},
  \bibinfo{author}{J.~M. Guedes}, \bibinfo{author}{H.~C. Rodrigues},
\newblock \bibinfo{title}{A hierarchical model for concurrent material and
  topology optimisation of three-dimensional structures},
\newblock \bibinfo{journal}{Structural and Multidisciplinary Optimization}
  \bibinfo{volume}{35} (\bibinfo{year}{2007}) \bibinfo{pages}{107--115}.
  \DOIprefix\doi{10.1007/s00158-007-0141-3}.
\bibitem[{Borrvall and Petersson(2002)}]{Borrvall2002}
\bibinfo{author}{T.~Borrvall}, \bibinfo{author}{J.~Petersson},
\newblock \bibinfo{title}{Topology optimization of fluids in stokes flow},
\newblock \bibinfo{journal}{International Journal for Numerical Methods in
  Fluids} \bibinfo{volume}{41} (\bibinfo{year}{2002}) \bibinfo{pages}{77--107}.
  \DOIprefix\doi{10.1002/fld.426}.
\bibitem[{Gersborg-Hansen et~al.(2005)Gersborg-Hansen, Sigmund, and
  Haber}]{GersborgHansen2005}
\bibinfo{author}{A.~Gersborg-Hansen}, \bibinfo{author}{O.~Sigmund},
  \bibinfo{author}{R.~Haber},
\newblock \bibinfo{title}{Topology optimization of channel flow problems},
\newblock \bibinfo{journal}{Structural and Multidisciplinary Optimization}
  \bibinfo{volume}{30} (\bibinfo{year}{2005}) \bibinfo{pages}{181--192}.
  \DOIprefix\doi{10.1007/s00158-004-0508-7}.
\bibitem[{Guest and Pr{\'{e}}vost(2006)}]{Guest2006}
\bibinfo{author}{J.~K. Guest}, \bibinfo{author}{J.~H. Pr{\'{e}}vost},
\newblock \bibinfo{title}{Topology optimization of creeping fluid flows using a
  darcy{\textendash}stokes finite element},
\newblock \bibinfo{journal}{International Journal for Numerical Methods in
  Engineering} \bibinfo{volume}{66} (\bibinfo{year}{2006})
  \bibinfo{pages}{461--484}. \DOIprefix\doi{10.1002/nme.1560}.
\bibitem[{Li et~al.(1999)Li, Steven, Querin, and Xie}]{Li1999}
\bibinfo{author}{Q.~Li}, \bibinfo{author}{G.~P. Steven}, \bibinfo{author}{O.~M.
  Querin}, \bibinfo{author}{Y.~Xie},
\newblock \bibinfo{title}{Shape and topology design for heat conduction by
  evolutionary structural optimization},
\newblock \bibinfo{journal}{International Journal of Heat and Mass Transfer}
  \bibinfo{volume}{42} (\bibinfo{year}{1999}) \bibinfo{pages}{3361--3371}.
  \DOIprefix\doi{10.1016/s0017-9310(99)00008-3}.
\bibitem[{Gao et~al.(2008)Gao, Zhang, Zhu, Xu, and Bassir}]{Gao2008}
\bibinfo{author}{T.~Gao}, \bibinfo{author}{W.~Zhang}, \bibinfo{author}{J.~Zhu},
  \bibinfo{author}{Y.~Xu}, \bibinfo{author}{D.~Bassir},
\newblock \bibinfo{title}{Topology optimization of heat conduction problem
  involving design-dependent heat load effect},
\newblock \bibinfo{journal}{Finite Elements in Analysis and Design}
  \bibinfo{volume}{44} (\bibinfo{year}{2008}) \bibinfo{pages}{805--813}.
  \DOIprefix\doi{10.1016/j.finel.2008.06.001}.
\bibitem[{Yamada et~al.(2011)Yamada, Izui, and Nishiwaki}]{Yamada2011}
\bibinfo{author}{T.~Yamada}, \bibinfo{author}{K.~Izui},
  \bibinfo{author}{S.~Nishiwaki},
\newblock \bibinfo{title}{A level set-based topology optimization method for
  maximizing thermal diffusivity in problems including design-dependent
  effects},
\newblock \bibinfo{journal}{Journal of Mechanical Design} \bibinfo{volume}{133}
  (\bibinfo{year}{2011}). \DOIprefix\doi{10.1115/1.4003684}.
\bibitem[{Sigmund and Jensen(2003)}]{Sigmund2003}
\bibinfo{author}{O.~Sigmund}, \bibinfo{author}{J.~S. Jensen},
\newblock \bibinfo{title}{Systematic design of phononic band{\textendash}gap
  materials and structures by topology optimization},
\newblock \bibinfo{journal}{Philosophical Transactions of the Royal Society of
  London. Series A: Mathematical, Physical and Engineering Sciences}
  \bibinfo{volume}{361} (\bibinfo{year}{2003}) \bibinfo{pages}{1001--1019}.
  \DOIprefix\doi{10.1098/rsta.2003.1177}.
\bibitem[{Du and Olhoff(2007)}]{Du2007}
\bibinfo{author}{J.~Du}, \bibinfo{author}{N.~Olhoff},
\newblock \bibinfo{title}{Minimization of sound radiation from vibrating
  bi-material structures using topology optimization},
\newblock \bibinfo{journal}{Structural and Multidisciplinary Optimization}
  \bibinfo{volume}{33} (\bibinfo{year}{2007}) \bibinfo{pages}{305--321}.
  \DOIprefix\doi{10.1007/s00158-006-0088-9}.
\bibitem[{Dong et~al.(2017)Dong, Zhao, Wang, and Zhang}]{Dong2017}
\bibinfo{author}{H.-W. Dong}, \bibinfo{author}{S.-D. Zhao},
  \bibinfo{author}{Y.-S. Wang}, \bibinfo{author}{C.~Zhang},
\newblock \bibinfo{title}{Topology optimization of anisotropic broadband
  double-negative elastic metamaterials},
\newblock \bibinfo{journal}{Journal of the Mechanics and Physics of Solids}
  \bibinfo{volume}{105} (\bibinfo{year}{2017}) \bibinfo{pages}{54--80}.
  \DOIprefix\doi{10.1016/j.jmps.2017.04.009}.
\bibitem[{fan Li et~al.(2016)fan Li, Huang, Meng, and Zhou}]{Li2016}
\bibinfo{author}{Y.~fan Li}, \bibinfo{author}{X.~Huang},
  \bibinfo{author}{F.~Meng}, \bibinfo{author}{S.~Zhou},
\newblock \bibinfo{title}{Evolutionary topological design for phononic band gap
  crystals},
\newblock \bibinfo{journal}{Structural and Multidisciplinary Optimization}
  \bibinfo{volume}{54} (\bibinfo{year}{2016}) \bibinfo{pages}{595--617}.
  \DOIprefix\doi{10.1007/s00158-016-1424-3}.
\bibitem[{Lu et~al.(2013)Lu, Yamamoto, Otomori, Yamada, Izui, and
  Nishiwaki}]{Lu2013}
\bibinfo{author}{L.~Lu}, \bibinfo{author}{T.~Yamamoto},
  \bibinfo{author}{M.~Otomori}, \bibinfo{author}{T.~Yamada},
  \bibinfo{author}{K.~Izui}, \bibinfo{author}{S.~Nishiwaki},
\newblock \bibinfo{title}{Topology optimization of an acoustic metamaterial
  with negative bulk modulus using local resonance},
\newblock \bibinfo{journal}{Finite Elements in Analysis and Design}
  \bibinfo{volume}{72} (\bibinfo{year}{2013}) \bibinfo{pages}{1--12}.
  \DOIprefix\doi{10.1016/j.finel.2013.04.005}.
\bibitem[{Roca et~al.(2019)Roca, Yago, Cante, Lloberas-Valls, and
  Oliver}]{Roca2019}
\bibinfo{author}{D.~Roca}, \bibinfo{author}{D.~Yago},
  \bibinfo{author}{J.~Cante}, \bibinfo{author}{O.~Lloberas-Valls},
  \bibinfo{author}{J.~Oliver},
\newblock \bibinfo{title}{Computational design of locally resonant acoustic
  metamaterials},
\newblock \bibinfo{journal}{Computer Methods in Applied Mechanics and
  Engineering} \bibinfo{volume}{345} (\bibinfo{year}{2019})
  \bibinfo{pages}{161--182}. \DOIprefix\doi{10.1016/j.cma.2018.10.037}.
\bibitem[{Huang et~al.(2012)Huang, Xie, Jia, Li, and Zhou}]{Huang2012}
\bibinfo{author}{X.~Huang}, \bibinfo{author}{Y.~M. Xie},
  \bibinfo{author}{B.~Jia}, \bibinfo{author}{Q.~Li}, \bibinfo{author}{S.~W.
  Zhou},
\newblock \bibinfo{title}{Evolutionary topology optimization of periodic
  composites for extremal magnetic permeability and electrical permittivity},
\newblock \bibinfo{journal}{Structural and Multidisciplinary Optimization}
  \bibinfo{volume}{46} (\bibinfo{year}{2012}) \bibinfo{pages}{385--398}.
  \DOIprefix\doi{10.1007/s00158-012-0766-8}.
\bibitem[{Zhou et~al.(2010)Zhou, Li, Sun, and Li}]{Zhou2010}
\bibinfo{author}{S.~Zhou}, \bibinfo{author}{W.~Li}, \bibinfo{author}{G.~Sun},
  \bibinfo{author}{Q.~Li},
\newblock \bibinfo{title}{A level-set procedure for the design of
  electromagnetic metamaterials},
\newblock \bibinfo{journal}{Optics Express} \bibinfo{volume}{18}
  (\bibinfo{year}{2010}) \bibinfo{pages}{6693}.
  \DOIprefix\doi{10.1364/oe.18.006693}.
\bibitem[{Zhou et~al.(2011)Zhou, Li, Chen, Sun, and Li}]{Zhou2011}
\bibinfo{author}{S.~Zhou}, \bibinfo{author}{W.~Li}, \bibinfo{author}{Y.~Chen},
  \bibinfo{author}{G.~Sun}, \bibinfo{author}{Q.~Li},
\newblock \bibinfo{title}{Topology optimization for negative permeability
  metamaterials using level-set algorithm},
\newblock \bibinfo{journal}{Acta Materialia} \bibinfo{volume}{59}
  (\bibinfo{year}{2011}) \bibinfo{pages}{2624--2636}.
  \DOIprefix\doi{10.1016/j.actamat.2010.12.049}.
\bibitem[{Sigmund and Torquato(1997)}]{Sigmund1997a}
\bibinfo{author}{O.~Sigmund}, \bibinfo{author}{S.~Torquato},
\newblock \bibinfo{title}{Design of materials with extreme thermal expansion
  using a three-phase topology optimization method},
\newblock \bibinfo{journal}{Journal of the Mechanics and Physics of Solids}
  \bibinfo{volume}{45} (\bibinfo{year}{1997}) \bibinfo{pages}{1037--1067}.
  \DOIprefix\doi{10.1016/s0022-5096(96)00114-7}.
\bibitem[{Sigmund(2001)}]{Sigmund2001a}
\bibinfo{author}{O.~Sigmund},
\newblock \bibinfo{title}{Design of multiphysics actuators using topology
  optimization {\textendash} part i: One-material structures},
\newblock \bibinfo{journal}{Computer Methods in Applied Mechanics and
  Engineering} \bibinfo{volume}{190} (\bibinfo{year}{2001})
  \bibinfo{pages}{6577--6604}. \DOIprefix\doi{10.1016/s0045-7825(01)00251-1}.
\bibitem[{Deng et~al.(2012)Deng, Yan, and Cheng}]{Deng2012}
\bibinfo{author}{J.~Deng}, \bibinfo{author}{J.~Yan},
  \bibinfo{author}{G.~Cheng},
\newblock \bibinfo{title}{Multi-objective concurrent topology optimization of
  thermoelastic structures composed of homogeneous porous material},
\newblock \bibinfo{journal}{Structural and Multidisciplinary Optimization}
  \bibinfo{volume}{47} (\bibinfo{year}{2012}) \bibinfo{pages}{583--597}.
  \DOIprefix\doi{10.1007/s00158-012-0849-6}.
\bibitem[{Yoon et~al.(2007)Yoon, Jensen, and Sigmund}]{Yoon2007}
\bibinfo{author}{G.~H. Yoon}, \bibinfo{author}{J.~S. Jensen},
  \bibinfo{author}{O.~Sigmund},
\newblock \bibinfo{title}{Topology optimization of
  acoustic{\textendash}structure interaction problems using a mixed finite
  element formulation},
\newblock \bibinfo{journal}{International Journal for Numerical Methods in
  Engineering} \bibinfo{volume}{70} (\bibinfo{year}{2007})
  \bibinfo{pages}{1049--1075}. \DOIprefix\doi{10.1002/nme.1900}.
\bibitem[{Maute and Allen(2004)}]{Maute2004}
\bibinfo{author}{K.~Maute}, \bibinfo{author}{M.~Allen},
\newblock \bibinfo{title}{Conceptual design of aeroelastic structures by
  topology optimization},
\newblock \bibinfo{journal}{Structural and Multidisciplinary Optimization}
  \bibinfo{volume}{27} (\bibinfo{year}{2004}) \bibinfo{pages}{27--42}.
  \DOIprefix\doi{10.1007/s00158-003-0362-z}.
\bibitem[{Andreasen and Sigmund(2013)}]{Andreasen2013}
\bibinfo{author}{C.~S. Andreasen}, \bibinfo{author}{O.~Sigmund},
\newblock \bibinfo{title}{Topology optimization of
  fluid{\textendash}structure-interaction problems in poroelasticity},
\newblock \bibinfo{journal}{Computer Methods in Applied Mechanics and
  Engineering} \bibinfo{volume}{258} (\bibinfo{year}{2013})
  \bibinfo{pages}{55--62}. \DOIprefix\doi{10.1016/j.cma.2013.02.007}.
\bibitem[{Jenkins and Maute(2015)}]{Jenkins2015}
\bibinfo{author}{N.~Jenkins}, \bibinfo{author}{K.~Maute},
\newblock \bibinfo{title}{Level set topology optimization of stationary
  fluid-structure interaction problems},
\newblock \bibinfo{journal}{Structural and Multidisciplinary Optimization}
  \bibinfo{volume}{52} (\bibinfo{year}{2015}) \bibinfo{pages}{179--195}.
  \DOIprefix\doi{10.1007/s00158-015-1229-9}.
\bibitem[{Alexandersen et~al.(2016)Alexandersen, Sigmund, and
  Aage}]{Alexandersen2016}
\bibinfo{author}{J.~Alexandersen}, \bibinfo{author}{O.~Sigmund},
  \bibinfo{author}{N.~Aage},
\newblock \bibinfo{title}{Large scale three-dimensional topology optimisation
  of heat sinks cooled by natural convection},
\newblock \bibinfo{journal}{International Journal of Heat and Mass Transfer}
  \bibinfo{volume}{100} (\bibinfo{year}{2016}) \bibinfo{pages}{876--891}.
  \DOIprefix\doi{10.1016/j.ijheatmasstransfer.2016.05.013}.
\bibitem[{Bruns(2007)}]{Bruns2007}
\bibinfo{author}{T.~Bruns},
\newblock \bibinfo{title}{Topology optimization of convection-dominated,
  steady-state heat transfer problems},
\newblock \bibinfo{journal}{International Journal of Heat and Mass Transfer}
  \bibinfo{volume}{50} (\bibinfo{year}{2007}) \bibinfo{pages}{2859--2873}.
  \DOIprefix\doi{10.1016/j.ijheatmasstransfer.2007.01.039}.
\bibitem[{Yaji et~al.(2015)Yaji, Yamada, Kubo, Izui, and Nishiwaki}]{Yaji2015}
\bibinfo{author}{K.~Yaji}, \bibinfo{author}{T.~Yamada},
  \bibinfo{author}{S.~Kubo}, \bibinfo{author}{K.~Izui},
  \bibinfo{author}{S.~Nishiwaki},
\newblock \bibinfo{title}{A topology optimization method for a coupled
  thermal{\textendash}fluid problem using level set boundary expressions},
\newblock \bibinfo{journal}{International Journal of Heat and Mass Transfer}
  \bibinfo{volume}{81} (\bibinfo{year}{2015}) \bibinfo{pages}{878--888}.
  \DOIprefix\doi{10.1016/j.ijheatmasstransfer.2014.11.005}.
\bibitem[{Marck et~al.(2013)Marck, Nemer, and Harion}]{Marck2013}
\bibinfo{author}{G.~Marck}, \bibinfo{author}{M.~Nemer}, \bibinfo{author}{J.-L.
  Harion},
\newblock \bibinfo{title}{Topology optimization of heat and mass transfer
  problems: Laminar flow},
\newblock \bibinfo{journal}{Numerical Heat Transfer, Part B: Fundamentals}
  \bibinfo{volume}{63} (\bibinfo{year}{2013}) \bibinfo{pages}{508--539}.
  \DOIprefix\doi{10.1080/10407790.2013.772001}.
\bibitem[{Rozvany(2008)}]{Rozvany2008}
\bibinfo{author}{G.~I.~N. Rozvany},
\newblock \bibinfo{title}{A critical review of established methods of
  structural topology optimization},
\newblock \bibinfo{journal}{Structural and Multidisciplinary Optimization}
  \bibinfo{volume}{37} (\bibinfo{year}{2008}) \bibinfo{pages}{217--237}.
  \DOIprefix\doi{10.1007/s00158-007-0217-0}.
\bibitem[{Eschenauer and Olhoff(2001)}]{Eschenauer2001}
\bibinfo{author}{H.~A. Eschenauer}, \bibinfo{author}{N.~Olhoff},
\newblock \bibinfo{title}{Topology optimization of continuum structures: a
  review},
\newblock \bibinfo{journal}{Applied Mechanics Review} \bibinfo{volume}{54}
  (\bibinfo{year}{2001}) \bibinfo{pages}{331--390}.
\bibitem[{Deaton and Grandhi(2013)}]{Deaton2013}
\bibinfo{author}{J.~D. Deaton}, \bibinfo{author}{R.~V. Grandhi},
\newblock \bibinfo{title}{A survey of structural and multidisciplinary
  continuum topology optimization: post 2000},
\newblock \bibinfo{journal}{Structural and Multidisciplinary Optimization}
  \bibinfo{volume}{49} (\bibinfo{year}{2013}) \bibinfo{pages}{1--38}.
  \DOIprefix\doi{10.1007/s00158-013-0956-z}.
\bibitem[{Sigmund and Maute(2013)}]{Sigmund2013}
\bibinfo{author}{O.~Sigmund}, \bibinfo{author}{K.~Maute},
\newblock \bibinfo{title}{Topology optimization approaches},
\newblock \bibinfo{journal}{Structural and Multidisciplinary Optimization}
  \bibinfo{volume}{48} (\bibinfo{year}{2013}) \bibinfo{pages}{1031--1055}.
  \DOIprefix\doi{10.1007/s00158-013-0978-6}.
\bibitem[{van Dijk et~al.(2013)van Dijk, Maute, Langelaar, and van
  Keulen}]{Dijk2013}
\bibinfo{author}{N.~P. van Dijk}, \bibinfo{author}{K.~Maute},
  \bibinfo{author}{M.~Langelaar}, \bibinfo{author}{F.~van Keulen},
\newblock \bibinfo{title}{Level-set methods for structural topology
  optimization: a review},
\newblock \bibinfo{journal}{Structural and Multidisciplinary Optimization}
  \bibinfo{volume}{48} (\bibinfo{year}{2013}) \bibinfo{pages}{437--472}.
  \DOIprefix\doi{10.1007/s00158-013-0912-y}.
\bibitem[{Munk et~al.(2015)Munk, Vio, and Steven}]{Munk2015}
\bibinfo{author}{D.~J. Munk}, \bibinfo{author}{G.~A. Vio},
  \bibinfo{author}{G.~P. Steven},
\newblock \bibinfo{title}{Topology and shape optimization methods using
  evolutionary algorithms: a review},
\newblock \bibinfo{journal}{Structural and Multidisciplinary Optimization}
  \bibinfo{volume}{52} (\bibinfo{year}{2015}) \bibinfo{pages}{613--631}.
  \DOIprefix\doi{10.1007/s00158-015-1261-9}.
\bibitem[{Hajela et~al.(1993)Hajela, Lee, and Lin}]{Hajela1993}
\bibinfo{author}{P.~Hajela}, \bibinfo{author}{E.~Lee}, \bibinfo{author}{C.-Y.
  Lin},
\newblock \bibinfo{title}{Genetic algorithms in structural topology
  optimization},
\newblock in: \bibinfo{booktitle}{Topology Design of Structures},
  \bibinfo{publisher}{Springer Netherlands}, \bibinfo{year}{1993}, pp.
  \bibinfo{pages}{117--133}. \DOIprefix\doi{10.1007/978-94-011-1804-0_10}.
\bibitem[{Adeli and Cheng(1994)}]{Adeli1994}
\bibinfo{author}{H.~Adeli}, \bibinfo{author}{N.-T. Cheng},
\newblock \bibinfo{title}{Augmented lagrangian genetic algorithm for structural
  optimization},
\newblock \bibinfo{journal}{Journal of Aerospace Engineering}
  \bibinfo{volume}{7} (\bibinfo{year}{1994}) \bibinfo{pages}{104--118}.
  \DOIprefix\doi{10.1061/(asce)0893-1321(1994)7:1(104)}.
\bibitem[{Chapman et~al.(1994)Chapman, Saitou, and Jakiela}]{Chapman1994}
\bibinfo{author}{C.~D. Chapman}, \bibinfo{author}{K.~Saitou},
  \bibinfo{author}{M.~J. Jakiela},
\newblock \bibinfo{title}{Genetic algorithms as an approach to configuration
  and topology design},
\newblock \bibinfo{journal}{Journal of Mechanical Design} \bibinfo{volume}{116}
  (\bibinfo{year}{1994}) \bibinfo{pages}{1005--1012}.
  \DOIprefix\doi{10.1115/1.2919480}.
\bibitem[{Hare et~al.(2013)Hare, Nutini, and Tesfamariam}]{Hare2013}
\bibinfo{author}{W.~Hare}, \bibinfo{author}{J.~Nutini},
  \bibinfo{author}{S.~Tesfamariam},
\newblock \bibinfo{title}{A survey of non-gradient optimization methods in
  structural engineering},
\newblock \bibinfo{journal}{Advances in Engineering Software}
  \bibinfo{volume}{59} (\bibinfo{year}{2013}) \bibinfo{pages}{19--28}.
  \DOIprefix\doi{10.1016/j.advengsoft.2013.03.001}.
\bibitem[{Wang et~al.(2005)Wang, Tai, and Wang}]{Wang2005}
\bibinfo{author}{S.~Y. Wang}, \bibinfo{author}{K.~Tai}, \bibinfo{author}{M.~Y.
  Wang},
\newblock \bibinfo{title}{An enhanced genetic algorithm for structural topology
  optimization},
\newblock \bibinfo{journal}{International Journal for Numerical Methods in
  Engineering} \bibinfo{volume}{65} (\bibinfo{year}{2005})
  \bibinfo{pages}{18--44}. \DOIprefix\doi{10.1002/nme.1435}.
\bibitem[{Luh and Lin(2009)}]{Luh2009}
\bibinfo{author}{G.-C. Luh}, \bibinfo{author}{C.-Y. Lin},
\newblock \bibinfo{title}{Structural topology optimization using ant colony
  optimization algorithm},
\newblock \bibinfo{journal}{Applied Soft Computing} \bibinfo{volume}{9}
  (\bibinfo{year}{2009}) \bibinfo{pages}{1343--1353}.
  \DOIprefix\doi{10.1016/j.asoc.2009.06.001}.
\bibitem[{Bends{\o}e(1989)}]{Bendsoe1989}
\bibinfo{author}{M.~P. Bends{\o}e},
\newblock \bibinfo{title}{Optimal shape design as a material distribution
  problem},
\newblock \bibinfo{journal}{Structural Optimization} \bibinfo{volume}{1}
  (\bibinfo{year}{1989}) \bibinfo{pages}{193--202}.
  \DOIprefix\doi{10.1007/bf01650949}.
\bibitem[{Xie and Steven(1993)}]{Xie1993}
\bibinfo{author}{Y.~M. Xie}, \bibinfo{author}{G.~P. Steven},
\newblock \bibinfo{title}{A simple evolutionary procedure for structural
  optimization},
\newblock \bibinfo{journal}{Computers {\&} Structures} \bibinfo{volume}{49}
  (\bibinfo{year}{1993}) \bibinfo{pages}{885--896}.
  \DOIprefix\doi{10.1016/0045-7949(93)90035-c}.
\bibitem[{Osher and Santosa(2001)}]{Osher2001}
\bibinfo{author}{S.~J. Osher}, \bibinfo{author}{F.~Santosa},
\newblock \bibinfo{title}{Level set methods for optimization problems involving
  geometry and constraints},
\newblock \bibinfo{journal}{Journal of Computational Physics}
  \bibinfo{volume}{171} (\bibinfo{year}{2001}) \bibinfo{pages}{272--288}.
  \DOIprefix\doi{10.1006/jcph.2001.6789}.
\bibitem[{Allaire et~al.(2005)Allaire, De~Gournay, Jouve, and
  Toader}]{Allaire2005}
\bibinfo{author}{G.~Allaire}, \bibinfo{author}{F.~De~Gournay},
  \bibinfo{author}{F.~Jouve}, \bibinfo{author}{A.~M. Toader},
\newblock \bibinfo{title}{Structural optimization using topological and shape
  sensitivity via a level set method},
\newblock \bibinfo{journal}{Control and cybernatics} \bibinfo{volume}{34}
  (\bibinfo{year}{2005}) \bibinfo{pages}{59}.
\bibitem[{Bourdin and Chambolle(2003)}]{Bourdin2003}
\bibinfo{author}{B.~Bourdin}, \bibinfo{author}{A.~Chambolle},
\newblock \bibinfo{title}{Design-dependent loads in topology optimization},
\newblock \bibinfo{journal}{{ESAIM}: Control, Optimisation and Calculus of
  Variations} \bibinfo{volume}{9} (\bibinfo{year}{2003})
  \bibinfo{pages}{19--48}. \DOIprefix\doi{10.1051/cocv:2002070}.
\bibitem[{Wang and Zhou(2004)}]{Wang2004a}
\bibinfo{author}{M.~Y. Wang}, \bibinfo{author}{S.~Zhou},
\newblock \bibinfo{title}{Phase field: A variationalmethod for structural
  topology optimization},
\newblock \bibinfo{journal}{Computer Modeling in Engineering and Sciences}
  \bibinfo{volume}{6} (\bibinfo{year}{2004}) \bibinfo{pages}{547--566}.
\bibitem[{Bends{\o}e and Kikuchi(1988)}]{Bendsoe1988}
\bibinfo{author}{M.~P. Bends{\o}e}, \bibinfo{author}{N.~Kikuchi},
\newblock \bibinfo{title}{Generating optimal topologies in structural design
  using a homogenization method},
\newblock \bibinfo{journal}{Computer Methods in Applied Mechanics and
  Engineering} \bibinfo{volume}{71} (\bibinfo{year}{1988})
  \bibinfo{pages}{197--224}. \DOIprefix\doi{10.1016/0045-7825(88)90086-2}.
\bibitem[{Mlejnek(1992)}]{Mlejnek1992}
\bibinfo{author}{H.~P. Mlejnek},
\newblock \bibinfo{title}{Some aspects of the genesis of structures},
\newblock \bibinfo{journal}{Structural Optimization} \bibinfo{volume}{5}
  (\bibinfo{year}{1992}) \bibinfo{pages}{64--69}.
  \DOIprefix\doi{10.1007/bf01744697}.
\bibitem[{Bends{\o}e and Sigmund(2004)}]{Bendsoe2004}
\bibinfo{author}{M.~P. Bends{\o}e}, \bibinfo{author}{O.~Sigmund},
  \bibinfo{title}{Topology Optimization}, \bibinfo{publisher}{Springer Berlin
  Heidelberg}, \bibinfo{year}{2004}. \DOIprefix\doi{10.1007/978-3-662-05086-6}.
\bibitem[{Xie and Steven(1997)}]{Xie1997}
\bibinfo{author}{Y.~M. Xie}, \bibinfo{author}{G.~P. Steven},
  \bibinfo{title}{Evolutionary Structural Optimization},
  \bibinfo{publisher}{Springer London}, \bibinfo{year}{1997}.
  \DOIprefix\doi{10.1007/978-1-4471-0985-3}.
\bibitem[{Yang et~al.(1999)Yang, Xie, Steven, and Querin}]{Yang1999}
\bibinfo{author}{X.~Y. Yang}, \bibinfo{author}{Y.~M. Xie},
  \bibinfo{author}{G.~P. Steven}, \bibinfo{author}{O.~M. Querin},
\newblock \bibinfo{title}{Bidirectional evolutionary method for stiffness
  optimization},
\newblock \bibinfo{journal}{{AIAA} Journal} \bibinfo{volume}{37}
  (\bibinfo{year}{1999}) \bibinfo{pages}{1483--1488}.
  \DOIprefix\doi{10.2514/3.14346}.
\bibitem[{Allaire et~al.(2002)Allaire, Jouve, and Toader}]{Allaire2002}
\bibinfo{author}{G.~Allaire}, \bibinfo{author}{F.~Jouve},
  \bibinfo{author}{A.-M. Toader},
\newblock \bibinfo{title}{A level-set method for shape optimization},
\newblock \bibinfo{journal}{Comptes Rendus Mathematique} \bibinfo{volume}{334}
  (\bibinfo{year}{2002}) \bibinfo{pages}{1125--1130}.
  \DOIprefix\doi{10.1016/s1631-073x(02)02412-3}.
\bibitem[{Sokolowski and Zochowski(1999)}]{Sokolowski1999}
\bibinfo{author}{J.~Sokolowski}, \bibinfo{author}{A.~Zochowski},
\newblock \bibinfo{title}{On the topological derivative in shape optimization},
\newblock \bibinfo{journal}{{SIAM} Journal on Control and Optimization}
  \bibinfo{volume}{37} (\bibinfo{year}{1999}) \bibinfo{pages}{1251--1272}.
  \DOIprefix\doi{10.1137/s0363012997323230}.
\bibitem[{Wang and Zhou(2004)}]{Wang2004}
\bibinfo{author}{M.~Y. Wang}, \bibinfo{author}{S.~Zhou},
\newblock \bibinfo{title}{Synthesis of shape and topology of multi-material
  structures with a phase-field method},
\newblock \bibinfo{journal}{Journal of Computer-Aided Materials Design}
  \bibinfo{volume}{11} (\bibinfo{year}{2004}) \bibinfo{pages}{117--138}.
  \DOIprefix\doi{10.1007/s10820-005-3169-y}.
\bibitem[{Takezawa et~al.(2010)Takezawa, Nishiwaki, and
  Kitamura}]{Takezawa2010}
\bibinfo{author}{A.~Takezawa}, \bibinfo{author}{S.~Nishiwaki},
  \bibinfo{author}{M.~Kitamura},
\newblock \bibinfo{title}{Shape and topology optimization based on the phase
  field method and sensitivity analysis},
\newblock \bibinfo{journal}{Journal of Computational Physics}
  \bibinfo{volume}{229} (\bibinfo{year}{2010}) \bibinfo{pages}{2697--2718}.
  \DOIprefix\doi{10.1016/j.jcp.2009.12.017}.
\bibitem[{Oliver et~al.(2019)Oliver, Yago, Cante, and
  Lloberas-Valls}]{Oliver2019}
\bibinfo{author}{J.~Oliver}, \bibinfo{author}{D.~Yago},
  \bibinfo{author}{J.~Cante}, \bibinfo{author}{O.~Lloberas-Valls},
\newblock \bibinfo{title}{Variational approach to relaxed topological
  optimization: Closed form solutions for structural problems in a sequential
  pseudo-time framework},
\newblock \bibinfo{journal}{Computer Methods in Applied Mechanics and
  Engineering} \bibinfo{volume}{355} (\bibinfo{year}{2019})
  \bibinfo{pages}{779--819}. \DOIprefix\doi{10.1016/j.cma.2019.06.038}.
\bibitem[{Yago et~al.(2020)Yago, Cante, Lloberas-Valls, and Oliver}]{Yago2020}
\bibinfo{author}{D.~Yago}, \bibinfo{author}{J.~Cante},
  \bibinfo{author}{O.~Lloberas-Valls}, \bibinfo{author}{J.~Oliver},
\newblock \bibinfo{title}{Topology optimization using the unsmooth variational
  topology optimization (unvartop) method: an educational implementation in
  matlab},
\newblock \bibinfo{journal}{Structural and Multidisciplinary Optimization}
  (\bibinfo{year}{2020}).
\bibitem[{Stolpe and Svanberg(2001)}]{Stolpe2001}
\bibinfo{author}{M.~Stolpe}, \bibinfo{author}{K.~Svanberg},
\newblock \bibinfo{title}{An alternative interpolation scheme for minimum
  compliance topology optimization},
\newblock \bibinfo{journal}{Structural and Multidisciplinary Optimization}
  \bibinfo{volume}{22} (\bibinfo{year}{2001}) \bibinfo{pages}{116--124}.
  \DOIprefix\doi{10.1007/s001580100129}.
\bibitem[{Bruns(2005)}]{Bruns2005}
\bibinfo{author}{T.~Bruns},
\newblock \bibinfo{title}{A reevaluation of the {SIMP} method with filtering
  and an alternative formulation for solid{\textendash}void topology
  optimization},
\newblock \bibinfo{journal}{Structural and Multidisciplinary Optimization}
  \bibinfo{volume}{30} (\bibinfo{year}{2005}) \bibinfo{pages}{428--436}.
  \DOIprefix\doi{10.1007/s00158-005-0537-x}.
\bibitem[{Zhou and Rozvany(1991)}]{Zhou1991}
\bibinfo{author}{M.~Zhou}, \bibinfo{author}{G.~Rozvany},
\newblock \bibinfo{title}{The {COC} algorithm, part {II}: Topological,
  geometrical and generalized shape optimization},
\newblock \bibinfo{journal}{Computer Methods in Applied Mechanics and
  Engineering} \bibinfo{volume}{89} (\bibinfo{year}{1991})
  \bibinfo{pages}{309--336}. \DOIprefix\doi{10.1016/0045-7825(91)90046-9}.
\bibitem[{Sigmund(1994)}]{Sigmund1994}
\bibinfo{author}{O.~Sigmund}, \bibinfo{title}{Design of material structures
  using topology optimization}, Ph.D. thesis, Technical University of Denmark,
  \bibinfo{year}{1994}.
\bibitem[{Sigmund(1997)}]{Sigmund1997}
\bibinfo{author}{O.~Sigmund},
\newblock \bibinfo{title}{On the design of compliant mechanisms using topology
  optimization},
\newblock \bibinfo{journal}{Mechanics of Structures and Machines}
  \bibinfo{volume}{25} (\bibinfo{year}{1997}) \bibinfo{pages}{493--524}.
  \DOIprefix\doi{10.1080/08905459708945415}.
\bibitem[{Sigmund and Petersson(1998)}]{Sigmund1998}
\bibinfo{author}{O.~Sigmund}, \bibinfo{author}{J.~Petersson},
\newblock \bibinfo{title}{Numerical instabilities in topology optimization: A
  survey on procedures dealing with checkerboards, mesh-dependencies and local
  minima},
\newblock \bibinfo{journal}{Structural Optimization} \bibinfo{volume}{16}
  (\bibinfo{year}{1998}) \bibinfo{pages}{68--75}.
  \DOIprefix\doi{10.1007/bf01214002}.
\bibitem[{Bourdin(2001)}]{Bourdin2001}
\bibinfo{author}{B.~Bourdin},
\newblock \bibinfo{title}{Filters in topology optimization},
\newblock \bibinfo{journal}{International Journal for Numerical Methods in
  Engineering} \bibinfo{volume}{50} (\bibinfo{year}{2001})
  \bibinfo{pages}{2143--2158}. \DOIprefix\doi{10.1002/nme.116}.
\bibitem[{Guest et~al.(2004)Guest, Pr{\'{e}}vost, and Belytschko}]{Guest2004}
\bibinfo{author}{J.~K. Guest}, \bibinfo{author}{J.~H. Pr{\'{e}}vost},
  \bibinfo{author}{T.~Belytschko},
\newblock \bibinfo{title}{Achieving minimum length scale in topology
  optimization using nodal design variables and projection functions},
\newblock \bibinfo{journal}{International Journal for Numerical Methods in
  Engineering} \bibinfo{volume}{61} (\bibinfo{year}{2004})
  \bibinfo{pages}{238--254}. \DOIprefix\doi{10.1002/nme.1064}.
\bibitem[{Guest et~al.(2011)Guest, Asadpoure, and Ha}]{Guest2011}
\bibinfo{author}{J.~K. Guest}, \bibinfo{author}{A.~Asadpoure},
  \bibinfo{author}{S.-H. Ha},
\newblock \bibinfo{title}{Eliminating beta-continuation from heaviside
  projection and density filter algorithms},
\newblock \bibinfo{journal}{Structural and Multidisciplinary Optimization}
  \bibinfo{volume}{44} (\bibinfo{year}{2011}) \bibinfo{pages}{443--453}.
  \DOIprefix\doi{10.1007/s00158-011-0676-1}.
\bibitem[{Sigmund(2007)}]{Sigmund2007}
\bibinfo{author}{O.~Sigmund},
\newblock \bibinfo{title}{Morphology-based black and white filters for topology
  optimization},
\newblock \bibinfo{journal}{Structural and Multidisciplinary Optimization}
  \bibinfo{volume}{33} (\bibinfo{year}{2007}) \bibinfo{pages}{401--424}.
  \DOIprefix\doi{10.1007/s00158-006-0087-x}.
\bibitem[{Wang et~al.(2010)Wang, Lazarov, and Sigmund}]{Wang2010}
\bibinfo{author}{F.~Wang}, \bibinfo{author}{B.~S. Lazarov},
  \bibinfo{author}{O.~Sigmund},
\newblock \bibinfo{title}{On projection methods, convergence and robust
  formulations in topology optimization},
\newblock \bibinfo{journal}{Structural and Multidisciplinary Optimization}
  \bibinfo{volume}{43} (\bibinfo{year}{2010}) \bibinfo{pages}{767--784}.
  \DOIprefix\doi{10.1007/s00158-010-0602-y}.
\bibitem[{Lazarov and Sigmund(2010)}]{Lazarov2010}
\bibinfo{author}{B.~S. Lazarov}, \bibinfo{author}{O.~Sigmund},
\newblock \bibinfo{title}{Filters in topology optimization based on
  helmholtz-type differential equations},
\newblock \bibinfo{journal}{International Journal for Numerical Methods in
  Engineering} \bibinfo{volume}{86} (\bibinfo{year}{2010})
  \bibinfo{pages}{765--781}. \DOIprefix\doi{10.1002/nme.3072}.
\bibitem[{Kawamoto et~al.(2010)Kawamoto, Matsumori, Yamasaki, Nomura, Kondoh,
  and Nishiwaki}]{Kawamoto2010}
\bibinfo{author}{A.~Kawamoto}, \bibinfo{author}{T.~Matsumori},
  \bibinfo{author}{S.~Yamasaki}, \bibinfo{author}{T.~Nomura},
  \bibinfo{author}{T.~Kondoh}, \bibinfo{author}{S.~Nishiwaki},
\newblock \bibinfo{title}{Heaviside projection based topology optimization by a
  {PDE}-filtered scalar function},
\newblock \bibinfo{journal}{Structural and Multidisciplinary Optimization}
  \bibinfo{volume}{44} (\bibinfo{year}{2010}) \bibinfo{pages}{19--24}.
  \DOIprefix\doi{10.1007/s00158-010-0562-2}.
\bibitem[{Haber et~al.(1996)Haber, Jog, and Bends{\o}e}]{Haber1996}
\bibinfo{author}{R.~B. Haber}, \bibinfo{author}{C.~S. Jog},
  \bibinfo{author}{M.~P. Bends{\o}e},
\newblock \bibinfo{title}{A new approach to variable-topology shape design
  using a constraint on perimeter},
\newblock \bibinfo{journal}{Structural Optimization} \bibinfo{volume}{11}
  (\bibinfo{year}{1996}) \bibinfo{pages}{1--12}.
  \DOIprefix\doi{10.1007/bf01279647}.
\bibitem[{Fernandes et~al.(1999)Fernandes, Guedes, and
  Rodrigues}]{Fernandes1999}
\bibinfo{author}{P.~Fernandes}, \bibinfo{author}{J.~Guedes},
  \bibinfo{author}{H.~Rodrigues},
\newblock \bibinfo{title}{Topology optimization of three-dimensional linear
  elastic structures with a constraint on
  {\textquotedblleft}perimeter{\textquotedblright}},
\newblock \bibinfo{journal}{Computers {\&} Structures} \bibinfo{volume}{73}
  (\bibinfo{year}{1999}) \bibinfo{pages}{583--594}.
  \DOIprefix\doi{10.1016/s0045-7949(98)00312-5}.
\bibitem[{Petersson and Sigmund(1998)}]{Petersson1998}
\bibinfo{author}{J.~Petersson}, \bibinfo{author}{O.~Sigmund},
\newblock \bibinfo{title}{Slope constrained topology optimization},
\newblock \bibinfo{journal}{International Journal for Numerical Methods in
  Engineering} \bibinfo{volume}{41} (\bibinfo{year}{1998})
  \bibinfo{pages}{1417--1434}.
  \DOIprefix\doi{10.1002/(sici)1097-0207(19980430)41:8<1417::aid-nme344>3.0.co;2-n}.
\bibitem[{Jog and Haber(1996)}]{Jog1996}
\bibinfo{author}{C.~S. Jog}, \bibinfo{author}{R.~B. Haber},
\newblock \bibinfo{title}{Stability of finite element models for
  distributed-parameter optimization and topology design},
\newblock \bibinfo{journal}{Computer Methods in Applied Mechanics and
  Engineering} \bibinfo{volume}{130} (\bibinfo{year}{1996})
  \bibinfo{pages}{203--226}. \DOIprefix\doi{10.1016/0045-7825(95)00928-0}.
\bibitem[{Suzuki and Kikuchi(1991)}]{Suzuki1991}
\bibinfo{author}{K.~Suzuki}, \bibinfo{author}{N.~Kikuchi},
\newblock \bibinfo{title}{A homogenization method for shape and topology
  optimization},
\newblock \bibinfo{journal}{Computer Methods in Applied Mechanics and
  Engineering} \bibinfo{volume}{93} (\bibinfo{year}{1991})
  \bibinfo{pages}{291--318}. \DOIprefix\doi{10.1016/0045-7825(91)90245-2}.
\bibitem[{Ma et~al.(1993)Ma, Kikuchi, and Hagiwara}]{Ma1993}
\bibinfo{author}{Z.~D. Ma}, \bibinfo{author}{N.~Kikuchi},
  \bibinfo{author}{I.~Hagiwara},
\newblock \bibinfo{title}{Structural topology and shape optimization for a
  frequency response problem},
\newblock \bibinfo{journal}{Computational Mechanics} \bibinfo{volume}{13}
  (\bibinfo{year}{1993}) \bibinfo{pages}{157--174}.
  \DOIprefix\doi{10.1007/bf00370133}.
\bibitem[{Svanberg(1987)}]{Svanberg1987}
\bibinfo{author}{K.~Svanberg},
\newblock \bibinfo{title}{The method of moving asymptotes{\textemdash}a new
  method for structural optimization},
\newblock \bibinfo{journal}{International Journal for Numerical Methods in
  Engineering} \bibinfo{volume}{24} (\bibinfo{year}{1987})
  \bibinfo{pages}{359--373}. \DOIprefix\doi{10.1002/nme.1620240207}.
\bibitem[{Schnack(1988)}]{Schnack1988}
\bibinfo{author}{E.~Schnack}, \bibinfo{title}{Gradientless shape optimization
  with FEM}, \bibinfo{publisher}{VDI-Verl}, \bibinfo{address}{D{\"u}sseldorf},
  \bibinfo{year}{1988}.
\bibitem[{Mattheck and Burkhardt(1990)}]{MATTHECK1990}
\bibinfo{author}{C.~Mattheck}, \bibinfo{author}{S.~Burkhardt},
\newblock \bibinfo{title}{A new method of structural shape optimization based
  on biological growth},
\newblock \bibinfo{journal}{International Journal of Fatigue}
  \bibinfo{volume}{12} (\bibinfo{year}{1990}) \bibinfo{pages}{185--190}.
  \DOIprefix\doi{10.1016/0142-1123(90)90094-u}.
\bibitem[{Zhou and Rozvany(2001)}]{Zhou2001}
\bibinfo{author}{M.~Zhou}, \bibinfo{author}{G.~Rozvany},
\newblock \bibinfo{title}{On the validity of {ESO} type methods in topology
  optimization},
\newblock \bibinfo{journal}{Structural and Multidisciplinary Optimization}
  \bibinfo{volume}{21} (\bibinfo{year}{2001}) \bibinfo{pages}{80--83}.
  \DOIprefix\doi{10.1007/s001580050170}.
\bibitem[{Xie and Steven(1994{\natexlab{a}})}]{Xie1994}
\bibinfo{author}{Y.~Xie}, \bibinfo{author}{G.~Steven},
\newblock \bibinfo{title}{Optimal design of multiple load case structures using
  an evolutionary procedure},
\newblock \bibinfo{journal}{Engineering Computations} \bibinfo{volume}{11}
  (\bibinfo{year}{1994}{\natexlab{a}}) \bibinfo{pages}{295--302}.
  \DOIprefix\doi{10.1108/02644409410799290}.
\bibitem[{Xie and Steven(1994{\natexlab{b}})}]{Xie1994a}
\bibinfo{author}{Y.~Xie}, \bibinfo{author}{G.~Steven},
\newblock \bibinfo{title}{A simple approach to structural frequency
  optimization},
\newblock \bibinfo{journal}{Computers {\&} Structures} \bibinfo{volume}{53}
  (\bibinfo{year}{1994}{\natexlab{b}}) \bibinfo{pages}{1487--1491}.
  \DOIprefix\doi{10.1016/0045-7949(94)90414-6}.
\bibitem[{Zhao et~al.(1997)Zhao, Steven, and Xie}]{Zhao1997}
\bibinfo{author}{C.~Zhao}, \bibinfo{author}{G.~Steven},
  \bibinfo{author}{Y.~Xie},
\newblock \bibinfo{title}{Evolutionary natural frequency optimization of
  two-dimensional structures with additional non-structural lumped masses},
\newblock \bibinfo{journal}{Engineering Computations} \bibinfo{volume}{14}
  (\bibinfo{year}{1997}) \bibinfo{pages}{233--251}.
  \DOIprefix\doi{10.1108/02644409710166208}.
\bibitem[{Chu et~al.(1996)Chu, Xie, Hira, and Steven}]{Chu1996}
\bibinfo{author}{D.~Chu}, \bibinfo{author}{Y.~Xie}, \bibinfo{author}{A.~Hira},
  \bibinfo{author}{G.~Steven},
\newblock \bibinfo{title}{Evolutionary structural optimization for problems
  with stiffness constraints},
\newblock \bibinfo{journal}{Finite Elements in Analysis and Design}
  \bibinfo{volume}{21} (\bibinfo{year}{1996}) \bibinfo{pages}{239--251}.
  \DOIprefix\doi{10.1016/0168-874x(95)00043-s}.
\bibitem[{Querin et~al.(1996)Querin, Steven, and Xie}]{Querin1996}
\bibinfo{author}{O.~Querin}, \bibinfo{author}{G.~Steven},
  \bibinfo{author}{Y.~Xie},
\newblock \bibinfo{title}{Topology optimisation of structures with material and
  geometric non-linearities},
\newblock in: \bibinfo{booktitle}{6th Symposium on Multidisciplinary Analysis
  and Optimization}, \bibinfo{publisher}{American Institute of Aeronautics and
  Astronautics}, \bibinfo{year}{1996}. \DOIprefix\doi{10.2514/6.1996-4116}.
\bibitem[{Manickarajah et~al.(1998)Manickarajah, Xie, and
  Steven}]{Manickarajah1998}
\bibinfo{author}{D.~Manickarajah}, \bibinfo{author}{Y.~Xie},
  \bibinfo{author}{G.~Steven},
\newblock \bibinfo{title}{An evolutionary method for optimization of plate
  buckling resistance},
\newblock \bibinfo{journal}{Finite Elements in Analysis and Design}
  \bibinfo{volume}{29} (\bibinfo{year}{1998}) \bibinfo{pages}{205--230}.
  \DOIprefix\doi{10.1016/s0168-874x(98)00012-2}.
\bibitem[{Li et~al.(2000)Li, Steven, Querin, and Xie}]{Li2000}
\bibinfo{author}{Q.~Li}, \bibinfo{author}{G.~P. Steven}, \bibinfo{author}{O.~M.
  Querin}, \bibinfo{author}{Y.~Xie},
\newblock \bibinfo{title}{Structural topology design with multiple thermal
  criteria},
\newblock \bibinfo{journal}{Engineering Computations} \bibinfo{volume}{17}
  (\bibinfo{year}{2000}) \bibinfo{pages}{715--734}.
  \DOIprefix\doi{10.1108/02644400010340642}.
\bibitem[{Qing~Li(2001)}]{Li2001a}
\bibinfo{author}{Y.~M.~X. Qing~Li, Grant P.~Steven},
\newblock \bibinfo{title}{Thermoelastic topology optimization for problems with
  varying temperature fields},
\newblock \bibinfo{journal}{Journal of Thermal Stresses} \bibinfo{volume}{24}
  (\bibinfo{year}{2001}) \bibinfo{pages}{347--366}.
  \DOIprefix\doi{10.1080/01495730151078153}.
\bibitem[{Li et~al.(1998)Li, Steven, and Xie}]{Li1998}
\bibinfo{author}{W.~Li}, \bibinfo{author}{G.~Steven}, \bibinfo{author}{Y.~Xie},
\newblock \bibinfo{title}{Shape design for elastic contact problems by
  evolutionary structural optimization},
\newblock in: \bibinfo{booktitle}{7th {AIAA}/{USAF}/{NASA}/{ISSMO} Symposium on
  Multidisciplinary Analysis and Optimization}, \bibinfo{publisher}{American
  Institute of Aeronautics and Astronautics}, \bibinfo{year}{1998}.
  \DOIprefix\doi{10.2514/6.1998-4851}.
\bibitem[{Li et~al.(2005)Li, Li, Steven, and Xie}]{Li2005}
\bibinfo{author}{W.~Li}, \bibinfo{author}{Q.~Li}, \bibinfo{author}{G.~P.
  Steven}, \bibinfo{author}{Y.~Xie},
\newblock \bibinfo{title}{An evolutionary shape optimization for elastic
  contact problems subject to multiple load cases},
\newblock \bibinfo{journal}{Computer Methods in Applied Mechanics and
  Engineering} \bibinfo{volume}{194} (\bibinfo{year}{2005})
  \bibinfo{pages}{3394--3415}. \DOIprefix\doi{10.1016/j.cma.2004.12.024}.
\bibitem[{Querin et~al.(1998)Querin, Steven, and Xie}]{Querin1998}
\bibinfo{author}{O.~Querin}, \bibinfo{author}{G.~Steven},
  \bibinfo{author}{Y.~Xie},
\newblock \bibinfo{title}{Evolutionary structural optimisation ({ESO}) using a
  bidirectional algorithm},
\newblock \bibinfo{journal}{Engineering Computations} \bibinfo{volume}{15}
  (\bibinfo{year}{1998}) \bibinfo{pages}{1031--1048}.
  \DOIprefix\doi{10.1108/02644409810244129}.
\bibitem[{Querin et~al.(2000)Querin, Young, Steven, and Xie}]{Querin2000}
\bibinfo{author}{O.~Querin}, \bibinfo{author}{V.~Young},
  \bibinfo{author}{G.~Steven}, \bibinfo{author}{Y.~Xie},
\newblock \bibinfo{title}{Computational efficiency and validation of
  bi-directional evolutionary structural optimisation},
\newblock \bibinfo{journal}{Computer Methods in Applied Mechanics and
  Engineering} \bibinfo{volume}{189} (\bibinfo{year}{2000})
  \bibinfo{pages}{559--573}. \DOIprefix\doi{10.1016/s0045-7825(99)00309-6}.
\bibitem[{Li et~al.(2001)Li, Steven, and Xie}]{Li2001}
\bibinfo{author}{Q.~Li}, \bibinfo{author}{G.~Steven}, \bibinfo{author}{Y.~Xie},
\newblock \bibinfo{title}{A simple checkerboard suppression algorithm for
  evolutionary structural optimization},
\newblock \bibinfo{journal}{Structural and Multidisciplinary Optimization}
  \bibinfo{volume}{22} (\bibinfo{year}{2001}) \bibinfo{pages}{230--239}.
  \DOIprefix\doi{10.1007/s001580100140}.
\bibitem[{Huang and Xie(2007)}]{Huang2007}
\bibinfo{author}{X.~Huang}, \bibinfo{author}{Y.~Xie},
\newblock \bibinfo{title}{Convergent and mesh-independent solutions for the
  bi-directional evolutionary structural optimization method},
\newblock \bibinfo{journal}{Finite Elements in Analysis and Design}
  \bibinfo{volume}{43} (\bibinfo{year}{2007}) \bibinfo{pages}{1039--1049}.
  \DOIprefix\doi{10.1016/j.finel.2007.06.006}.
\bibitem[{Yang et~al.(2002)Yang, Xie, Liu, Parks, and Clarkson}]{Yang2002}
\bibinfo{author}{X.~Yang}, \bibinfo{author}{Y.~Xie}, \bibinfo{author}{J.~Liu},
  \bibinfo{author}{G.~Parks}, \bibinfo{author}{P.~Clarkson},
\newblock \bibinfo{title}{Perimeter control in the bidirectional evolutionary
  optimization method},
\newblock \bibinfo{journal}{Structural and Multidisciplinary Optimization}
  \bibinfo{volume}{24} (\bibinfo{year}{2002}) \bibinfo{pages}{430--440}.
  \DOIprefix\doi{10.1007/s00158-002-0256-5}.
\bibitem[{Zhu et~al.(2006)Zhu, Zhang, and Qiu}]{Zhu2006}
\bibinfo{author}{J.~H. Zhu}, \bibinfo{author}{W.~H. Zhang},
  \bibinfo{author}{K.~P. Qiu},
\newblock \bibinfo{title}{Bi-directional evolutionary topology optimization
  using element replaceable method},
\newblock \bibinfo{journal}{Computational Mechanics} \bibinfo{volume}{40}
  (\bibinfo{year}{2006}) \bibinfo{pages}{97--109}.
  \DOIprefix\doi{10.1007/s00466-006-0087-0}.
\bibitem[{Huang and Xie(2008)}]{Huang2008}
\bibinfo{author}{X.~Huang}, \bibinfo{author}{Y.~M. Xie},
\newblock \bibinfo{title}{Bi-directional evolutionary topology optimization of
  continuum structures with one or multiple materials},
\newblock \bibinfo{journal}{Computational Mechanics} \bibinfo{volume}{43}
  (\bibinfo{year}{2008}) \bibinfo{pages}{393--401}.
  \DOIprefix\doi{10.1007/s00466-008-0312-0}.
\bibitem[{Huang and Xie(2010)}]{Huang2010}
\bibinfo{author}{X.~Huang}, \bibinfo{author}{Y.~M. Xie},
  \bibinfo{title}{Evolutionary Topology Optimization of Continuum Structures},
  \bibinfo{publisher}{John Wiley {\&} Sons, Ltd}, \bibinfo{year}{2010}.
  \DOIprefix\doi{10.1002/9780470689486}.
\bibitem[{Osher and Sethian(1988)}]{Osher1988}
\bibinfo{author}{S.~Osher}, \bibinfo{author}{J.~A. Sethian},
\newblock \bibinfo{title}{Fronts propagating with curvature-dependent speed:
  Algorithms based on hamilton-jacobi formulations},
\newblock \bibinfo{journal}{Journal of Computational Physics}
  \bibinfo{volume}{79} (\bibinfo{year}{1988}) \bibinfo{pages}{12--49}.
  \DOIprefix\doi{10.1016/0021-9991(88)90002-2}.
\bibitem[{Sethian(1999)}]{Sethian1999}
\bibinfo{author}{J.~A. Sethian}, \bibinfo{title}{Level Set Methods and Fast
  Marching Methods}, \bibinfo{publisher}{Cambridge University Press},
  \bibinfo{year}{1999}. \URLprefix
  \url{https://www.ebook.de/de/product/2965913/j_a_sethian_level_set_methods_and_fast_marching_methods.html}.
\bibitem[{Osher and Fedkiw(2003)}]{Osher2003}
\bibinfo{author}{S.~Osher}, \bibinfo{author}{R.~Fedkiw}, \bibinfo{title}{Level
  Set Methods and Dynamic Implicit Surfaces}, \bibinfo{publisher}{Springer New
  York}, \bibinfo{year}{2003}. \DOIprefix\doi{10.1007/b98879}.
\bibitem[{Haber and Bendsoe(1998)}]{Haber1998}
\bibinfo{author}{R.~Haber}, \bibinfo{author}{M.~Bendsoe},
\newblock \bibinfo{title}{Problem formulation, solution procedures and
  geometric modeling - key issues in variable-topology optimization},
\newblock in: \bibinfo{booktitle}{7th {AIAA}/{USAF}/{NASA}/{ISSMO} Symposium on
  Multidisciplinary Analysis and Optimization}, \bibinfo{publisher}{American
  Institute of Aeronautics and Astronautics}, \bibinfo{year}{1998}.
  \DOIprefix\doi{10.2514/6.1998-4948}.
\bibitem[{Sethian and Wiegmann(2000)}]{Sethian2000}
\bibinfo{author}{J.~Sethian}, \bibinfo{author}{A.~Wiegmann},
\newblock \bibinfo{title}{Structural boundary design via level set and immersed
  interface methods},
\newblock \bibinfo{journal}{Journal of Computational Physics}
  \bibinfo{volume}{163} (\bibinfo{year}{2000}) \bibinfo{pages}{489--528}.
  \DOIprefix\doi{10.1006/jcph.2000.6581}.
\bibitem[{Sokolowski and Zolesio(1992)}]{Sokolowski1992}
\bibinfo{author}{J.~Sokolowski}, \bibinfo{author}{J.-P. Zolesio},
  \bibinfo{title}{Introduction to Shape Optimization},
  \bibinfo{publisher}{Springer Berlin Heidelberg}, \bibinfo{year}{1992}.
  \DOIprefix\doi{10.1007/978-3-642-58106-9}.
\bibitem[{Eschenauer et~al.(1994)Eschenauer, Kobelev, and
  Schumacher}]{Eschenauer1994}
\bibinfo{author}{H.~A. Eschenauer}, \bibinfo{author}{V.~V. Kobelev},
  \bibinfo{author}{A.~Schumacher},
\newblock \bibinfo{title}{Bubble method for topology and shape optimization of
  structures},
\newblock \bibinfo{journal}{Structural Optimization} \bibinfo{volume}{8}
  (\bibinfo{year}{1994}) \bibinfo{pages}{42--51}.
  \DOIprefix\doi{10.1007/bf01742933}.
\bibitem[{Schumacher(1996)}]{Schumacher1996}
\bibinfo{author}{A.~Schumacher}, \bibinfo{title}{Topologieoptimierung von
  Bauteilstrukturen unter Verwendung von Lochpositionierungskriterien}, Ph.D.
  thesis, Forschungszentrum für Multidisziplinäre Analysen und Angewandte
  Strukturoptimierung, Institut für Mechanik und Regelungstechnik,
  \bibinfo{address}{Siegen, Germany}, \bibinfo{year}{1996}.
\bibitem[{Duysinx et~al.(2006)Duysinx, Miegroet, Jacobs, and
  Fleury}]{Duysinx2006}
\bibinfo{author}{P.~Duysinx}, \bibinfo{author}{L.~V. Miegroet},
  \bibinfo{author}{T.~Jacobs}, \bibinfo{author}{C.~Fleury},
\newblock \bibinfo{title}{Generalized shape optimization using x-{FEM} and
  level set methods},
\newblock in: \bibinfo{booktitle}{Solid Mechanics and Its Applications},
  \bibinfo{publisher}{Springer Netherlands}, \bibinfo{year}{2006}, pp.
  \bibinfo{pages}{23--32}. \DOIprefix\doi{10.1007/1-4020-4752-5_3}.
\bibitem[{Wein et~al.(2020)Wein, Dunning, and Norato}]{Wein2020}
\bibinfo{author}{F.~Wein}, \bibinfo{author}{P.~D. Dunning},
  \bibinfo{author}{J.~A. Norato},
\newblock \bibinfo{title}{A review on feature-mapping methods for structural
  optimization},
\newblock \bibinfo{journal}{Structural and Multidisciplinary Optimization}
  \bibinfo{volume}{62} (\bibinfo{year}{2020}) \bibinfo{pages}{1597--1638}.
  \DOIprefix\doi{10.1007/s00158-020-02649-6}.
\bibitem[{Allaire et~al.(1997)Allaire, Bonnetier, Francfort, and
  Jouve}]{Allaire1997}
\bibinfo{author}{G.~Allaire}, \bibinfo{author}{E.~Bonnetier},
  \bibinfo{author}{G.~Francfort}, \bibinfo{author}{F.~Jouve},
\newblock \bibinfo{title}{Shape optimization by the homogenization method},
\newblock \bibinfo{journal}{Numerische Mathematik} \bibinfo{volume}{76}
  (\bibinfo{year}{1997}) \bibinfo{pages}{27--68}.
  \DOIprefix\doi{10.1007/s002110050253}.
\bibitem[{Dambrine and Kateb(2009)}]{Dambrine2009}
\bibinfo{author}{M.~Dambrine}, \bibinfo{author}{D.~Kateb},
\newblock \bibinfo{title}{On the ersatz material approximation in level-set
  methods},
\newblock \bibinfo{journal}{{ESAIM}: Control, Optimisation and Calculus of
  Variations} \bibinfo{volume}{16} (\bibinfo{year}{2009})
  \bibinfo{pages}{618--634}. \DOIprefix\doi{10.1051/cocv/2009023}.
\bibitem[{Wang and Wang(2006)}]{Wang2006}
\bibinfo{author}{S.~Wang}, \bibinfo{author}{M.~Y. Wang},
\newblock \bibinfo{title}{Radial basis functions and level set method for
  structural topology optimization},
\newblock \bibinfo{journal}{International Journal for Numerical Methods in
  Engineering} \bibinfo{volume}{65} (\bibinfo{year}{2006})
  \bibinfo{pages}{2060--2090}. \DOIprefix\doi{10.1002/nme.1536}.
\bibitem[{Wang et~al.(2007)Wang, Lim, Khoo, and Wang}]{Wang2007}
\bibinfo{author}{S.~Wang}, \bibinfo{author}{K.~Lim}, \bibinfo{author}{B.~Khoo},
  \bibinfo{author}{M.~Wang},
\newblock \bibinfo{title}{An extended level set method for shape and topology
  optimization},
\newblock \bibinfo{journal}{Journal of Computational Physics}
  \bibinfo{volume}{221} (\bibinfo{year}{2007}) \bibinfo{pages}{395--421}.
  \DOIprefix\doi{10.1016/j.jcp.2006.06.029}.
\bibitem[{Luo et~al.(2007)Luo, Tong, Wang, and Wang}]{Luo2007}
\bibinfo{author}{Z.~Luo}, \bibinfo{author}{L.~Tong}, \bibinfo{author}{M.~Y.
  Wang}, \bibinfo{author}{S.~Wang},
\newblock \bibinfo{title}{Shape and topology optimization of compliant
  mechanisms using a parameterization level set method},
\newblock \bibinfo{journal}{Journal of Computational Physics}
  \bibinfo{volume}{227} (\bibinfo{year}{2007}) \bibinfo{pages}{680--705}.
  \DOIprefix\doi{10.1016/j.jcp.2007.08.011}.
\bibitem[{Luo et~al.(2008)Luo, Wang, Wang, and Wei}]{Luo2008}
\bibinfo{author}{Z.~Luo}, \bibinfo{author}{M.~Y. Wang},
  \bibinfo{author}{S.~Wang}, \bibinfo{author}{P.~Wei},
\newblock \bibinfo{title}{A level set-based parameterization method for
  structural shape and topology optimization},
\newblock \bibinfo{journal}{International Journal for Numerical Methods in
  Engineering} \bibinfo{volume}{76} (\bibinfo{year}{2008})
  \bibinfo{pages}{1--26}. \DOIprefix\doi{10.1002/nme.2092}.
\bibitem[{Allaire and Jouve(2005)}]{Allaire2005a}
\bibinfo{author}{G.~Allaire}, \bibinfo{author}{F.~Jouve},
\newblock \bibinfo{title}{A level-set method for vibration and multiple loads
  structural optimization},
\newblock \bibinfo{journal}{Computer Methods in Applied Mechanics and
  Engineering} \bibinfo{volume}{194} (\bibinfo{year}{2005})
  \bibinfo{pages}{3269--3290}. \DOIprefix\doi{10.1016/j.cma.2004.12.018}.
\bibitem[{Ha and Cho(2005)}]{Ha2005}
\bibinfo{author}{S.-H. Ha}, \bibinfo{author}{S.~Cho},
\newblock \bibinfo{title}{Topological shape optimization of heat conduction
  problems using level set approach},
\newblock \bibinfo{journal}{Numerical Heat Transfer, Part B: Fundamentals}
  \bibinfo{volume}{48} (\bibinfo{year}{2005}) \bibinfo{pages}{67--88}.
  \DOIprefix\doi{10.1080/10407790590935966}.
\bibitem[{Michael Yu~Wang(2004)}]{MichaelYuWang2004}
\bibinfo{author}{X.~W. Michael Yu~Wang},
\newblock \bibinfo{title}{Pde-driven level sets, shape sensitivity and
  curvature flow for structural topology optimization},
\newblock \bibinfo{journal}{Computer Modeling in Engineering \& Sciences}
  \bibinfo{volume}{6} (\bibinfo{year}{2004}) \bibinfo{pages}{373--396}.
  \URLprefix \url{http://www.techscience.com/CMES/v6n4/24856}.
  \DOIprefix\doi{10.3970/cmes.2004.006.373}.
\bibitem[{C{\'{e}}a et~al.(2000)C{\'{e}}a, Garreau, Guillaume, and
  Masmoudi}]{Cea2000}
\bibinfo{author}{J.~C{\'{e}}a}, \bibinfo{author}{S.~Garreau},
  \bibinfo{author}{P.~Guillaume}, \bibinfo{author}{M.~Masmoudi},
\newblock \bibinfo{title}{The shape and topological optimizations connection},
\newblock \bibinfo{journal}{Computer Methods in Applied Mechanics and
  Engineering} \bibinfo{volume}{188} (\bibinfo{year}{2000})
  \bibinfo{pages}{713--726}. \DOIprefix\doi{10.1016/s0045-7825(99)00357-6}.
\bibitem[{Garreau et~al.(2001)Garreau, Guillaume, and Masmoudi}]{Garreau2001}
\bibinfo{author}{S.~Garreau}, \bibinfo{author}{P.~Guillaume},
  \bibinfo{author}{M.~Masmoudi},
\newblock \bibinfo{title}{The topological asymptotic for {PDE} systems: The
  elasticity case},
\newblock \bibinfo{journal}{{SIAM} Journal on Control and Optimization}
  \bibinfo{volume}{39} (\bibinfo{year}{2001}) \bibinfo{pages}{1756--1778}.
  \DOIprefix\doi{10.1137/s0363012900369538}.
\bibitem[{Burger et~al.(2004)Burger, Hackl, and Ring}]{Burger2004}
\bibinfo{author}{M.~Burger}, \bibinfo{author}{B.~Hackl},
  \bibinfo{author}{W.~Ring},
\newblock \bibinfo{title}{Incorporating topological derivatives into level set
  methods},
\newblock \bibinfo{journal}{Journal of Computational Physics}
  \bibinfo{volume}{194} (\bibinfo{year}{2004}) \bibinfo{pages}{344--362}.
  \DOIprefix\doi{10.1016/j.jcp.2003.09.033}.
\bibitem[{Giusti et~al.(2016)Giusti, Ferrer, and Oliver}]{Giusti2016}
\bibinfo{author}{S.~Giusti}, \bibinfo{author}{A.~Ferrer},
  \bibinfo{author}{J.~Oliver},
\newblock \bibinfo{title}{Topological sensitivity analysis in heterogeneous
  anisotropic elasticity problem. theoretical and computational aspects},
\newblock \bibinfo{journal}{Computer Methods in Applied Mechanics and
  Engineering} \bibinfo{volume}{311} (\bibinfo{year}{2016})
  \bibinfo{pages}{134--150}. \DOIprefix\doi{10.1016/j.cma.2016.08.004}.
\bibitem[{Novotny et~al.(2007)Novotny, Feij{\'{o}}o, Taroco, and
  Padra}]{Novotny2007}
\bibinfo{author}{A.~Novotny}, \bibinfo{author}{R.~Feij{\'{o}}o},
  \bibinfo{author}{E.~Taroco}, \bibinfo{author}{C.~Padra},
\newblock \bibinfo{title}{Topological sensitivity analysis for
  three-dimensional linear elasticity problem},
\newblock \bibinfo{journal}{Computer Methods in Applied Mechanics and
  Engineering} \bibinfo{volume}{196} (\bibinfo{year}{2007})
  \bibinfo{pages}{4354--4364}. \DOIprefix\doi{10.1016/j.cma.2007.05.006}.
\bibitem[{Giusti et~al.(2009)Giusti, Novotny, and Soko{\l}owski}]{Giusti2009}
\bibinfo{author}{S.~M. Giusti}, \bibinfo{author}{A.~A. Novotny},
  \bibinfo{author}{J.~Soko{\l}owski},
\newblock \bibinfo{title}{Topological derivative for steady-state orthotropic
  heat diffusion problem},
\newblock \bibinfo{journal}{Structural and Multidisciplinary Optimization}
  \bibinfo{volume}{40} (\bibinfo{year}{2009}) \bibinfo{pages}{53--64}.
  \DOIprefix\doi{10.1007/s00158-009-0359-3}.
\bibitem[{Marczak(2007)}]{Marczak2007}
\bibinfo{author}{R.~J. Marczak},
\newblock \bibinfo{title}{Topology optimization and boundary
  elements{\textemdash}a preliminary implementation for linear heat transfer},
\newblock \bibinfo{journal}{Engineering Analysis with Boundary Elements}
  \bibinfo{volume}{31} (\bibinfo{year}{2007}) \bibinfo{pages}{793--802}.
  \DOIprefix\doi{10.1016/j.enganabound.2007.01.005}.
\bibitem[{Barbarosie and Toader(2009)}]{Barbarosie2009}
\bibinfo{author}{C.~Barbarosie}, \bibinfo{author}{A.-M. Toader},
\newblock \bibinfo{title}{Shape and topology optimization for periodic
  problems},
\newblock \bibinfo{journal}{Structural and Multidisciplinary Optimization}
  \bibinfo{volume}{40} (\bibinfo{year}{2009}) \bibinfo{pages}{381--391}.
  \DOIprefix\doi{10.1007/s00158-009-0378-0}.
\bibitem[{Amstutz et~al.(2010)Amstutz, Giusti, Novotny, and
  de~Souza~Neto}]{Amstutz2010}
\bibinfo{author}{S.~Amstutz}, \bibinfo{author}{S.~M. Giusti},
  \bibinfo{author}{A.~A. Novotny}, \bibinfo{author}{E.~A. de~Souza~Neto},
\newblock \bibinfo{title}{Topological derivative for multi-scale linear
  elasticity models applied to the synthesis of microstructures},
\newblock \bibinfo{journal}{International Journal for Numerical Methods in
  Engineering} \bibinfo{volume}{84} (\bibinfo{year}{2010})
  \bibinfo{pages}{733--756}. \DOIprefix\doi{10.1002/nme.2922}.
\bibitem[{Challis and Guest(2009)}]{Challis2009}
\bibinfo{author}{V.~J. Challis}, \bibinfo{author}{J.~K. Guest},
\newblock \bibinfo{title}{Level set topology optimization of fluids in stokes
  flow},
\newblock \bibinfo{journal}{International Journal for Numerical Methods in
  Engineering} \bibinfo{volume}{79} (\bibinfo{year}{2009})
  \bibinfo{pages}{1284--1308}. \DOIprefix\doi{10.1002/nme.2616}.
\bibitem[{Fulma{\'{n}}ski et~al.(2007)Fulma{\'{n}}ski, Laurain, Scheid, and
  Soko{\l}owski}]{Fulmanski2007}
\bibinfo{author}{P.~Fulma{\'{n}}ski}, \bibinfo{author}{A.~Laurain},
  \bibinfo{author}{J.-F. Scheid}, \bibinfo{author}{J.~Soko{\l}owski},
\newblock \bibinfo{title}{A level set method in shape and topology optimization
  for variational inequalities},
\newblock \bibinfo{journal}{International Journal of Applied Mathematics and
  Computer Science} \bibinfo{volume}{17} (\bibinfo{year}{2007})
  \bibinfo{pages}{413--430}. \DOIprefix\doi{10.2478/v10006-007-0034-z}.
\bibitem[{Norato et~al.(2007)Norato, Bends{\o}e, Haber, and
  Tortorelli}]{Norato2007}
\bibinfo{author}{J.~A. Norato}, \bibinfo{author}{M.~P. Bends{\o}e},
  \bibinfo{author}{R.~B. Haber}, \bibinfo{author}{D.~A. Tortorelli},
\newblock \bibinfo{title}{A topological derivative method for topology
  optimization},
\newblock \bibinfo{journal}{Structural and Multidisciplinary Optimization}
  \bibinfo{volume}{33} (\bibinfo{year}{2007}) \bibinfo{pages}{375--386}.
  \DOIprefix\doi{10.1007/s00158-007-0094-6}.
\bibitem[{Amstutz and Andrä(2006)}]{Amstutz2006}
\bibinfo{author}{S.~Amstutz}, \bibinfo{author}{H.~Andrä},
\newblock \bibinfo{title}{A new algorithm for topology optimization using a
  level-set method},
\newblock \bibinfo{journal}{Journal of Computational Physics}
  \bibinfo{volume}{216} (\bibinfo{year}{2006}) \bibinfo{pages}{573--588}.
  \DOIprefix\doi{10.1016/j.jcp.2005.12.015}.
\bibitem[{He et~al.(2007)He, Kao, and Osher}]{He2007}
\bibinfo{author}{L.~He}, \bibinfo{author}{C.-Y. Kao},
  \bibinfo{author}{S.~Osher},
\newblock \bibinfo{title}{Incorporating topological derivatives into shape
  derivatives based level set methods},
\newblock \bibinfo{journal}{Journal of Computational Physics}
  \bibinfo{volume}{225} (\bibinfo{year}{2007}) \bibinfo{pages}{891--909}.
  \DOIprefix\doi{10.1016/j.jcp.2007.01.003}.
\bibitem[{Allen and Cahn(1979)}]{Allen1979}
\bibinfo{author}{S.~M. Allen}, \bibinfo{author}{J.~W. Cahn},
\newblock \bibinfo{title}{A microscopic theory for antiphase boundary motion
  and its application to antiphase domain coarsening},
\newblock \bibinfo{journal}{Acta Metallurgica} \bibinfo{volume}{27}
  (\bibinfo{year}{1979}) \bibinfo{pages}{1085--1095}.
  \DOIprefix\doi{10.1016/0001-6160(79)90196-2}.
\bibitem[{Cahn and Hilliard(1958)}]{Cahn1958}
\bibinfo{author}{J.~W. Cahn}, \bibinfo{author}{J.~E. Hilliard},
\newblock \bibinfo{title}{Free energy of a nonuniform system. i. interfacial
  free energy},
\newblock \bibinfo{journal}{The Journal of Chemical Physics}
  \bibinfo{volume}{28} (\bibinfo{year}{1958}) \bibinfo{pages}{258--267}.
  \DOIprefix\doi{10.1063/1.1744102}.
\bibitem[{Eyre(1993)}]{Eyre1993}
\bibinfo{author}{D.~J. Eyre},
\newblock \bibinfo{title}{Systems of cahn{\textendash}hilliard equations},
\newblock \bibinfo{journal}{{SIAM} Journal on Applied Mathematics}
  \bibinfo{volume}{53} (\bibinfo{year}{1993}) \bibinfo{pages}{1686--1712}.
  \DOIprefix\doi{10.1137/0153078}.
\bibitem[{Burger and Stainko(2006)}]{Burger2006}
\bibinfo{author}{M.~Burger}, \bibinfo{author}{R.~Stainko},
\newblock \bibinfo{title}{Phase-field relaxation of topology optimization with
  local stress constraints},
\newblock \bibinfo{journal}{{SIAM} Journal on Control and Optimization}
  \bibinfo{volume}{45} (\bibinfo{year}{2006}) \bibinfo{pages}{1447--1466}.
  \DOIprefix\doi{10.1137/05062723x}.
\bibitem[{Yamada et~al.(2010)Yamada, Izui, Nishiwaki, and
  Takezawa}]{Yamada2010}
\bibinfo{author}{T.~Yamada}, \bibinfo{author}{K.~Izui},
  \bibinfo{author}{S.~Nishiwaki}, \bibinfo{author}{A.~Takezawa},
\newblock \bibinfo{title}{A topology optimization method based on the level set
  method incorporating a fictitious interface energy},
\newblock \bibinfo{journal}{Computer Methods in Applied Mechanics and
  Engineering} \bibinfo{volume}{199} (\bibinfo{year}{2010})
  \bibinfo{pages}{2876--2891}. \DOIprefix\doi{10.1016/j.cma.2010.05.013}.
\bibitem[{Lim et~al.(2011)Lim, Yamada, Min, and Nishiwaki}]{Lim2011}
\bibinfo{author}{S.~Lim}, \bibinfo{author}{T.~Yamada},
  \bibinfo{author}{S.~Min}, \bibinfo{author}{S.~Nishiwaki},
\newblock \bibinfo{title}{Topology optimization of a magnetic actuator based on
  a level set and phase-field approach},
\newblock \bibinfo{journal}{{IEEE} Transactions on Magnetics}
  \bibinfo{volume}{47} (\bibinfo{year}{2011}) \bibinfo{pages}{1318--1321}.
  \DOIprefix\doi{10.1109/tmag.2010.2097583}.
\bibitem[{Yago et~al.(2020)Yago, Cante, Lloberas-Valls, and Oliver}]{Yago2020a}
\bibinfo{author}{D.~Yago}, \bibinfo{author}{J.~Cante},
  \bibinfo{author}{O.~Lloberas-Valls}, \bibinfo{author}{J.~Oliver},
\newblock \bibinfo{title}{Topology optimization of thermal problems in a
  nonsmooth variational setting: closed-form optimality criteria},
\newblock \bibinfo{journal}{Computational Mechanics} \bibinfo{volume}{66}
  (\bibinfo{year}{2020}) \bibinfo{pages}{259--286}.
  \DOIprefix\doi{10.1007/s00466-020-01850-0}.
\bibitem[{Rozvany(2001)}]{Rozvany2001}
\bibinfo{author}{G.~Rozvany},
\newblock \bibinfo{title}{Aims, scope, methods, history and unified terminology
  of computer-aided topology optimization in structural mechanics},
\newblock \bibinfo{journal}{Structural and Multidisciplinary Optimization}
  \bibinfo{volume}{21} (\bibinfo{year}{2001}) \bibinfo{pages}{90--108}.
  \DOIprefix\doi{10.1007/s001580050174}.
\bibitem[{Andreassen et~al.(2010)Andreassen, Clausen, Schevenels, Lazarov, and
  Sigmund}]{Andreassen2010}
\bibinfo{author}{E.~Andreassen}, \bibinfo{author}{A.~Clausen},
  \bibinfo{author}{M.~Schevenels}, \bibinfo{author}{B.~S. Lazarov},
  \bibinfo{author}{O.~Sigmund},
\newblock \bibinfo{title}{Efficient topology optimization in {MATLAB} using 88
  lines of code},
\newblock \bibinfo{journal}{Structural and Multidisciplinary Optimization}
  \bibinfo{volume}{43} (\bibinfo{year}{2010}) \bibinfo{pages}{1--16}.
  \DOIprefix\doi{10.1007/s00158-010-0594-7}.
\bibitem[{Huang et~al.(2014)Huang, Li, Zhou, and Xie}]{Huang2014}
\bibinfo{author}{X.~Huang}, \bibinfo{author}{Y.~Li}, \bibinfo{author}{S.~Zhou},
  \bibinfo{author}{Y.~Xie},
\newblock \bibinfo{title}{Topology optimization of compliant mechanisms with
  desired structural stiffness},
\newblock \bibinfo{journal}{Engineering Structures} \bibinfo{volume}{79}
  (\bibinfo{year}{2014}) \bibinfo{pages}{13--21}.
  \DOIprefix\doi{10.1016/j.engstruct.2014.08.008}.
\bibitem[{Aage et~al.(2014)Aage, Andreassen, and Lazarov}]{Aage2014}
\bibinfo{author}{N.~Aage}, \bibinfo{author}{E.~Andreassen},
  \bibinfo{author}{B.~S. Lazarov},
\newblock \bibinfo{title}{Topology optimization using {PETSc}: An easy-to-use,
  fully parallel, open source topology optimization framework},
\newblock \bibinfo{journal}{Structural and Multidisciplinary Optimization}
  \bibinfo{volume}{51} (\bibinfo{year}{2014}) \bibinfo{pages}{565--572}.
  \DOIprefix\doi{10.1007/s00158-014-1157-0}.
\bibitem[{Liu and Tovar(2014)}]{Liu2014}
\bibinfo{author}{K.~Liu}, \bibinfo{author}{A.~Tovar},
\newblock \bibinfo{title}{An efficient 3d topology optimization code written in
  matlab},
\newblock \bibinfo{journal}{Structural and Multidisciplinary Optimization}
  \bibinfo{volume}{50} (\bibinfo{year}{2014}) \bibinfo{pages}{1175--1196}.
  \DOIprefix\doi{10.1007/s00158-014-1107-x}.
\bibitem[{Sigmund(2001)}]{Sigmund2001}
\bibinfo{author}{O.~Sigmund},
\newblock \bibinfo{title}{A 99 line topology optimization code written in
  matlab},
\newblock \bibinfo{journal}{Structural and Multidisciplinary Optimization}
  \bibinfo{volume}{21} (\bibinfo{year}{2001}) \bibinfo{pages}{120--127}.
  \DOIprefix\doi{10.1007/s001580050176}.
\bibitem[{{TOPOPT Group}(2018)}]{TOPOPT2018}
\bibinfo{author}{{TOPOPT Group}}, \bibinfo{title}{Efficient topology
  optimization in matlab using 88 lines of code},
  \bibinfo{howpublished}{Webpage}, \bibinfo{year}{2018}. \URLprefix
  \url{http://www.topopt.mek.dtu.dk/Apps-and-software/Efficient-topology-optimization-in-MATLAB},
  \bibinfo{note}{accessed: 2020-04-28}.

\end{thebibliography}

\end{document}